\documentclass[numberedappendix,natbib209]{emulateapj}
\usepackage{longtable}
\usepackage{natbib}

\begin{document}

\title{The Evolution of the Stellar Mass Functions of Star-Forming and Quiescent Galaxies to \lowercase{z} $=$ 4 from the COSMOS/UltraVISTA Survey\altaffilmark{1}}
\author{Adam Muzzin\altaffilmark{2}, Danilo Marchesini\altaffilmark{3}, Mauro Stefanon\altaffilmark{4}, Marijn Franx\altaffilmark{2}, Henry J. McCracken\altaffilmark{5}, Bo Milvang-Jensen\altaffilmark{6}, James S. Dunlop\altaffilmark{7}, J. P. U. Fynbo\altaffilmark{6}, Gabriel Brammer\altaffilmark{8}, Ivo Labb\'{e}\altaffilmark{2}, Pieter G. van Dokkum\altaffilmark{9}}

\altaffiltext{1}{Based on data products from observations made with ESO Telescopes at the La Silla Paranal Observatory under ESO programme ID 179.A-2005 and on data products produced by TERAPIX and the Cambridge Astronomy Survey Unit on behalf of the UltraVISTA consortium.}
\altaffiltext{2}{Leiden Observatory, Leiden University, PO Box 9513,
  2300 RA Leiden, The Netherlands}
\altaffiltext{3}{Department of Physics and Astronomy, Tufts University, Medford, MA 06520, USA}
\altaffiltext{4}{Physics and Astronomy Department, University of Missouri, Columbia, MO 65211}
\altaffiltext{5}{Institut d'Astrophysique de Paris, UMR7095 CNRS, Universit\'{e} Pierre et Marie Curie, 98 bis Boulevard Arago, 75014 Paris, France}
\altaffiltext{6}{Dark Cosmology Centre, Niels Bohr Institute, University of Copenhagen, Juliane Maries Vej 30, 2100 Copenhagen, Denmark}
\altaffiltext{7}{SUPA, Institute for Astronomy, University of Edinburgh, Royal Observatory, Edinburgh EH9 3HJ, UK}
\altaffiltext{8}{European Southern Observatory, Alonso de C\'{o}rdova 3107, Casilla 19001, Vitacura, Santiago, Chile} 
\altaffiltext{9}{Department of Astronomy, Yale University, New Haven, CT, 06520-8101} 
\begin{abstract}
We present measurements of the stellar mass functions (SMFs) of star-forming and quiescent galaxies to $z$ = 4 using a sample of 95 675 galaxies in the COSMOS/UltraVISTA field.  Sources have been selected from the DR1 UltraVISTA K$_{s}$-band imaging which covers a unique combination of a wide area (1.62 deg$^2$), to a significant depth (K$_{s,tot}$ $=$ 23.4, 90\% completeness).  The SMFs of the combined population are in good agreement with previous measurements and show that the stellar mass density of the universe was only 50\%, 10\% and 1\% of its current value at $z \sim$ 0.75, 2.0, and 3.5, respectively.  The quiescent population drives most of the overall growth, with the stellar mass density of these galaxies increasing as $\rho_{star}$ $\propto$ (1 + $z$)$^{-4.7\pm0.4}$ since $z =$ 3.5, whereas the mass density of star-forming galaxies increases as $\rho_{star}$ $\propto$ (1 + $z$)$^{-2.3\pm0.2}$.  At $z >$ 2.5, star-forming galaxies dominate the total SMF at all stellar masses, although a nonzero population of quiescent galaxies persists to $z = $ 4.  Comparisons of the K$_{s}$-selected star-forming galaxy SMFs to UV-selected SMFs at 2.5 $< z < $ 4 show reasonable agreement and suggests UV-selected samples are representative of the majority of the stellar mass density at $z >$ 3.5.  We estimate the average mass growth of individual galaxies by selecting galaxies at fixed cumulative number density.  The average galaxy with Log(M$_{*}$/M$_{\odot}$) = 11.5 at $z =$ 0.3 has grown in mass by only 0.2 dex (0.3 dex) since $z =$ 2.0(3.5), whereas those with Log(M$_{*}$/M$_{\odot}$) = 10.5 have grown by $>$ 1.0 dex since $z = 2$.  At $z <$ 2, the time derivatives of the mass growth are always larger for lower-mass galaxies, which demonstrates that the mass growth in galaxies since that redshift is mass-dependent and primarily bottom-up.  Lastly, we examine potential sources of systematic uncertainties on the SMFs and find that those from photo-$z$ templates, SPS modeling, and the definition of quiescent galaxies dominate the total error budget in the SMFs.  
\end{abstract}
\section{Introduction}
In the current $\Lambda$CDM paradigm, the dominant structures in the universe are dark matter halos which grow out of an initial field of density perturbations via gravitational collapse \citep{White1978}.  Simulations and analytical models show that this process proceeds primarily in a hierarchical, bottom-up manner, with low-mass halos forming early and subsequently growing via continued accretion and merging to form more massive halos at later times \citep{White1991,Kauffmann1993,Kauffmann1999}.  
\newline\indent
In contrast to the predicted hierarchical growth of the dark matter halos, observational studies suggest that the stellar baryonic component of the halos (i.e., galaxies), may grow in an anti-hierarchical, top-down manner.  It appears that many of the most massive galaxies (Log(M$_{star}$/M$_{\odot}$) $>$ 11) in the local universe assembled their stellar mass rapidly and at early times ($z >$ 2), whereas lower-mass galaxies grew more gradually over cosmic time \citep[e.g.,][]{Marchesini2009,Marchesini2010,Ilbert2010,Caputi2011,Brammer2011}.
\newline\indent
Understanding these apparently contrasting evolutionary paths for the dark matter assembly and stellar mass assembly of galaxies is a significant challenge for current models of galaxy formation \citep[e.g.,][]{Marchesini2009,Fontanot2009}.  
In particular, the differential evolution between the baryonic and non-baryonic components of galaxies makes it clear that the baryonic physics of galaxy formation must be more complex than the cooling of gas onto halos at a rate dictated by gravity.  Indeed, it implies that there is a tenuous balance between gas accretion rates, gas consumption rates (in both star formation events and black hole growth), mergers, as well as feedback processes such as AGN activity, supernovae, or stellar winds.  It is also clear that the efficiency of these processes must scale with halo mass and evolve with redshift \citep[e.g.,][]{Schaye2010,Weinmann2012,Henriques2013}.  Given the complex, non-linear interplay between these processes, and the various possible prescriptions of implementing them within models, it is important to have a benchmark for the models so that we can evaluate if progress is being made.
\newline\indent
For cosmological simulations, the benchmark that has been most widely adopted is the ability of models to match the volume density of galaxies as function of their stellar mass, also known as the stellar mass function (hereafter SMF).  If a model can reproduce the SMFs at various redshifts it suggests (although does not prove) that it may be a better description of the baryonic physics of galaxy formation than those that do not.  Given that it is a key benchmark for models, the most precise and accurate measurements of the SMF possible over as large a range in redshift and stellar mass are valuable quantities.
\newline\indent
In recent years, with the growth of deep and wide-field near infrared (NIR) imaging surveys, there have been myriad measurements of the evolution of the SMFs from the local universe \citep[e.g.,][]{Cole2001,Bell2003,Li2009,Baldry2012} up to $z =$ 2 -- 5 \citep[e.g.,][]{Drory2005,Bundy2006,Pozzetti2007,Arnouts2007,Perezgonzalez2008,Drory2009,Marchesini2009,Marchesini2010,Ilbert2010,Pozzetti2010,Dominguezsanchez2011,Bielby2012,Moustakas2013,Ilbert2013}.  In this paper we present an improved measurement of the SMF of galaxies over the redshift range 0.2 $< z <$ 4.0.  These measurements are made from a new K$_{s}$-selected catalog of the COSMOS field which uses data from the DR1 UltraVISTA survey \citep[see][]{McCracken2012}.  The UltraVISTA catalog is unique in its combination of covering a wide area (1.62 deg$^2$), to a relatively deep depth (K$_{s,tot}$ $<$ 23.4, 90\% completeness).  This combination allows the most accurate measurement of the high mass end of the SMFs up to $z =$ 4.0 to date.  Details of the catalog and a public release of all catalog data products are presented in a companion paper by \cite{Muzzin2013a}.
\newline\indent
We note that an independent analysis of the SMFs out to $z =$ 4 using the UltraVISTA data has also recently been performed by \cite{Ilbert2013}.  That analysis is based on a different catalog than that of the \cite{Muzzin2013a}, and uses different photometric redshift and stellar mass fitting techniques.  In an appendix we make a more detailed comparison between our SMFs and those derived by \cite{Ilbert2013}.
\newline\indent
The layout of this paper is as follows.  In $\S$ 2 we present details of the COSMOS/UltraVISTA dataset and discuss the stellar mass and photometric redshift measurements.  In $\S$ 3 we detail how the SMFs and the uncertainties are calculated.  In $\S$ 4 we derive the SMFs of star forming and quiescent galaxies and the stellar mass density and number density evolution up to $z =$ 4.  In $\S$ 5 we present a discussion of our results, including a comparison to UV-selected SMFs at $z >$ 3 and an estimation of the typical mass growth of galaxies using a fixed cumulative number density approach.  We conclude in $\S$ 6 with a summary of our results.  In an appendix we present a detailed look at possible sources of systematic error and their effect on the derived SMFs.  Throughout this paper we assume a $\Omega_{\Lambda}$ = 0.7, $\Omega_{m}$ = 0.3, and H$_{0}$ = 70 km s$^{-1}$ Mpc$^{-1}$ cosmology.  All magnitudes are in the AB system.  
\begin{figure*}
\plotone{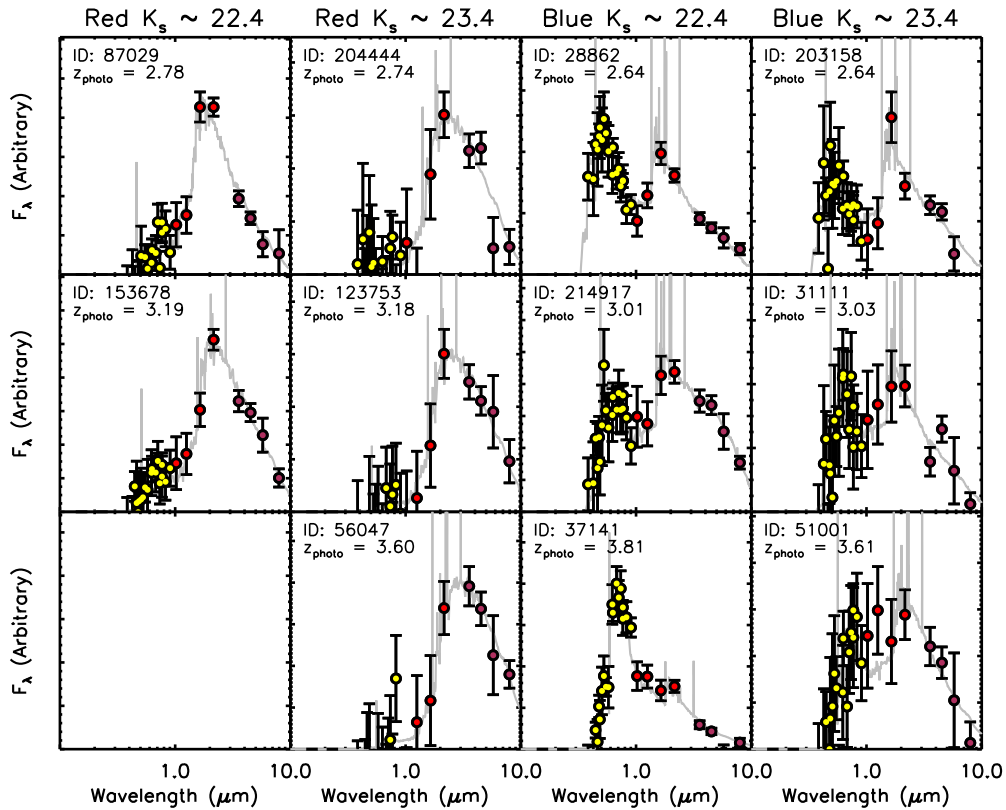}
\caption{\footnotesize Example SEDs from EAZY of red and blue galaxies in three redshift ranges: 2.5 $< z <$ 3.0 (top row), 3.0 $< z <$ 3.5 (middle row), 3.5 $< z <$ 4.0 (bottom row).  The second and fourth columns show galaxies that have magnitudes near the limiting magnitude of the SMFs (K$_{s}$ $\sim$ 23.4), and the first and third columns show galaxies that are $\sim$ 1 mag brighter.  There are no red galaxies with K$_{s}$ $\sim$ 22.4 and 3.5 $< z <$ 4.0.  The SEDs of galaxies at K$_{s}$ $=$ 23.4 limit typically have S/N $\sim$ 5 and therefore we have limited the SMFs to a limiting M$_{star}$ that corresponds to this limit. }
\end{figure*}
\section{The Dataset}
\indent
This study is based on a K$_{s}$-selected catalog of the COSMOS/UltraVISTA field from \cite{Muzzin2013a}.  The catalog contains PSF-matched photometry in 30 photometric bands covering the wavelength range 0.15$\micron$ $\rightarrow$ 24$\micron$ and includes the available $GALEX$ \citep{Martin2005}, CFHT/Subaru \citep{Capak2007}, UltraVISTA \citep{McCracken2012}, and S-COSMOS \citep{Sanders2007} datasets.  Sources are selected from the DR1 UltraVISTA K$_{s}$-band imaging \citep{McCracken2012} which reaches a depth of K$_{s,tot}$ $<$ 23.4 at 90\% completeness.  A detailed description of the photometric catalog construction, photometric redshift ($z_{phot}$) measurements, and stellar mass (hereafter, M$_{star}$) estimates is presented in \cite{Muzzin2013a}.  A public release of all data products from the catalog is also presented with that paper.  Here we briefly describe the aspects of the catalog relevant to the measurement of the SMFs.  
\subsection{Photometric Redshifts and Stellar Masses}
Each galaxy in the catalog has a $z_{phot}$ determined by fitting the photometry in the 0.15$\micron$ $\rightarrow$ 8.0$\micron$ bands to template Spectral Energy Distributions (SEDs) using the EAZY code \citep{Brammer2008}.  In default mode, EAZY fits photometric redshifts using linear combinations of 6 templates from the PEGASE models \citep{Fioc1999} as well as an additional red template from the \cite{Maraston2005} models.  In order to improve the accuracy of the $z_{phot}$ for high-redshift galaxies we added two new templates to the default set, a $\sim$ 1 Gyr old poststarburst template, as well as a slightly dust-reddened Lyman Break template \citep[see][]{Muzzin2013a}.  Comparison of the $z_{phot}$ to 5100 spectroscopic redshifts from the zCOSMOS-bright 10k sample \citep{Lilly2007}, as well as 19 spectroscopic redshifts for red galaxies at $z >$ 1 \citep{vandesande2011,Onodera2012,Bezanson2013,vandeSande2013} shows that the $z_{phot}$ have an rms dispersion of $\delta$$z$/(1 + $z$) = 0.013 and a $>$ 3$\sigma$ catastrophic outlier fraction of 1.6\%.
\newline\indent
Stellar masses for all galaxies have been determined by fitting the SEDs of galaxies to stellar population synthesis (SPS) models using the FAST code \citep{Kriek2009}.  It is well-known that the M$_{star}$ derived from SED fitting depends on the assumptions made (metallicity, SPS model, dust law, IMF) in this process \citep[e.g.,][]{Marchesini2009,Muzzin2009b,Muzzin2009c,Conroy2009}.  These assumptions typically result in systematic changes to the SMFs, rather than larger random errors \citep[e.g.,][]{Marchesini2009}.  Given the complexity of these systematic dependencies, in this paper we base the majority of the analysis on a default set of assumptions for the SED modeling and then in the Appendix we expand the range of SED modeling parameter space and explore their effects on the SMFs.  In the Appendix we also explore the effects of expanding the EAZY template set to include an old-and-dusty template, which provides a good fit for some of the bright high-redshift population \citep[see also,][]{Marchesini2010}.
\newline\indent
For the default set of M$_{star}$ we fit the SEDs to a set of models with exponentially-declining star formation histories of the form SFR $\propto$ $e^{-t/\tau}$, where $t$ is the time since the onset of star formation, and $\tau$ sets the timescale of the decline in the SFR.  We use the models of \cite{Bruzual2003}, hereafter BC03, with solar metallicity, a \cite{Calzetti2000} dust law, and assume a \cite{Kroupa2001} IMF\footnote{The stellar masses in the \cite{Muzzin2013a} catalog are computed with a \cite{Chabrier2003} IMF.  For easy comparison with the literature we have converted them to a \cite{Kroupa2001} IMF by increasing them by 0.04 dex}.  We allow log($\tau$/Gyr) to range between 7.0 and 10.0 Gyr, log($t$/Gyr) between 7.0 and 10.1 Gyr, and A$_{v}$ between 0 and 4.  The maximum allowed age of galaxies is set by the age of the universe at their $z_{phot}$.  Further details on the default model set and the fitting process are discussed in \cite{Muzzin2013a}.
\begin{figure}
\plotone{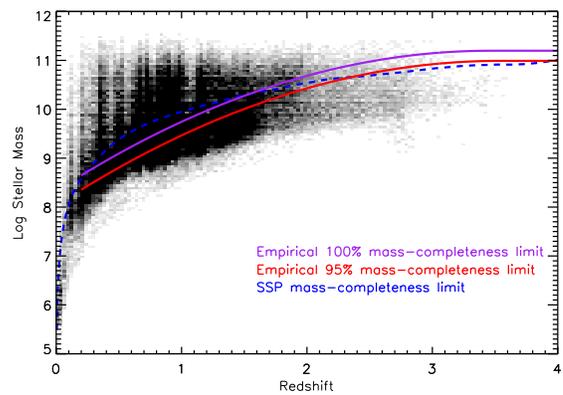}
\caption{\footnotesize Grayscale representation of the density of galaxy stellar masses as a function of redshift in the K$_{s}$-selected catalog.  The 100\% and 95\% mass-completeness limits determined using the deeper datasets are shown as the purple and red curves, respectively.  Also shown is the 100\% completeness limit for an SSP formed at $z =$ 10.}
\end{figure}
\begin{figure*}
\plotone{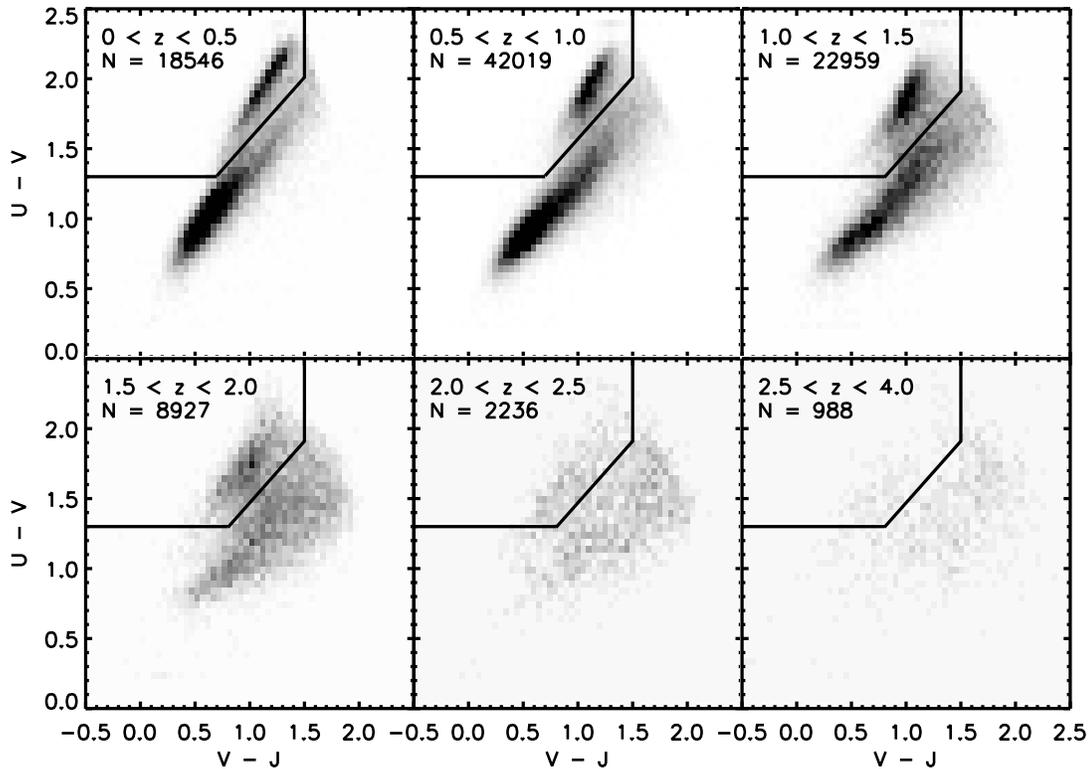}
\caption{\footnotesize UVJ color-color diagram at various redshifts for galaxies more massive than the 95\% mass-completeness limits.  The bimodality in the galaxy population is clearly visible up to $z =$ 2.  The cuts used to separate star forming from quiescent galaxies for the SMFs are shown as the solid lines.}
\end{figure*}
\section{Construction of the Stellar Mass Functions}
\indent
Here we outline how the SMFs for the quiescent, star forming, and combined populations are constructed.  
\subsection{Galaxy Sample and Completeness}
The \cite{Muzzin2013a} catalog contains a total of 262 615 objects down to a 3$\sigma$ limit of K$_{s}$ $<$ 24.35 in a 2.1$^{\prime\prime}$ aperture.  From that parent sample we define a mass-complete sample for computing the SMFs by applying various cuts to the catalog.  
\newline\indent
Simulations of the catalog completeness \citep[see][, Figure 4]{Muzzin2013a}, show that the 90\% point-source completeness limit in total magnitudes for the UltraVISTA data is K$_{s,tot}$ $=$ 23.4 after the blending of sources is accounted for.  This limit in K$_{s,tot}$ also corresponds to the $\sim$ 5$\sigma$ limit for the photometry in the 2.1$^{\prime\prime}$ color aperture, and therefore is a sensible limiting magnitude for computing the SMFs. 
\newline\indent
As a demonstration of the quality of the SEDs near the 90\% completeness limit, in Figure 1 we plot some randomly-chosen examples of red and blue galaxy SEDs in three redshift bins: 2.5 $< z <$ 3.0 (top row), 3.0 $< z <$ 3.5 (middle row), 3.5 $< z <$ 4.0 (bottom row).  We plot SEDs of galaxies that have fluxes near the 90\% completeness limit (K$_{s,tot}$ $\sim$ 23.4), as well as SEDs of galaxies that are $\sim$ 1 magnitude brighter (K$_{s,tot}$ $\sim$ 22.4).  Figure 1 shows that the SEDs of both red and blue galaxies at K$_{s,tot}$ $\sim$ 22.4 are very well constrained.  It also shows that at K$_{s,tot}$ $\sim$ 23.4, the SEDs are also reasonably well-constrained; however, the typical S/N in a 2.1$^{\prime\prime}$ aperture is $\sim$ 5.  
\newline\indent
It is possible to include galaxies fainter than the 90\% K$_{s,tot}$ completeness limit in the SMFs and correct for this incompleteness; however, given that the quality of the SEDs near K$_{s,tot}$ $\sim$ 23.4 becomes marginal, we have chosen to restrict the sample to galaxies with good S/N photometry.  This ensures that all galaxies included in the SMFs have reasonable well-determined M$_{star}$ and $z_{phot}$.  
\newline\indent
When constructing the SMFs we also exclude objects flagged as stars (\texttt{star} = 1) based on a color-color cut, as well as those with badly contaminated photometry (SExtractor flag \texttt{K\_flag} $>$ 4).  Objects nearby very bright stars (\texttt{contamination} = 1) or bad regions (\texttt{nan\_contam} $>$ 3) are also excluded, and the reduction in area from these effects is taken account of in the total survey volume.
\newline\indent
Once these cuts are applied, the final sample of galaxies available for the analysis is 160 070.  In Figure 2 we plot a grayscale representation of the M$_{star}$ of this sample as a function of $z_{phot}$.   In general, the sample is dominated by objects at $z <$ 2; however, there are reliable sources out to $z =$ 4.
\newline\indent
\subsection{Stellar Mass Completeness vs. z}
\indent
Figure~2 shows the $M_{star}$ of galaxies down to 90\% K$_{s}$-band 
completeness limit of the survey; however, in order to construct the SMFs, the 
limiting $M_{star}$ above which the magnitude-limited sample is complete 
needs to be determined. In order to estimate the redshift-dependent 
completeness limit in M$_{star}$ we adopt the approach developed in 
\cite{Marchesini2009}, which exploits the availability of other survey 
data that are deeper than UltraVISTA. Specifically, we employed the 
$K$-selected FIRES \citep{Labbe2003,Forsterschreiber2006} and 
the FIREWORKS \citep{Wuyts2008} catalogs, already used in \cite{Marchesini2009,Marchesini2010}, and the $H_{\rm 160}$-selected catalogs over the Hubble 
Ultra-Deep Field (H-UDF) used in \cite{Marchesini2012}. The FIRES-HDFS, 
FIRES-MS1054, FIREWORKS, and HUDF reach limiting magnitudes of 
$K_{\rm S,tot}=25.6$, 24.1, 23.7, and 25.6, respectively.  The M$_{star}$ in those catalogs has been calculated using the same SED modeling assumptions as in the UltraVISTA catalog.
\newline\indent
Briefly, to estimate the redshift-dependent stellar mass completeness limit 
of the UltraVISTA sample at $K_{s,tot}=23.4$, we first selected galaxies 
belonging to the available deeper samples.  We then scaled their fluxes and 
M$_{star}$ to match the $K$-band completeness limit of the UltraVISTA 
sample. The upper envelope of points in the ($M_{star,scaled}$ -- $z$) space, 
encompassing 100\% of the points, represents the most massive galaxies at K$_{s}$ = 23.4, 
and so provides a redshift-dependent M$_{star}$ completeness limit 
for the UltraVISTA sample. We refer to \cite{Marchesini2009} for a more detailed 
description of this method. In Figure~2 we show this empirically-derived 
100\% mass-completeness limit as the purple curve. Also, for reference we show 
the mass-completeness limit for a simple stellar population (SSP) formed at 
$z=10$ which is extreme but indicative of a maximally-old population.
\newline\indent
Figure 2 shows that for galaxies at $z <$ 1.5 and K$_{s,tot}$ $\sim$ 23.4, the most extreme M/L ratios are less extreme than an SSP.  Around $z =$ 1.5 the SSP curve and the empirical 100\% completeness curve cross each other which implies that there exist galaxies that have larger M/L ratios than an SSP.  Such galaxies are typically galaxies with intermediate-to-old ages (for their redshift) with up to several magnitudes of dust extinction.  More detailed SED modeling for galaxies with spectroscopic redshifts shows that these dusty and old galaxies are not uncommon among the massive galaxy population at $z >$ 1.5 \citep[see e.g.,][]{Kriek2006,Kriek2008,Muzzin2009b}.  
\newline\indent
As Figure 2 shows, the empirically-derived 100\% mass-completeness limits are high due to the old and dusty population.  Adopting the empirical 100\% completeness limit for the SMFs therefore requires the exclusion of 57\% of the magnitude-limited sample.  
\newline\indent
The 100\% completeness limit is set by most extreme M/L ratio at any given redshift, regardless of the frequency of its occurrence.  In principle, if only a small fraction of objects have these extreme M/L ratios, then adopting the 100\% mass-completeness limit is an inefficient use of the data.  In order to try to make better use of the dataset we also derived 95\% mass-completeness limits for the sample and this limit is also plotted in Figure 2.  
\newline\indent
At all redshifts the 95\% mass-completeness limits are 0.2 -- 0.3 dex lower showing that it is only a small fraction of the overall population of galaxies that have extreme M/L ratios.  If we adopt the 100\% mass-completeness limits, the resulting sample of galaxies is 67 942.  Adopting the 95\% mass-completeness limits increases the sample by a factor of 1.4 to 95 675 galaxies.  Given this substantial increase in statistics, and the advantage gained by probing further down the SMFs at higher redshift, we have adopted the 95\% mass-completeness limits for the SMFs, but correct the lowest-mass bin in each SMF by 5\% in order to account for this.
\subsection{Separation of Quiescent and Star-Forming Galaxies}
\indent
It is well-known that the overall galaxy population is bi-modal in the distribution of colors and SFRs \citep[e.g.,][]{Kauffmann2003,Hogg2004,Balogh2004,Blanton2009} and that this bi-modality persists out to high redshift \citep[e.g.,][]{Bell2004,Taylor2009,Williams2009,Brammer2009,Brammer2011}.  Given the bi-modality, separating the evolution of the SMFs of star-forming and quiescent galaxies as a function of redshift is useful for understanding the relationship between the two populations. 
\newline\indent
In recent years, several methods have been developed to classify galaxies into these categories.  In this analysis we perform classification between the types using the rest-frame U - V vs. V - J color-color diagram (hereafter the UVJ diagram).  The UVJ classification has been used in many previous studies \citep[e.g.,][]{Labbe2005,Wuyts2007,Williams2009,Brammer2011,Patel2012a}.  These previous studies have shown that separation of star-forming and quiescent galaxies in this color-color space is well-correlated with separation using UV+IR determined SSFRs \citep[e.g.,][]{Williams2009}, and SED-fitting determined SSFRs \citep[e.g.,][]{Williams2010} up to $z =$ 2.5.  Separation in this color space is also correlated with the detection and non-detection of galaxies at 24$\micron$, down to implied SFRs of $\sim$ 40 M$_{\odot}$ yr$^{-1}$ at $z \sim$ 2 \citep{Wuyts2007,Brammer2011}.  We choose to separate galaxies based on a rest-frame color-cut as opposed to a cut in a derived quantity such as specific star formation rate (SSFR), because rest-frame colors can be calculated in a straightforward way for each galaxy in the sample.  UV+IR SFRs can only be calculated for the most strongly star-forming galaxies at high-redshift due to the limited depth of the 24$\micron$ data. 
\newline\indent
In Figure 3, we plot the U - V vs. V - J diagram for galaxies more massive than the 95\% mass-completeness limits in several redshift bins.  The galaxy bi-modality is clearly visible in the UVJ diagram up to $z =$ 2, but thereafter becomes less pronounced at the M$_{star}$ completeness limits probed by the K$_{s}$-selected UltraVISTA catalog.
\newline\indent
To distinguish between star forming and quiescent galaxies we use box regions in the UVJ diagram that are similar, although not identical to those defined in \cite{Williams2009, Whitaker2011} and \cite{Brammer2011}.  These regions are plotted as the solid lines in Figure 3.  Quiescent galaxies are defined as,
\begin{eqnarray}
U - V > 1.3, V - J < 1.5, \mbox{[all redshifts]}\\
U - V > (V - J)\times0.88 + 0.69, \mbox{[0.0} < z < \mbox{1.0]} \\
U - V > (V - J)\times0.88 + 0.59, \mbox{[1.0} < z < \mbox{4.0]}
\end{eqnarray} 
We note that these boxes are chosen arbitrarily, with the main criteria being that they lie roughly between the two modes of the population seen in Figure 3.  They were originally defined by \cite{Williams2009} who defined them in such a way to maximize the difference in SSFRs between the regions; however, our rest-frame color distribution is slightly different than \cite{Williams2009} which is the reason that we have adjusted the box locations.  In the appendix we explore the effect on the SMFs of moving the location of the boxes in UVJ space.  In general, we find that changing the UVJ box has little effect on the high-mass end of the quiescent SMF, and the low-mass end of the star-forming SMF because those galaxies are dominated by very red and very blue galaxies, respectively.  It does have a larger effect on the SMF for intermediate mass galaxies, which is not unexpected given that these are typically the transition population at most redshifts.
\subsection{Stellar Mass Function Construction and Fitting}
With well-defined mass-completeness limits as a function of redshift and criteria for separating star-forming and quiescent galaxies, SMFs can now be computed.  We employ two methods to determine the SMFs, the 1/V$_{max}$ method and a maximum-likelihood method.  These methods have different strengths and weaknesses. The 1/V$_{max}$ method has the advantage that it does not assume a parametric form of the SMF, allowing a direct visualization of the data; however, it is a fully normalized solution and is susceptible to the effects of clustering.  Conversely, the maximum-likelihood method has the advantage that it is not affected by density inhomogeneities \citep[e.g.,][]{Efstathiou1988}; however, it does assume a functional form for the fit.
\subsubsection{The 1/V$_{max}$ Method}
To measure the SMFs for the sample we have applied an extended version of the 1/V$_{max}$ algorithm \citep{Schmidt1968} as defined in \cite{Avni1980}.  The method has been used to determine the rest-frame optical luminosity functions and SMFs by \cite{Marchesini2007,Marchesini2009,Marchesini2010,Marchesini2012} and we refer to those papers for an in-detail description of the method.  
\newline\indent
In brief, for each M$_{star}$ we determine the maximum volume within which an object of that M$_{star}$ could be detected.  This volume is determined as a function of M$_{star}$ using the maximum redshift that the survey is complete for objects of that M$_{star}$.  The SMF is then calculated by counting galaxies in bins of M$_{star}$ and correcting those bins with 1/V$_{max}$.  Poisson error bars are determined for each bin using the prescription of \cite{Gehrels1986}, which is valid for small number statistics.
\begin{figure}
\plotone{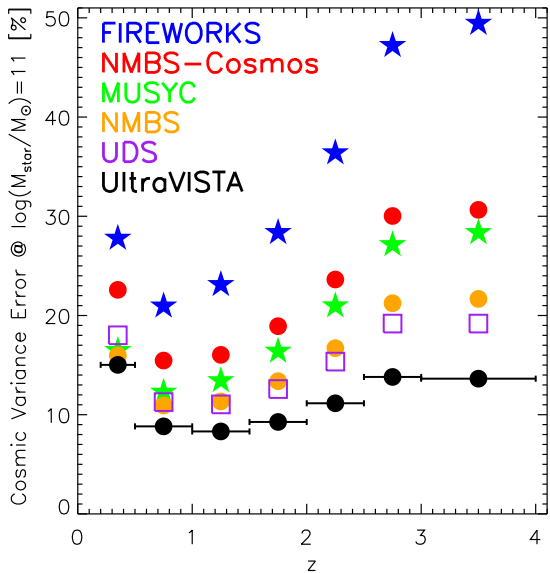}
\caption{\footnotesize Uncertainty in the number density of galaxies with Log(M$_{star}$/M$_{\odot}$) $=$ 11.0 due to cosmic variance as as a function of redshift calculated using the prescription of \cite{Moster2011}.  Other surveys with smaller areas but also more independent sight lines are shown for comparison (see text for details).  The uncertainties in UltraVISTA due to cosmic variance are $\sim$ 8 - 15\% at Log(M$_{star}$/M$_{\odot}$) $=$ 11.0 over the full redshift range.}
\end{figure}
\subsubsection{The Maximum Likelihood Method} 
The SMFs are also determined using the maximum-likelihood method outlined by \cite{Sandage1979}.  For this method it is assumed that the number density of galaxies ($\Phi$(M$_{star}$)) is described by a \cite{Schechter1976} function of the form,
\begin{equation}
\Phi(M) = (ln 10)\Phi^{*}[10^{(M - M^{*})(1+\alpha)}] \times exp[-10^{(M - M^{*})}],
\end{equation}
where M = log(M$_{star}$/M$_{\odot}$), $\alpha$ is the low-mass-end slope, M$^{*}$ = log(M$^{*}_{star}$/M$_{\odot}$) is the characteristic mass, and $\Phi^{*}$ is the normalization.  For each possible combination of $\alpha$ and M$^{*}_{star}$ the likelihood that each galaxy would be found in the survey is calculated.   The best-fit solution for $\alpha$ and M$^{*}_{star}$ in each redshift bin is obtained by maximizing the combined likelihoods of all galaxies ($\Lambda$) with respect to these parameters.  The $\Phi^{*}$ is determined by requiring that the total number of observed galaxies is reproduced.  The errors in $\Phi^{*}$ are then determined from the minimum and maximum values of $\Phi^{*}$ allowed by the confidence contours in the $\alpha$ vs. M$^{*}_{star}$ plane.  Further details of the fitting process can be found in \cite{Marchesini2007,Marchesini2009}.
\subsubsection{The Low-Mass-End Slope $\alpha$}
The SMFs are computed over a large redshift range, and as was shown in Figure 2, the limiting 95\% completeness limit in M$_{star}$ is a strong function of redshift.  The SMFs reach $\sim$ 1.5 dex deeper than M$^{*}_{star}$ at $z <$ 0.5, but only to $\sim$ M$^{*}_{star}$ itself at $z =$ 3.5.  This means that $\alpha$ is well constrained at $z \leq$ 2, but is poorly constrained at $z \geq$ 2.  Given the well-known correlation between $\alpha$ and M$^{*}_{star}$, it is important to be aware that the true uncertainties in quantities such as M$^{*}_{star}$, or the stellar mass density can be systematically larger than the random uncertainties due to the data not reaching a mass limit sufficiently low to constrain $\alpha$.  In order to quantify the uncertainties in all parameters better, in all redshift ranges we have performed the Schechter function fits both with $\alpha$ as a free parameter, and fixing it to a known value.  
\newline\indent
In the fits with $\alpha$ held fixed, we have chosen values of $\alpha$ = $-$1.2, $-$0.4, and $-$1.3 for the total, quiescent, and star forming populations, respectively.  As discussed in $\S$ 4, these values are similar to those derived at $z <$ 1  when $\alpha$ is fit as a free parameter.  They are also consistent with values at $z >$ 1 derived from studies that probe the low-mass end better than UltraVISTA \citep[e.g.,][]{Fontana2006,Perezgonzalez2008,Marchesini2009,Stark2009,Lee2012}.  
\newline\indent
In addition to fits with $\alpha$ fixed and free, we have also performed fits to a ``double" Schechter function.  Several recent studies have shown that at low-redshift, the low-mass end of the SMF (Log(M$_{star}$/M$_{\odot}$) $<$ 9.5) is better described by the sum of two Schechter functions with identical M$^{*}_{star}$, but different $\Phi^{*}$ and $\alpha$ \citep[e.g.,][]{Li2009,Baldry2012}.  The UltraVISTA SMFs reach the limiting M$_{star}$ where a clear departure from a single Schechter function fit is seen at $z <$ 1, and so for the SMFs in that redshift range fits are also performed with a double Schechter function.  
\begin{figure*}
\plotone{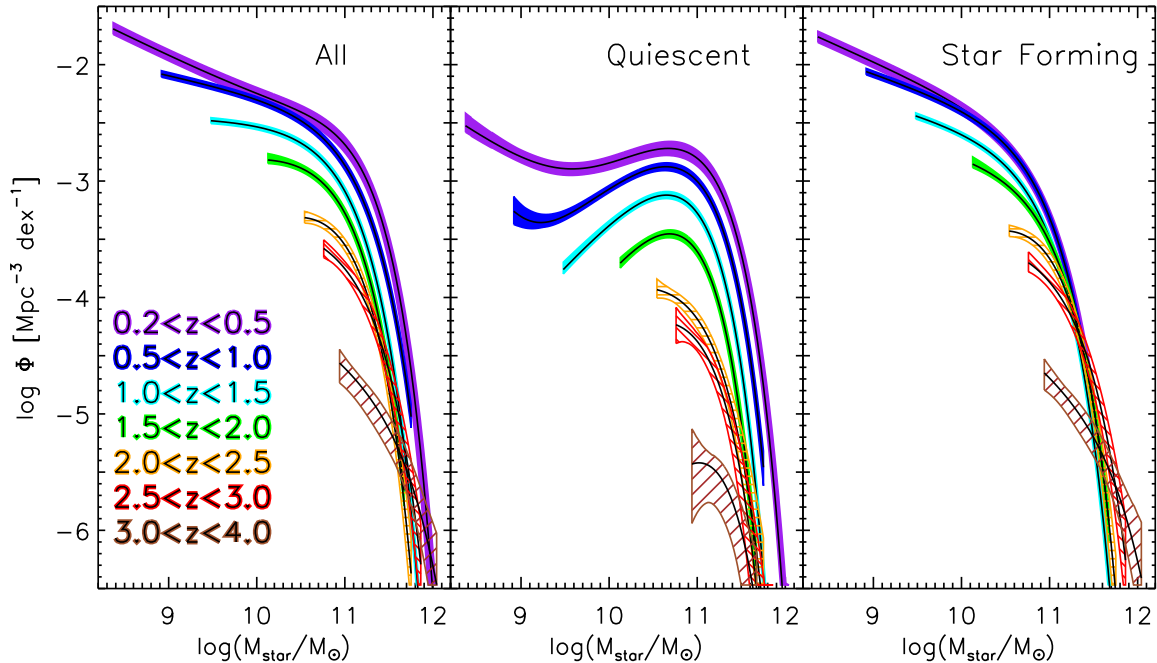}
\caption{\footnotesize Stellar mass functions of all galaxies, quiescent galaxies, and star-forming galaxies in different redshift intervals.  The shaded/hatched regions represent the total 1$\sigma$ uncertainties of the maximum-likelihood analysis, including cosmic variance and the errors from photometric uncertainties as derived using the MC realizations.  The normalization of the SMF of quiescent galaxies evolves rapidly with redshift, whereas the normalization for star-forming galaxies evolves relatively slowly.  In particular, there is almost no change at the high-mass end of the star forming SMF, whereas there is clear growth at the high-mass end of the quiescent population.  There is also evidence for evolution of the low-mass end slope for quiescent galaxies.  At low-redshift a double Schechter function fit is required to reproduce the total SMF.}
\end{figure*}
\begin{figure*}
\plotone{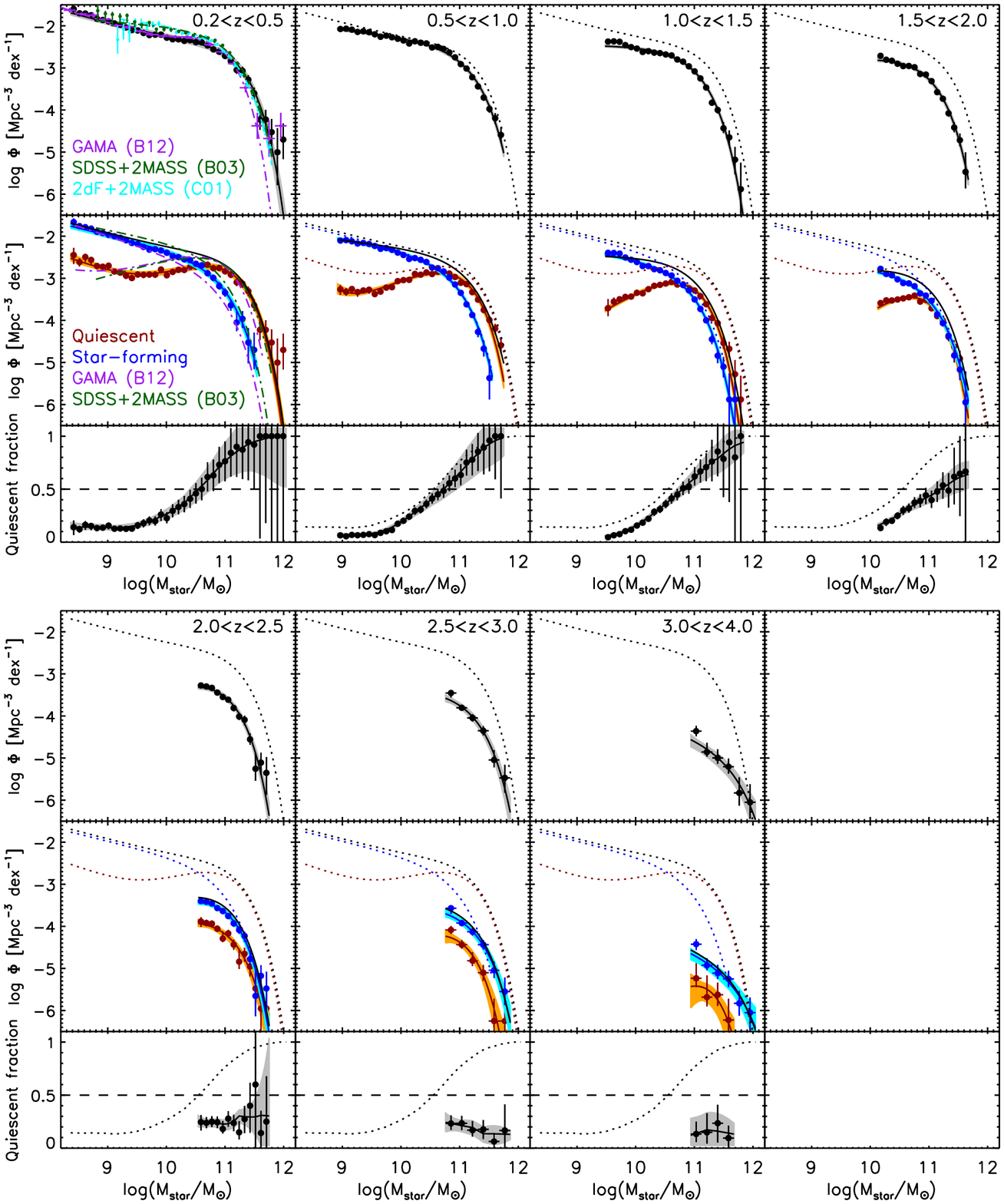}
\caption{\footnotesize Top panels: The SMFs of all galaxies in different redshift bins from 0.2 $< z <$ 4.0.  The black points represent the SMFs determined using the 1/V$_{max}$ method and the black solid curves are the SMFs determined using the maximum-likelihood method.  The gray shaded regions represent the total 1$\sigma$ uncertainties of the maximum-likelihood analysis, including cosmic variance and the errors from photometric uncertainties as derived from the MC simulations.  Overplotted in the 0.2 $< z <$ 0.5 bin are the SMFs from Cole et al. (2001), Bell et al. (2003), and Baldry et al. (2012).  In the remaining redshift bins the dotted curve is the total SMF from UltraVISTA in the 0.2 $< z <$ 0.5 bin.  Middle panels: SMFs as with the top panels, but for the quiescent galaxies (red points, red solid curves) and star-forming galaxies (blue points, blue solid curve).  The orange and cyan shaded regions represent the total 1$\sigma$ uncertainties of the maximum-likelihood analysis for quiescent and star-forming galaxies, respectively.  Bottom panels: Fraction of quiescent galaxies as a function of M$_{star}$. }
\end{figure*}
\subsubsection{Determination of Uncertainties in the SMFs}
In addition to the Poisson uncertainties, there are several other sources of uncertainty in the construction of SMFs that need to be taken into account.  The largest are caused by the fact that M$_{star}$ itself is not an observable quantity, but is derived from observables (i.e., multiwavelength photometry) using a set of models.  The effect of photometric uncertainties on the derived $z_{phot}$ and M$_{star}$ is a non-trivial function of color, magnitude, and redshift caused by a range of data depths in various bands within the survey.  
\newline\indent
In order to calculate uncertainties in the SMFs due to photometric uncertainties we perform 100 Monte Carlo (MC) realizations of the catalog.  Within each realization the photometry in the catalog is perturbed using the measured photometric uncertainties.  New $z_{phot}$ and M$_{star}$ are calculated for each galaxy using the perturbed catalog.  The 100 MC catalogs are then used to recalculate the SMFs and the range of values gives an empirical estimate of the uncertainties in the SMFs due to uncertainties in M$_{star}$ and $z_{phot}$ that propagate from the photometric uncertainties.
\newline\indent
In addition to these $z_{phot}$ and M$_{star}$ uncertainties, the uncertainty from cosmic variance is also included using the prescriptions of \cite{Moster2011}.  In Figure 4 we plot the uncertainty in the abundance of galaxies with Log(M$_{star}$/M$_{\odot}$) $=$ 11.0 due to cosmic variance as a function of redshift.  Cosmic variance is most pronounced at the high-mass end where galaxies are more clustered, and at low redshift, where the survey volume is smallest.  Also plotted in Figure 4 are the cosmic variance uncertainties from other NIR surveys such as FIREWORKS \citep{Wuyts2008}, MUYSC \citep{Quadri2007,Marchesini2009}, NMBS \citep{Whitaker2011}, and the UDS \citep{Williams2009}.  These surveys cover areas that are factor of $\sim$ 50, 16, 4, and 2 smaller than UltraVISTA, respectively.  Figure 4 shows that the improved area from UltraVISTA offers a factor of 1.5 improvement in the uncertainties in cosmic variance compared to even the best previous surveys, and that over the full redshift range the uncertainty from cosmic variance is $\sim$ 8 - 15\% at Log(M$_{star}$/M$_{\odot}$) $=$ 11.0.
\newline\indent
The total uncertainties in the determination of the SMFs are derived as 
follows. For the $1/V_{max}$ method, the total 1$\sigma$ random error 
in each mass bin is the quadrature sum of the Poisson error, the error from 
photometric uncertainties as derived using the MC realizations, and the error 
due to cosmic variance. For the maximum-likelihood method, the total 
1$\sigma$ random errors of the Schechter function parameters $\alpha$, 
$M^{*}_{star}$, and $\Phi^{*}$ are the quadrature sum of the errors 
from the maximum-likelihood analysis, the errors from photometric 
uncertainties as derived using the MC realizations, and the error due to 
cosmic variance (affecting only the normalization $\Phi^{*}$). 
\begin{figure*}
\plotone{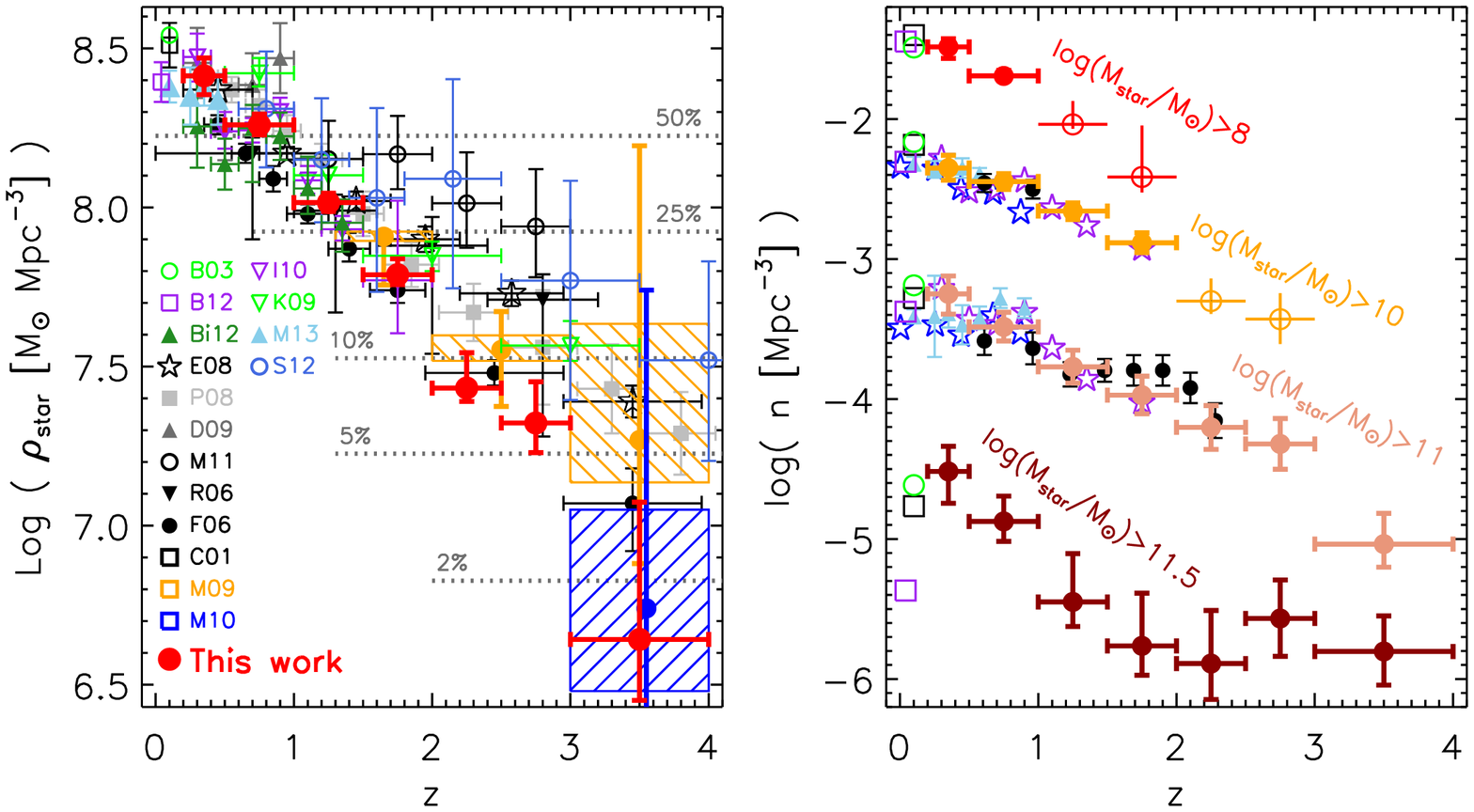}
\caption{\footnotesize Left panel:  The evolution of the stellar mass density of galaxies from $z =$ 4 to $z =$ 0 down to a limit of log(M$_{star}$/M$_{\odot}$) = 8.0.  The UltraVISTA measurements are shown in red with error bars representing total 1$\sigma$ random errors inclusive of cosmic variance and the errors from photometric uncertainties as derived using the MC simulations.  Other measurements from the literature are shown (see text for the definition of references) and agree well with the UltraVISTA measurements within the uncertainties.  Right panel: The evolution of the number density of galaxies above a fixed mass limit from UltraVISTA.  Open circles denote extrapolations of the Schechter function beyond the data.  The black points, light blue triangles, purple stars, and blue stars are from the NMBS \cite{Brammer2011}, PRIMUS \citep{Moustakas2013}, S-COSMOS \citep{Ilbert2010} and zCOSMOS \citep{Pozzetti2010} data and all agree well with the UltraVISTA measurements. }
\end{figure*}
\begin{figure*}
\plotone{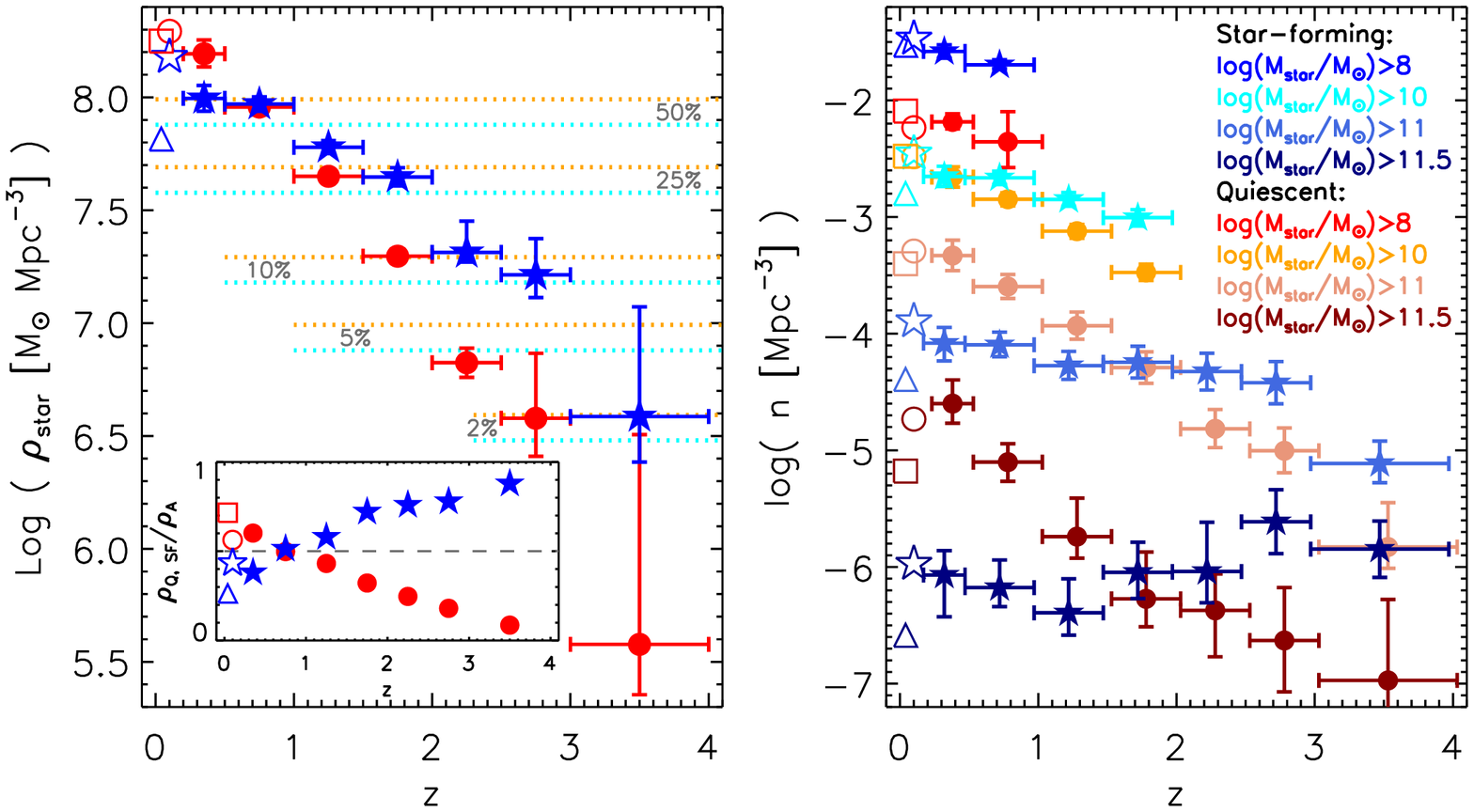}
\caption{\footnotesize Left panel: The evolution of the stellar mass density of star forming (blue) and quiescent (red) galaxies as a function of redshift with error bars representing total 1$\sigma$ random errors inclusive of cosmic variance and the errors from photometric uncertainties as derived using the MC simulations.  At low-redshift the measurements from \cite{Bell2003} (circle and star) and \cite{Baldry2012} (square and triangle) are shown.   The mass density in quiescent galaxies evolves faster than the mass density in star-forming galaxies, particularly at high redshift.  Although they dominate the high-mass end of the mass function at $z <$ 2.5, quiescent galaxies do not dominate the overall mass density of the universe until $z <$ 0.75 due to a much shallower low-mass-end slope.  Right panel: Evolution of the number densities of star forming and quiescent galaxies at a fixed mass limit as a function of redshift.   }
\end{figure*}
\section{The Stellar Mass Functions, Mass Densities and Number Densities to $z$ = 4}
\subsection{The Stellar Mass Functions}
\indent
In Figure 5 we plot the best-fit maximum-likelihood SMFs for the star-forming, quiescent, and combined populations of galaxies.  Figure 5 illustrates the redshift evolution of the SMFs of the individual populations, which we discuss in detail in $\S$ 5.  To better illustrate the relative contribution of both star-forming and quiescent galaxies to the combined SMF, 
in Figure 6 we plot the SMFs derived using the 1/V$_{max}$ method (points), as well as the fits from the maximum-likelihood method (filled regions) in the same redshift bins.  The SMFs of the combined population are plotted in the top panels, and the SMFs of the star-forming and quiescent populations are plotted in the middle panels.  Within each of the higher redshift bins, the SMFs from the lowest-redshift bin (0.2 $< z <$ 0.5) are shown as the dotted line as a fiducial to demonstrate the relative evolution of the SMFs.  The fraction of quiescent galaxies as a function of M$_{star}$ is shown in the bottom panels and the best-fit Schechter function parameters for these redshift ranges are listed in Table 1.  
\newline\indent
For reference, in the lowest-redshift panel (0.2 $< z <$ 0.5) of Figure 6 we plot the SMFs at $z \sim$ 0.1 for the total population from the of studies of \cite{Cole2001}, \cite{Bell2003}, and \cite{Baldry2012}.  Plotted in the middle panel of the lowest-redshift bin are the SMFs of star forming and quiescent galaxies from \cite{Bell2003} and \cite{Baldry2012}.  Qualitatively, these are similar to our measurements, although we note that the selection of star-forming and quiescent galaxies is done differently than the UVJ selection in UltraVISTA. 
\subsection{The Stellar Mass Density and Number Density}
In the left panel of Figure 7 we plot the integrated stellar mass density of all galaxies as a function of redshift.  For consistency with other studies in the literature, the stellar mass densities have been calculated by integrating the maximum-likelihood Schechter function fits down to a limit of Log(M$_{star}$/M$_{\odot}$) = 8.0 at each redshift.  We perform this integration using the full maximum-likelihood fit, even though at $z >$ 2 the $\alpha$ is not well-constrained.  In order to account for possible systematic errors caused by an underestimate of $\alpha$ due to the limited data depth, at all redshifts we also include the uncertainties from the maximum likelihood fits with $\alpha$ = -1.2.  Therefore, the quoted uncertainties in Figure 7, and all subsequent figures that use integration of the SMFs span the full range of uncertainties for both the $\alpha$-free and $\alpha$-fixed SMFs.  
\newline\indent
Overplotted in Figure 7 are measurements of the stellar mass density at various redshifts from previous studies by \cite{Cole2001} (C01), \cite{Bell2003} (B03), \cite{Drory2005} (D05), \cite{Rudnick2006} (R06), \cite{Fontana2006} (F06), \cite{Elsner2008} (E08), \cite{Perezgonzalez2008} (P08), \cite{Marchesini2009} (M09), \cite{Kajisawa2009} (K09), \cite{Drory2009} (D09), \cite{Marchesini2010} (M10), \cite{Ilbert2010} (I10), \cite{Mortlock2011} (M11), \cite{Baldry2012} (B12), \cite{Bielby2012} (Bi12), \cite{Santini2012} (S12), and \cite{Moustakas2013} (M13).  The substantially larger volume covered by UltraVISTA allows for an impressive improvement in the uncertainties in the evolution of the stellar mass densities.  Within the uncertainties, the majority of previous measurements agree reasonably well with the UltraVISTA measurements, particularly at $z <$ 2.  At $z >$ 2 there is less agreement with previous datasets, with the UltraVISTA stellar mass densities being lower than some previous works such as \cite{Elsner2008,Perezgonzalez2008} and \cite{Santini2012}.  The disagreement with \cite{Elsner2008} and \cite{Perezgonzalez2008} is because those studies measure a larger $\phi^{*}$ than UltraVISTA.  The discrepancy with \cite{Santini2012} is primarily because they measure a steep $\alpha$ at $z >$ 2.
\newline\indent
Stellar mass densities and their uncertainties for star-forming and quiescent galaxies have also been computed using the same integration method as for the total stellar mass density.  These are plotted in the left panel of Figure 8 as a function of redshift.  In general, it is clear that at $z <$ 3.5, the stellar mass density of quiescent galaxies grows faster than that of star-forming galaxies.  We explore the implications of this further in $\S$ 5.  All stellar mass densities are listed in Table 2.
\newline\indent
In the right panel of Figure 7 we plot the evolution of the integrated number densities of galaxies calculated using the same Schechter function fits.  The number densities are determined down to limiting masses of Log(M$_{star}$/M$_{\odot}$) = 11.5, 11.0, 10.0, and 8.0, and these points are labeled in the figure.  Plotted as black points are the number densities measured from the NMBS survey \citep{Brammer2011} for mass limits of Log(M$_{star}$/M$_{\odot}$) = 11.0 and 10.0.  These points were measured using a catalog constructed in a similar way to the UltraVISTA catalog and agree well with the number densities in UltraVISTA.  Also shown in Figure 7 are the integrated number densities at $z <$ 1 calculated from the PRIMUS survey \citep{Moustakas2013}, which covers an area $\sim$ 3$\times$ larger than UltraVISTA.  These are also consistent with the UltraVISTA measurements.  Similar integrated number densities for both the star-forming and quiescent populations are shown in the right panel of Figure 8, and the values are listed in Table 2.
\begin{deluxetable*}{lcccllccc}
\centering
\tablecaption{Best-fit Schechter Function Parameters for the SMFs}
\tablehead{\colhead{Redshift} & \colhead{Sample} & \colhead{Number} &  \colhead{$\log{M^{\rm lim}_{\rm star}}$} & 
           \colhead{$\log{M^{\star}_{\rm star}}$} & \colhead{$\Phi^{\star}$} & 
           \colhead{$\alpha$} &  \colhead{$\Phi^{\star}_{\rm 2}$} &  \colhead{$\alpha_{\rm 2}$} \\
                     &   &   &  ($M_{\odot}$) & ($M_{\odot}$) & (10$^{-4}$~Mpc$^{-3}$) 
                     &   &  (10$^{-4}$~Mpc$^{-3}$)  &   }
\startdata
$0.2\leq z <0.5$ & All          & 18546 & 8.37 & 11.22$^{+0.03}_{-0.03}$(0.03) & 12.16$^{+0.52}_{-0.50}$($^{+1.75}_{-1.55}$) & -1.29$\pm$0.01(0.01)    & \nodata & \nodata \\
$0.2\leq z <0.5$ & All          & 18546 & 8.37 & 11.06$\pm$0.01(0.01)     & 19.02$\pm$0.14(2.31)     & -1.2              & \nodata & \nodata \\
$0.2\leq z <0.5$ & All          & 18546 & 8.37 & 10.97$\pm$0.06(0.06)     & 16.27$^{+3.12}_{-1.39}$($^{+3.88}_{-2.41}$)     & -0.53$^{+0.16}_{-0.27}$($^{+0.16}_{-0.28}$)   & 9.47$^{+2.02}_{-3.64}$($^{+2.32}_{-3.83}$) & -1.37$^{+0.01}_{-0.06}$($^{+0.01}_{-0.06}$) \\
$0.2\leq z <0.5$ & Quiescent    &  4364 & 8.37 & 11.21$\pm$0.03(0.04)     & 10.09$\pm$0.54($^{+1.69}_{-1.33}$)     & -0.92$\pm$0.02($^{+0.04}_{-0.02}$)    & \nodata & \nodata \\
$0.2\leq z <0.5$ & Quiescent    &  4364 & 8.37 & 10.75$\pm$0.01(0.01)     & 30.65$\pm$0.01(3.71)     & -0.4              & \nodata & \nodata \\
$0.2\leq z <0.5$ & Quiescent    &  4364 & 8.37 & 10.92$^{+0.06}_{-0.02}$($^{+0.06}_{-0.02}$)     & 19.68$^{+1.09}_{-1.73}$($^{+2.64}_{-2.96}$)     & -0.38$^{+0.06}_{-0.11}$($^{+0.06}_{-0.12}$)    & 0.58$^{+0.26}_{-0.31}$($^{+0.27}_{-0.32}$) & -1.52$^{+0.06}_{-0.16}$($^{+0.06}_{-0.16}$) \\
$0.2\leq z <0.5$ & Star-forming & 14182 & 8.37 & 10.81$\pm$0.03(0.03)     & 11.35$^{+0.76}_{-0.67}$($^{+1.60}_{-1.51}$) & -1.34$\pm$0.01(0.01)    & \nodata & \nodata \\
$0.2\leq z <0.5$ & Star-forming & 14182 & 8.37 & 10.75$\pm$0.01(0.01) & 13.58$^{+0.15}_{-0.13}$($^{+1.62}_{-1.59}$) & -1.3              & \nodata & \nodata \\
 & & & & & & & & \\
$0.5\leq z <1.0$ & All          & 42019 & 8.92 & 11.00$^{+0.02}_{-0.01}$($^{+0.02}_{-0.01}$) & 16.25$^{+0.28}_{-0.62}$($^{+1.17}_{-1.28}$) & -1.17$\pm$0.01(0.01)    & \nodata & \nodata \\
$0.5\leq z <1.0$ & All          & 42019 & 8.92 & 11.04$\pm$0.01(0.01)     & 14.48$\pm$0.07(1.00)     & -1.2              & \nodata & \nodata \\
$0.5\leq z <1.0$ & Quiescent    &  9127 & 8.92 & 10.87$^{+0.02}_{-0.01}$($^{+0.02}_{-0.01}$)     & 13.68$^{+0.26}_{-0.36}$($^{+1.09}_{-1.01}$) & -0.44$\pm$0.02($^{+0.04}_{-0.02}$)    & \nodata & \nodata \\
$0.5\leq z <1.0$ & Quiescent    &  9127 & 8.92 & 10.84$\pm$0.01(0.02)     & 14.38$\pm$0.02(0.99)     & -0.4               & \nodata & \nodata \\
$0.5\leq z <1.0$ & Quiescent    &  9127 & 8.92 & 10.84$\pm$0.02(0.03)     & 14.55$^{+0.56}_{-0.42}$($^{+1.21}_{-1.09}$)     & -0.36$^{+0.06}_{-0.03}$($^{+0.06}_{-0.04}$)     & 0.005$^{+0.021}_{-0.003}$($^{+0.021}_{-0.004}$) & -2.32$^{+0.40}_{-0.32}$($^{+0.41}_{-0.38}$) \\
$0.5\leq z <1.0$ & Star-forming & 32892 & 8.92 & 10.78$^{+0.01}_{-0.02}$     & 12.71$^{+0.51}_{-0.29}$($^{+1.07}_{-0.91}$) & -1.26$\pm$0.01($^{+0.02}_{-0.01}$)     & \nodata & \nodata \\
$0.5\leq z <1.0$ & Star-forming & 32892 & 8.92 & 10.82$\pm$0.01(0.02)     & 10.95$^{+0.06}_{-0.08}$(0.73) & -1.3               & \nodata & \nodata \\
 & & & & & & & & \\
$1.0\leq z <1.5$ & All          & 22959 & 9.48 & 10.87$^{+0.02}_{-0.01}$($^{+0.02}_{-0.01}$)     & 13.91$^{+0.43}_{-0.59}$($^{+1.05}_{-1.23}$) & -1.02$\pm$0.02(0.02)     & \nodata & \nodata \\
$1.0\leq z <1.5$ & All          & 22959 & 9.48 & 10.99$\pm$0.01(0.01)     &  9.30$\pm$0.06(0.66)     & -1.2               & \nodata & \nodata \\
$1.0\leq z <1.5$ & Quiescent    &  6455 & 9.48 & 10.73$\pm$0.02(0.02)     &  8.81$\pm$0.19(0.63)     & -0.17$\pm$0.04($^{+0.06}_{-0.04}$)     & \nodata & \nodata \\
$1.0\leq z <1.5$ & Quiescent    &  6455 & 9.48 & 10.83$\pm$0.01(0.02)     &  7.48$\pm$0.02($^{+0.51}_{-0.56}$)     & -0.4               & \nodata & \nodata \\
$1.0\leq z <1.5$ & Star-forming & 16504 & 9.48 & 10.76$\pm$0.02(0.02) & 8.87$^{+0.50}_{-0.54}$($^{+0.79}_{-0.86}$) & -1.21$\pm$0.03(0.03)     & \nodata & \nodata \\
$1.0\leq z <1.5$ & Star-forming & 16504 & 9.48 & 10.82$\pm$0.01(0.01)     &  7.20$^{+0.06}_{-0.11}$(0.49) & -1.3               & \nodata & \nodata \\
 & & & & & & & & \\
$1.5\leq z <2.0$ & All          &  8927 & 10.03 & 10.81$\pm$0.02($^{+0.02}_{-0.03}$) & 10.13$^{+0.58}_{-0.56}$($^{+1.11}_{-1.02}$) & -0.86$\pm$0.06($^{+0.11}_{-0.06}$)     & \nodata & \nodata \\
$1.5\leq z <2.0$ & All          &  8927 & 10.03 & 10.96$\pm$0.01($^{+0.03}_{-0.01}$)     &  6.33$^{+0.06}_{-0.11}$($^{+0.53}_{-0.71}$) & -1.2               & \nodata & \nodata \\
$1.5\leq z <2.0$ & Quiescent    &  2656 & 10.03 & 10.67$\pm$0.03(0.04)     &  4.15$^{+0.06}_{-0.08}$($^{+0.35}_{-0.36}$)     &  0.03$\pm$0.11(0.12)    & \nodata & \nodata \\
$1.5\leq z <2.0$ & Quiescent    &  2656 & 10.03 & 10.80$\pm$0.01(0.02)     &  3.61$^{+0.02}_{-0.04}$(0.30)     & -0.4               & \nodata & \nodata \\
$1.5\leq z <2.0$ & Star-forming &  6271 & 10.03 & 10.85$^{+0.02}_{-0.03}$($^{+0.02}_{-0.04}$)     &  5.68$^{+0.60}_{-0.40}$($^{+0.89}_{-0.65}$)     & -1.16$^{+0.07}_{-0.05}$($^{+0.14}_{-0.06}$)     & \nodata & \nodata \\
$1.5\leq z <2.0$ & Star-forming &  6271 & 10.03 & 10.91$\pm$0.01($^{+0.04}_{-0.01}$)     &  4.49$\pm$0.09($^{+0.39}_{-0.60}$) & -1.3               & \nodata & \nodata \\
 & & & & & & & & \\
$2.0\leq z <2.5$ & All          &  2236 & 10.54 & 10.81$\pm$0.05($^{+0.06}_{-0.05}$) &  4.79$^{+0.28}_{-0.41}$($^{+0.70}_{-0.76}$) & -0.55$^{+0.16}_{-0.19}$($^{+0.22}_{-0.24}$)     & \nodata & \nodata \\
$2.0\leq z <2.5$ & All          &  2236 & 10.54 & 11.00$\pm$0.02(0.02)     &  2.94$^{+0.07}_{-0.11}$(0.40) & -1.2               & \nodata & \nodata \\
$2.0\leq z <2.5$ & Quiescent    &   528 & 10.54 & 10.87$\pm$0.10($^{+0.11}_{-0.17}$) &  1.02$^{+0.17}_{-0.23}$($^{+0.30}_{-0.27}$) & -0.71$^{+0.37}_{-0.33}$($^{+0.67}_{-0.39}$) & \nodata & \nodata \\
$2.0\leq z <2.5$ & Quiescent    &   528 & 10.54 & 10.79$\pm$0.02(0.03)     &  1.14$\pm$0.05(0.18)     & -0.4                & \nodata & \nodata \\
$2.0\leq z <2.5$ & Star-forming &  1708 & 10.54 & 10.80$\pm$0.05(0.06) &  3.72$^{+0.25}_{-0.32}$($^{+0.58}_{-0.66}$) & -0.53$^{+0.22}_{-0.19}$($^{+0.28}_{-0.24}$) & \nodata & \nodata \\
$2.0\leq z <2.5$ & Star-forming &  1708 & 10.54 & 11.03$\pm$0.02($^{+0.03}_{-0.02}$) &  2.01$^{+0.08}_{-0.10}$($^{+0.28}_{-0.30}$) & -1.3               & \nodata & \nodata \\
 & & & & & & & & \\
$2.5\leq z <3.0$ & All          &   814 & 10.76 & 11.03$^{+0.10}_{-0.09}$($^{+0.12}_{-0.11}$) &  1.93$^{+0.43}_{-0.51}$($^{+0.62}_{-0.68}$() & -1.01$^{+0.37}_{-0.34}$($^{+0.45}_{-0.41}$) & \nodata & \nodata \\
$2.5\leq z <3.0$ & All          &   814 & 10.76 & 11.09$\pm$0.02(0.03) &  1.66$\pm$0.10(0.34)     & -1.2               & \nodata & \nodata \\
$2.5\leq z <3.0$ & Quiescent    &   178 & 10.76 & 10.80$^{+0.23}_{-0.17}$($^{+0.27}_{-0.21}$) &  0.65$^{+0.12}_{-0.24}$($^{+0.18}_{-0.27}$) & -0.39$^{+1.03}_{-0.95}$($^{+1.18}_{-1.11}$)   & \nodata & \nodata \\
$2.5\leq z <3.0$ & Quiescent    &   178 & 10.76 & 10.81$\pm$0.04(0.05)     &  0.66$^{+0.08}_{-0.07}$($^{+0.17}_{-0.14}$) & -0.4               & \nodata & \nodata \\
$2.5\leq z <3.0$ & Star-forming &   636 & 10.76 & 11.06$^{+0.13}_{-0.10}$($^{+0.14}_{-0.12}$) &  1.39$^{+0.35}_{-0.48}$($^{+0.47}_{-0.57}$) & -1.03$\pm$0.39(0.47) & \nodata & \nodata \\
$2.5\leq z <3.0$ & Star-forming &   636 & 10.76 & 11.14$\pm$0.03($^{+0.05}_{-0.03}$)     &  1.09$\pm$0.09($^{+0.22}_{-0.26}$) & -1.3               & \nodata & \nodata \\
 & & & & & & & & \\
$3.0\leq z <4.0$ & All          &    174 & 10.94 & 11.49$^{+0.36}_{-0.22}$($^{+0.37}_{-0.28}$) &  0.09$^{+0.09}_{-0.07}$($^{+0.15}_{-0.08}$) & -1.45$^{+0.59}_{-0.54}$($^{+0.70}_{-0.60}$) & \nodata & \nodata \\
$3.0\leq z <4.0$ & All          &    174 & 10.94 & 11.40$\pm$0.06(0.08)     &  0.13$\pm$0.02($^{+0.08}_{-0.04}$)     & -1.2             & \nodata & \nodata \\
$3.0\leq z <4.0$ & Quiescent    &     28 & 10.94 & 10.85$^{+0.64}_{-0.32}$($^{+0.76}_{-0.43}$) &  0.04$^{+0.03}_{-0.04}$($^{+0.05}_{-0.04}$)     &  0.46$^{+3.16}_{-2.41}$($^{+3.30}_{-2.78}$) & \nodata & \nodata \\
$3.0\leq z <4.0$ & Quiescent    &     28 & 10.94 & 11.00$\pm$0.10($^{+0.14}_{-0.11}$) &  0.05$^{+0.02}_{-0.01}$($^{+0.05}_{-0.02}$) & -0.4            & \nodata & \nodata \\
$3.0\leq z <4.0$ & Star-forming &    146 & 10.94 & 11.56$^{+0.44}_{-0.25}$($^{+0.45}_{-0.33}$) &  0.06$^{+0.08}_{-0.05}$($^{+0.12}_{-0.05}$) & -1.51$^{+0.61}_{-0.55}$($^{+0.77}_{-0.62}$) & \nodata & \nodata \\
$3.0\leq z <4.0$ & Star-forming &    146 & 10.94 & 11.47$\pm$0.07($^{+0.08}_{-0.10}$) &  0.08$\pm$0.02($^{+0.05}_{-0.03}$)     & -1.3             & \nodata & \nodata \\
\enddata
\tablecomments{The listed errors are the 1$\sigma$ Poisson errors, 
whereas the values in parenthesis list the total 1$\sigma$ errors, 
including Poisson uncertainties, the uncertainties from photometric 
redshift random errors, and cosmic variance.} 
\end{deluxetable*}
\begin{deluxetable*}{llrrrrr}
\tabletypesize{\scriptsize}
\centering
\tablecaption{Number and Stellar Mass Densities:\label{density}}
\tablehead{\colhead{Redshift} & \colhead{Density} & 
                                \colhead{$\log{\big(\frac{M_{\rm star}}{M_{\odot}}\big)}>8$}  & 
                                \colhead{$\log{\big(\frac{M_{\rm star}}{M_{\odot}}\big)}>9$}  &  
                                \colhead{$\log{\big(\frac{M_{\rm star}}{M_{\odot}}\big)}>10$} & 
                                \colhead{$\log{\big(\frac{M_{\rm star}}{M_{\odot}}\big)}>11$} & 
                                \colhead{$\log{\big(\frac{M_{\rm star}}{M_{\odot}}\big)}>11.5$}\\}
\startdata
$0.2\leq z <0.5$ & $\log{(\eta_{\rm A})}$  & -1.49$\pm$0.06     & -1.89$\pm$0.07 & -2.35$\pm$0.09 & -3.25$\pm$0.13 & -4.52$^{+0.18}_{-0.23}$ \\ 
                 & $\log{(\eta_{\rm Q})}$  & -2.19$^{+0.07}_{-0.06}$ & -2.45$\pm$0.07 & -2.66$\pm$0.09 & -3.33$\pm$0.13 & -4.60$^{+0.20}_{-0.17}$ \\ 
                 & $\log{(\eta_{\rm SF})}$  & -1.58$\pm$0.06 & -2.03$\pm$0.07 & -2.65$\pm$0.09 & -4.08$^{+0.13}_{-0.15}$ & -6.07$^{+0.21}_{-0.36}$ \\ 
                 & $\log{(\rho_{\rm A})}$  & 8.41$\pm$0.06 & 8.40$\pm$0.07 & 8.35$\pm$0.09 & 7.98$^{+0.13}_{-0.14}$ & 7.08$^{+0.18}_{-0.23}$ \\ 
                 & $\log{(\rho_{\rm Q})}$  & 8.19$\pm$0.06 & 8.19$\pm$0.07 & 8.18$\pm$0.09 & 7.91$\pm$0.13     & 7.00$^{+0.21}_{-0.17}$ \\ 
                 & $\log{(\rho_{\rm SF})}$  & 7.99$\pm$0.06 & 7.97$\pm$0.07 & 7.85$\pm$0.09 & 7.08$^{+0.14}_{-0.16}$  & 5.50$^{+0.21}_{-0.35}$ \\ 
$0.5\leq z <1.0$ & $\log{(\eta_{\rm A})}$  & -1.69$\pm$0.04 & -2.00$\pm$0.04 & -2.45$\pm$0.06 & -3.49$\pm$0.11 & -4.88$^{+0.18}_{-0.14}$ \\ 
                 & $\log{(\eta_{\rm Q})}$  & -2.36$^{+0.26}_{-0.22}$ & -2.70$\pm$0.04 & -2.85$\pm$0.06 & -3.60$\pm$0.10 & -5.10$^{+0.16}_{-0.17}$ \\ 
                 & $\log{(\eta_{\rm SF})}$  & -1.70$^{+0.04}_{-0.03}$ & -2.09$\pm$0.04 & -2.66$\pm$0.06 & -4.10$^{+0.11}_{-0.10}$ & -6.18$^{+0.24}_{-0.16}$ \\ 
                 & $\log{(\rho_{\rm A})}$  & 8.26$\pm$0.03 & 8.25$\pm$0.04 & 8.19$\pm$0.06 & 7.73$^{+0.11}_{-0.10}$ & 6.72$^{+0.19}_{-0.14}$ \\ 
                 & $\log{(\rho_{\rm Q})}$  & 7.96$\pm$0.03 & 7.96$\pm$0.04 & 7.94$\pm$0.06 & 7.61$\pm$0.10     & 6.48$^{+0.16}_{-0.17}$ \\ 
                 & $\log{(\rho_{\rm SF})}$  & 7.97$\pm$0.03 & 7.95$\pm$0.04 & 7.84$\pm$0.06 & 7.06$\pm$0.11     & 5.39$^{+0.24}_{-0.16}$     \\ 
$1.0\leq z <1.5$ & $\log{(\eta_{\rm A})}$  & -2.04$^{+0.17}_{-0.03}$ & -2.26$^{+0.07}_{-0.04}$ & -2.66$\pm$0.06 & -3.77$\pm$0.12 & -5.45$^{+0.35}_{-0.18}$ \\ 
                 & $\log{(\eta_{\rm Q})}$  & -3.01$^{+0.05}_{-0.03}$ & -3.01$^{+0.05}_{-0.04}$ & -3.12$\pm$0.06 & -3.93$\pm$0.12 & -5.74$^{+0.33}_{-0.19}$ \\ 
                 & $\log{(\eta_{\rm SF})}$  & -1.94$^{+0.10}_{-0.04}$ & -2.30$^{+0.05}_{-0.04}$ & -2.85$\pm$0.06 & -4.27$\pm$0.12 & -6.39$^{+0.29}_{-0.19}$ \\ 
                 & $\log{(\rho_{\rm A})}$  & 8.02$\pm$0.03 & 8.01$\pm$0.04 & 7.95$\pm$0.06 & 7.41$^{+0.13}_{-0.12}$ & 6.13$^{+0.36}_{-0.18}$ \\ 
                 & $\log{(\rho_{\rm Q})}$  & 7.65$\pm$0.03 & 7.65$\pm$0.04 & 7.64$\pm$0.06 & 7.25$\pm$0.12      & 5.83$^{+0.34}_{-0.19}$ \\ 
                 & $\log{(\rho_{\rm SF})}$  & 7.78$\pm$0.03 & 7.76$\pm$0.04 & 7.66$\pm$0.06 & 6.88$^{+0.13}_{-0.12}$ & 5.17$^{+0.30}_{-0.19}$ \\ 
$1.5\leq z <2.0$ & $\log{(\eta_{\rm A})}$  & -2.41$^{+0.37}_{-0.11}$ & -2.56$^{+0.19}_{-0.08}$ & -2.88$\pm$0.07 & -3.97$\pm$0.14 & -5.76$^{+0.38}_{-0.21}$ \\ 
                 & $\log{(\eta_{\rm Q})}$  & -3.39$^{+0.12}_{-0.04}$ & -3.40$^{+0.10}_{-0.05}$ & -3.48$\pm$0.07 & -4.29$\pm$0.14 & -6.27$^{+0.40}_{-0.24}$ \\ 
                 & $\log{(\eta_{\rm SF})}$  & -2.21$^{+0.21}_{-0.19}$ & -2.52$\pm$0.11 & -3.01$\pm$0.07 & -4.25$\pm$0.14 & -6.05$^{+0.26}_{-0.23}$ \\ 
                 & $\log{(\rho_{\rm A})}$  & 7.79$^{+0.05}_{-0.03}$ & 7.79$^{+0.05}_{-0.04}$ & 7.74$\pm$0.07 & 7.20$^{+0.14}_{-0.13}$ & 5.81$^{+0.39}_{-0.21}$ \\ 
                 & $\log{(\rho_{\rm Q})}$  & 7.29$\pm$0.03     & 7.30$\pm$0.04      & 7.29$\pm$0.07 & 6.88$\pm$0.14     & 5.29$^{+0.42}_{-0.24}$ \\ 
                 & $\log{(\rho_{\rm SF})}$  & 7.65$\pm$0.04    & 7.64$\pm$0.05      & 7.56$\pm$0.07 & 6.93$\pm$0.14      & 5.53$^{+0.26}_{-0.23}$ \\ 
$2.0\leq z <2.5$ & $\log{(\eta_{\rm A})}$  & -3.06$^{+0.70}_{-0.13}$ & -3.11$^{+0.43}_{-0.11}$ & -3.30$^{+0.16}_{-0.09}$ & -4.20$\pm$0.16 & -5.89$^{+0.38}_{-0.26}$ \\ 
                 & $\log{(\eta_{\rm Q})}$  & -3.58$^{+0.35}_{-0.41}$ & -3.67$^{+0.22}_{-0.30}$ & -3.91$^{+0.11}_{-0.14}$ & -4.82$\pm$0.17 & -6.37$^{+0.31}_{-0.40}$ \\ 
                 & $\log{(\eta_{\rm SF})}$  & -3.18$^{+0.87}_{-0.16}$ & -3.23$^{+0.52}_{-0.12}$ & -3.42$^{+0.18}_{-0.09}$ & -4.32$\pm$0.16 & -6.04$^{+0.42}_{-0.27}$ \\ 
                 & $\log{(\rho_{\rm A})}$  & 7.43$^{+0.11}_{-0.04}$ & 7.43$^{+0.11}_{-0.06}$ & 7.41$^{+0.10}_{-0.08}$ & 6.98$\pm$0.16 & 5.69$^{+0.40}_{-0.26}$ \\ 
                 & $\log{(\rho_{\rm Q})}$  & 6.83$\pm$0.07     & 6.82$\pm$0.07      & 6.80$\pm$0.09     & 6.38$\pm$0.17 & 5.21$^{+0.32}_{-0.41}$ \\ 
                 & $\log{(\rho_{\rm SF})}$  & 7.31$^{+0.14}_{-0.04}$ & 7.31$^{+0.13}_{-0.06}$ & 7.29$^{+0.10}_{-0.08}$ & 6.86$\pm$0.16 & 5.53$^{+0.45}_{-0.27}$ \\ 
$2.5\leq z <3.0$ & $\log{(\eta_{\rm A})}$  & -2.88$^{+0.69}_{-0.52}$ & -3.09$^{+0.41}_{-0.35}$ & -3.43$^{+0.19}_{-0.18}$ & -4.32$\pm$0.18 & -5.57$^{+0.28}_{-0.27}$ \\ 
                 & $\log{(\eta_{\rm Q})}$  & -4.02$^{+1.29}_{-0.68}$ & -4.05$^{+0.84}_{-0.52}$ & -4.19$^{+0.39}_{-0.29}$ & -5.00$\pm$0.20 & -6.63$^{+0.45}_{-0.44}$ \\ 
                 & $\log{(\eta_{\rm SF})}$  & -2.99$^{+0.84}_{-0.53}$ & -3.20$^{+0.49}_{-0.35}$ & -3.55$^{+0.21}_{-0.18}$ & -4.42$\pm$0.18 & -5.61$^{+0.28}_{-0.27}$ \\ 
                 & $\log{(\rho_{\rm A})}$  & 7.32$^{+0.13}_{-0.09}$ & 7.32$^{+0.13}_{-0.10}$ & 7.28$^{+0.12}_{-0.11}$ & 6.91$\pm$0.18 & 6.04$\pm$0.28 \\ 
                 & $\log{(\rho_{\rm Q})}$  & 6.58$^{+0.29}_{-0.17}$ & 6.58$^{+0.27}_{-0.17}$ & 6.57$^{+0.22}_{-0.15}$ & 6.19$\pm$0.20 & 4.95$^{+0.48}_{-0.46}$ \\ 
                 & $\log{(\rho_{\rm SF})}$  & 7.21$^{+0.16}_{-0.10}$ & 7.21$^{+0.15}_{-0.11}$ & 7.17$^{+0.13}_{-0.12}$ & 6.82$\pm$0.18 & 6.00$\pm$0.28 \\ 
$3.0\leq z <4.0$ & $\log{(\eta_{\rm A})}$  & -3.14$^{+1.41}_{-1.25}$ & -3.65$^{+0.84}_{-0.74}$ & -4.20$^{+0.36}_{-0.34}$ & -5.04$^{+0.22}_{-0.16}$ & -5.80$^{+0.25}_{-0.24}$ \\ 
                 & $\log{(\eta_{\rm Q})}$  & -5.44$^{+3.26}_{-0.94}$ & -5.44$^{+2.19}_{-0.67}$ & -5.46$^{+1.15}_{-0.42}$ & -5.83$^{+0.38}_{-0.18}$ & -6.97$^{+0.69}_{-0.49}$ \\ 
                 & $\log{(\eta_{\rm SF})}$  & -3.11$^{+1.47}_{-1.38}$ & -3.67$^{+0.88}_{-0.81}$ & -4.27$^{+0.38}_{-0.35}$ & -5.11$^{+0.19}_{-0.17}$ & -5.85$\pm$0.24 \\ 
                 & $\log{(\rho_{\rm A})}$  & 6.64$^{+0.43}_{-0.19}$ & 6.63$^{+0.31}_{-0.18}$ & 6.57$^{+0.20}_{-0.15}$ & 6.31$^{+0.21}_{-0.17}$ & 5.89$\pm$0.26 \\ 
                 & $\log{(\rho_{\rm Q})}$  & 5.58$^{+0.93}_{-0.22}$ & 5.58$^{+0.81}_{-0.22}$ & 5.58$^{+0.62}_{-0.20}$ & 5.44$^{+0.42}_{-0.19}$ & 4.62$^{+0.77}_{-0.53}$ \\ 
                 & $\log{(\rho_{\rm SF})}$  & 6.59$^{+0.49}_{-0.20}$ & 6.58$^{+0.34}_{-0.19}$ & 6.50$^{+0.19}_{-0.15}$ & 6.25$^{+0.19}_{-0.17}$ & 5.86$\pm$0.27 \\ 
\enddata
\tablecomments{Number densities $\eta$ are in units of Mpc$^{-3}$, while stellar 
mass densities $\rho$ are in units of M$_{\odot}$~Mpc$^{-3}$. Both the number and stellar 
mass densities are calculated for $\log{(M_{\rm star}/M_{\odot})}<13$. The listed 
errors are the total 1$\sigma$ errors, including Poisson uncertainties, the 
uncertainties from photometric redshift random errors, and cosmic variance.}
\end{deluxetable*}
\section{Discussion}
\begin{figure}
\plotone{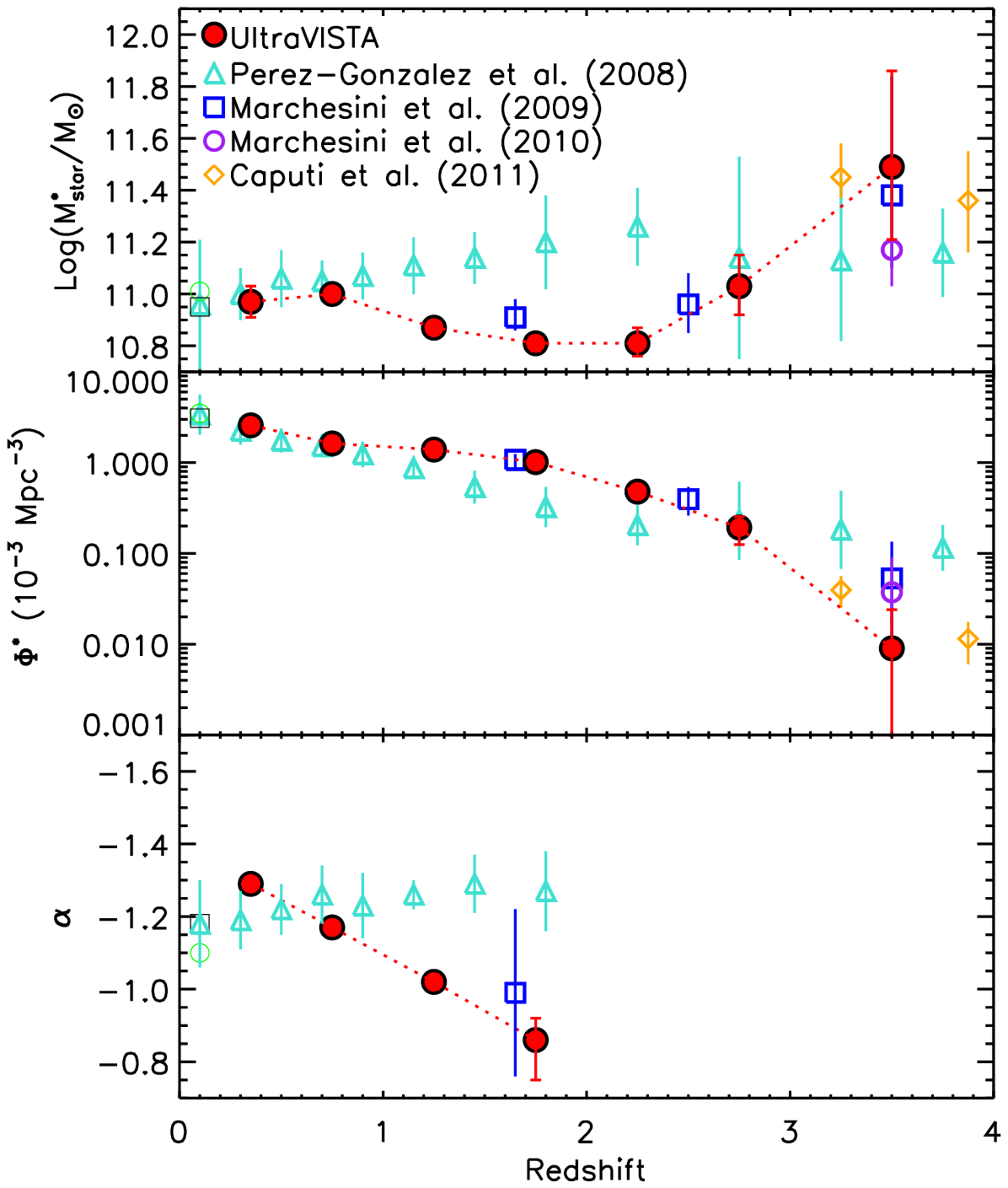}
\caption{\footnotesize Best-fit Schechter parameters as a function of redshift for combined SMFs determined using the UltraVISTA data (red circles).  The squares and circles are the $z =$ 0 measurements from \cite{Cole2001}, and \cite{Bell2003}.  Also shown are parameters from previous rest-frame-optical-selected SMFs from \cite{Perezgonzalez2008}, \cite{Marchesini2009}, \cite{Marchesini2010}, and \cite{Caputi2011}.  The SMFs show little evolution in M$^{*}_{star}$ with redshift, but an evolution of 2.70$^{+0.96}_{-0.42}$ dex in $\Phi^{*}$ since $z =$ 3.5.  It appears that $\alpha$ flattens with increasing redshift; however, this is more likely to be a result of depth of the data rather than a true flattening (see text).}
\end{figure}
\subsection{The Combined Population}
In Figure 9 we plot the evolution of the Schechter parameters M$^{*}_{star}$, $\Phi^{*}$, and $\alpha$ as a function of redshift for the combined population of both star-forming and quiescent galaxies.  
All points plotted are from the single Schechter function fits except the 0.2 $< z <$ 0.5 bin where we have plotted the M$^{*}_{star}$ and combined $\Phi^{*}$s of the double Schechter function fits.  We do not plot $\alpha$ at $z >$ 2 in Figure 9 because the limiting mass of the data at these redshifts is too high to provide constraints.  The lack of constraints combined with the correlation between M$^{*}_{star}$ and $\alpha$ means that the uncertainties in $\alpha$ at $z >$ 2 are likely to be underestimated.  Also plotted in Figure 9 are the best-fit Schechter parameters from the low-redshift SMFs from \cite{Cole2001}, and \cite{Bell2003}, and the high-redshift rest-frame-optical-selected SMFs from \cite{Perezgonzalez2008}, \cite{Marchesini2009}, \cite{Marchesini2010}, and \cite{Caputi2011}.  
\newline\indent
Comparison of the parameters in our lowest-redshift bin (0.2 $< z <$ 0.5) to those from the \cite{Cole2001} and \cite{Bell2003} parameters shows good agreement for both $M_{star}$ and $\Phi^{*}$.  There is some disagreement in $\alpha$, with our data having a steeper low-mass-end than the local SMFs.  It is unclear why this is, as the UltraVISTA data reach a comparable depth in M$_{star}$ as the local studies.  Part of the discrepancy may be because \cite{Cole2001} and \cite{Bell2003} fit a single Schechter function when a double is required.  If we compare our double Schechter function fits ($\alpha_{1}$ = -0.53$^{+0.16}_{-0.28}$, $\alpha_{2}$ = -1.37$^{+0.01}_{-0.06}$) to those derived from \cite{Baldry2012} ($\alpha_{1}$ = -0.35 $\pm$ 0.18, $\alpha_{2}$ = -1.47 $\pm$ 0.05), we find good agreement.
\newline\indent
Figure 9 also shows that there is good agreement between our SMFs and previous high-redshift SMFs in the literature.  There is a significant improvement in the uncertainties in the SMFs derived from the UltraVISTA catalog, mostly due to the fact it covers an area that is a factor of 8.8 and 11.4 larger than the area used in the \cite{Perezgonzalez2008} and \cite{Marchesini2009} studies, respectively.  Figure 9 confirms the results of those previous works and shows that within the substantially smaller uncertainties, there is still no significant evolution in M$^{*}_{star}$ out to $z \sim$ 3.0.  This lack of evolution implies that whatever process causes the exponential tail of the Schechter function, does so in a consistent way over much of cosmic time.  At $z >$ 3.5 we find some evidence for an change in M$^{*}_{star}$; however, given the lack of constraints on $\alpha$ at this redshift and the correlation between M$^{*}_{star}$ and $\alpha$ the uncertainties are still large.
\newline\indent
Although there is no significant evolution in M$^{*}_{star}$, there is a substantial evolution in $\Phi^{*}$ from $z =$ 3.5 to $z =$ 0.0.  If we compare to the $\Phi^{*}$ at $z =$ 0 from \cite{Cole2001}, we find that it evolves by 2.58$^{+1.01}_{-0.37}$ dex between $z \sim$ 3.5 and $z \sim$ 0.0.  As Figure 9 shows, this evolution is stronger than the values of 1.22$\pm$0.43, 1.76$^{+0.40}_{-0.82}$, 1.92$^{+0.39}_{-0.36}$, 1.89$^{+0.14}_{-0.19}$ dex measured previously by \cite{Perezgonzalez2008}, \cite{Marchesini2009}, \cite{Marchesini2010}, and \cite{Caputi2011}, respectively.  
\newline\indent
Interestingly, it appears that there is a statistically significant evolution in $\alpha$ up to $z =$ 2, the redshift where the data is deep enough that there is still reasonable constraints on $\alpha$.  A flattening of the slope with redshift was also seen in the SMFs of \cite{Marchesini2009} which probe to slightly lower M$_{star}$.  Such a flattening in the combined population is a natural consequence of the fact that there appears to be little evolution in the $\alpha$ of the star-forming population, but a flattening in $\alpha$ for the quiescent population with increasing redshift (e.g., Figure 5, see $\S$ 5.2).  UV-selected samples suggest a steep $\alpha$ at high-redshift \citep[e.g.,][]{Reddy2009,Stark2009,Gonzalez2011,Lee2012}, which taken at face value disagree with the flattening of the slope observed for the K$_{s}$-selected SMF.  As we show in $\S$ 5.3, UV-selection misses a fraction of the massive galaxy population due to their quiescence and/or dustiness, so may overestimate $\alpha$ for the combined population.  
\newline\indent
Recently \cite{Santini2012} have measured the faint-end slope using ultra-deep HAWK-I K$_{s}$ data which is a better comparison to the UltraVISTA SMFs than UV-selected SMFs.  They also find a steep faint-end slope ($\alpha$ = $-$1.84 $\pm$ 0.06) at $z \sim$ 2, which taken at face value does not agree well with our measurement.  Of course, $\alpha$ and M$^{*}_{star}$ are correlated, and \cite{Santini2012} cover an area that is two orders-of-magnitude smaller than UltraVISTA and therefore have poor constraints on M$^{*}_{star}$.  At $z \sim$ 2, they find a value of Log(M$^{*}_{star}$/M$_{\odot}$) = 11.82 $\pm$ 0.28 which is almost an order-of-magnitude larger than the UltraVISTA value of Log(M$^{*}_{star}$/M$_{\odot}$) = 10.81$^{+0.06}_{-0.05}$.  This makes it clear that in order to simultaneously fit the \cite{Santini2012} measurements and the UltraVISTA measurements, a double Schechter function is required.  
\newline\indent
This implies that the decline in $\alpha$ with increasing redshift in the UltraVISTA SMFs is more a reflection of the chugging limiting M$_{star}$ with increasing redshift.  At higher-redshift the UltraVISTA Schechter function fits become increasingly dominated by the high-mass end and does not contrain the double Schechter function upturn.  Probing further down the low-mass end will most likely result in steeper $\alpha$ than have been measured with the current depth.  We note that given the limiting M$_{star}$ of the UltraVISTA SMFs, and that $\alpha$ is not well-constrained at high-redshift we have also performed all the fits with fixed $\alpha$ = $-$1.2, which is closer to the \cite{Santini2012} and UV-selected SMF measurements.  These fixed-$\alpha$ fits are incorporated into the uncertainties in quantities such as the stellar mass densities and so these should still be representative.
\newline\indent
If the $\alpha$ for star forming and quiescent galaxy SMFs shown in Figure 5 are representative of the $\alpha$ at lower masses than are probed by the current catalog, then it suggests that the single Schechter function fit we have used will be a poor representation of faint end at increasing redshift.  Because the $\alpha$ of the star forming and quiescent SMFs are so different, and the mix of the two populations is a strong function of mass (e.g., Figure 6), a double Schechter function parameterization will be necessary to describe the faint end of the combined population below the mass limit of the current data.  If so, then the flattening in $\alpha$ with redshift that we measure for the combined population is most likely a consequence of the fact that the mass limit of the survey is near the M$_{star}$ where the number densities transition between being dominated by the quiescent population, to being dominated by the star-forming population.  A more conclusive understanding of the differences in the slope of rest-frame-optical-selected and UV-selected SMFs will have to wait until deeper rest-frame-optical data is available and can select galaxies to comparable limits in stellar mass as UV-selected samples.
\subsection{The Quiescent Population}
The SMF of quiescent population shows significant evolution since $z =$ 3.5.  Although there is little change in M$^{*}_{star}$, the $\Phi^{*}$ increases by 0.57$^{+0.03}_{-0.04}$, 1.39$^{+0.07}_{-0.07}$, and 2.75$^{+0.36}_{-0.23}$ dex since $z \sim$ 1.0, 2.0, and 3.5, respectively.  With the improved uncertainties from UltraVISTA, there is also evidence for a steady increase in the number density of even the most massive quiescent galaxies (Log(M$_{star}$/M$_{\odot}$ $>$ 11.5) since $z =$ 3.5 (Figure 8).  The uncertainties from earlier studies \citep[e.g.,][]{Brammer2011} were large enough that they could accommodate no growth in the number densities of these galaxies since $z =$ 2.2.
\newline\indent
At $z <$ 1 the data is complete to a sufficiently small M$_{star}$ that the upturn in the number density of quiescent galaxies at Log(M$_{star}$/M$_{\odot}$) $<$ 9.5 can be clearly seen.  This upturn is also seen at $z =$ 0 in the quiescent population \cite[e.g.,][]{Bell2003,Baldry2012}, and the UltraVISTA data now confirms that it persists to at least $z =$ 1.  In the appendix we show that the existence of the upturn is robust to the definition of quiescent galaxies, although its prominence can be increased or decreased depending on the strictness of the selection.  Interestingly, the location of the upturn seems to evolve in mass, from Log(M$_{star}$/M$_{\odot}$) $\sim$ 9.2 at $z =$ 0.75, to Log(M$_{star}$/M$_{\odot}$) $\sim$ 9.5 at $z =$ 0.35.  
\newline\indent
It has been suggested that the upturn is the result of a population of star-forming satellite galaxies that have been quenched in high-density environments \citep[e.g.,][]{Peng2010,Peng2012}.  The \cite{Peng2010} model separates the quenching process into two forms: ``environmental quenching", and ``mass quenching", the former of which is caused by high-density environments, the latter of which is caused by processes internal to the galaxy itself.  In their model, the M$_{star}$ where the upturn occurs is determined by the relative $\alpha$'s of the mass-quenched quiescent population, and the environmentally-quenched quiescent population (formerly a recently star-forming population).  In the case that the $\alpha$ of the self-quenched quiescent populations and the star-forming populations do not evolve with redshift -- which is consistent with our measurement -- an evolution in the M$_{star}$ of the upturn would imply a change in the fraction of galaxies that self-quench as compared to environmentally-quench.  An evolution to higher masses with decreasing redshift would imply an increase in the fraction of galaxies that are environmental-quenched with decreasing redshift, i.e., that environmental-quenching becomes increasingly more important at lower redshift.  
\newline\indent
It is also interesting to examine the evolution of the fraction of quiescent galaxies as a function of both M$_{star}$ and $z$.  These fractions are plotted in the bottom panels of Figure 6.  At $z <$ 1 we recover the well-known result that quiescent galaxies dominate the high-mass-end of the SMF \citep[e.g.,][]{Bundy2006,Ilbert2010,Pozzetti2010,Brammer2011}, but thereafter some interesting trends emerge.  
\newline\indent
The fraction of quiescent galaxies with Log(M$_{star}$/M$_{\odot}$) $>$ 11.0 continues to decline slowly with increasing redshift up to $z =$ 1.5.  After $z >$ 1.5 that decline accelerates significantly, and we find that by $z =$ 2.5 the fraction of quiescent galaxies has decreased to the point where star-forming galaxies dominate the SMF at all stellar masses.  This result is depicted in Figure 10 where we plot the M$_{star}$ at which quiescent galaxies dominate the SMF (denoted as M$_{cross}$) as a function of redshift.  Also plotted in Figure 10 are measurements of M$_{cross}$ from other studies \citep{Bell2003,Bundy2006,Pozzetti2010,Baldry2012}.  These studies use different definitions of star-forming and quiescent galaxies, so a direct comparison to the UltraVISTA measurements is difficult; however, we note that they are largely consistent with the UltraVISTA measurements in the redshift range where the data overlap.
\newline\indent
As Figure 10 shows, the crossing mass at $z =$ 0.35 is Log(M$_{star}$/M$_{\odot}$) $=$ 10.55, and the crossing mass evolves as Log(M$_{cross}$/M$_{\odot}$) $\propto$ (1+$z$)$^{0.9}$ up to $z =$ 1.5.  Thereafter it evolves much more rapidly Log(M$_{cross}$/M$_{\odot}$) $\propto$ (1+$z$)$^{5.6}$ up to $z =$ 2.5, after which star-forming galaxies dominate the full population.  
\newline\indent
Despite the sharp change in the evolution of M$_{cross}$ at $z =$ 1.5, the overall increase in the number density of the quiescent population (i.e., $\Phi^{*}$) is fairly smooth with redshift.  Therefore, the rapid evolution of M$_{cross}$ at $z >$ 1.5 is really a reflection of the fact that M$_{cross}$ moves onto the exponential part of the Schechter function at high-redshift, whereas it occurs on the power-law part of the Schechter function at low redshift. 
\newline\indent
Although their fraction diminishes substantially at high-redshift, it is noteworthy that we find a non-zero fraction of quiescent galaxies ($\sim$ 10-20\%) up to $z =$ 3.5.  A similarly small, but non-zero fraction was also found by \cite{Marchesini2010} in the NMBS.  We will discuss the SEDs of the quiescent population in more detail in a future paper (D. Marchesini, in preparation), but we note that for some, the best-fit ages are $\leq$ 1.0 Gyr.  If they formed most of their stars in a rapid burst, as suggested by the best-fit $\tau$-model, it suggests that there may be a non-zero population of quiescent galaxies that extends to as high as $z =$ 6 -- 8.  Substantially deeper wide-field data (or longer-wavelength selection) will be needed to find such galaxies at $z >$ 4, as the K$_{s}$-band moves blueward of the Balmer break and quickly becomes inefficient at detecting red galaxies.
\begin{figure}
\plotone{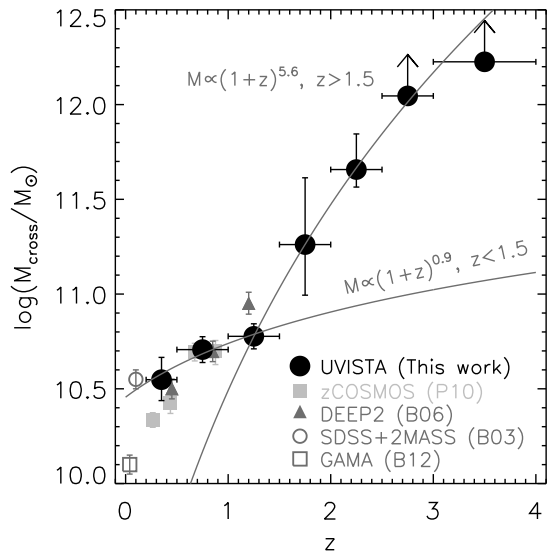}
\caption{\footnotesize The M$_{star}$ at which quiescent galaxies dominate over star-forming galaxies (M$_{cross}$) as a function of redshift.  Measurements from other surveys are shown and agree reasonably well with the UltraVISTA measurement.  Quiescent galaxies dominate the high-mass end of the SMF up to $z \sim$ 1.5.  Thereafter star-forming galaxies quickly become dominant at all M$_{star}$.}
\end{figure}
\subsection{The Star Forming Population}
\begin{figure}
\plotone{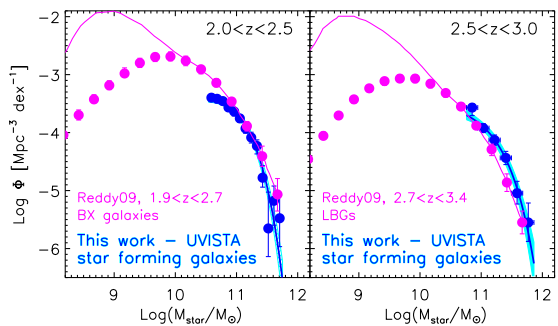}
\caption{\footnotesize Left Panel: Comparison of the UltraVISTA star forming SMF at 2.0 $< z <$ 2.5 and the BM/BX-selected SMF from \cite{Reddy2009}.  The SMFs show good agreement at the high-mass end but the UltraVISTA SMF suggests a shallower low-mass-end slope than the BM/BX SMF.  Right Panel: Comparison of the UltraVISTA star forming SMF at 2.5 $< z <$ 3.0 and the LBG-selected SMF from \cite{Reddy2009}.  These show reasonable agreement at the high-mass end, although the UltraVISTA SMF suggests that the LBG selection may miss $\sim$ 50\% of the massive galaxy population in this redshift range.}
\end{figure}
Figure 5 shows that in contrast to the SMF of quiescent galaxies, the evolution of the SMF of star-forming galaxies from $z =$ 3.5 to $z =$ 0 is fairly modest.  The M$^{*}_{star}$ and $\alpha$ show no significant evolution up to $z =$ 3.5, albeit with large uncertainties in the latter.  Unlike the quiescent population, the number density of Log(M$_{star}$/M$_{\odot}$) $>$ 11.5 galaxies shows almost no evolution up to $z =$ 3.5 (Figure 8).  The only significant evolution is in $\Phi^{*}$, which evolves by 0.45$^{+0.03}_{-0.03}$, 1.01$^{+0.06}_{-0.06}$, and 2.40$^{+0.21}_{-0.21}$ dex, since $z \sim$ 1.0, 2.0, 3.5.  If we compare this to the evolution of the quiescent population at the same redshifts, we find that the quiescent population has grown faster in $\Phi^{*}$ by factors of $\sim$ 1.3, 2.4, and 2.2 since $z \sim$ 1.0, 2.0, and 3.5, respectively.  This shows that at all redshifts the majority of the growth in the combined SMF is due to the increase in the quiescent population.  
\newline\indent
The non-evolution in M$^{*}_{star}$ and $\alpha$ and the rather slow evolution in $\Phi^{*}$ for the star-forming population is remarkable if considered in the context of the evolution of the star formation rate per unit M$_{star}$ (specific star formation rates, hereafter SSFR) over the same redshift range.  The SSFR of star-forming galaxies with Log(M$_{star}$/M$_{\odot}$) $=$ 10.0(11.0) declines by a factor of $\sim$ 20(25) since $z \sim$ 2 \citep[e.g.,][]{Muzzin2013a}, which means the growth rate of these galaxies is evolving substantially with redshift.  That the process of galaxy quenching evolves in such a way to keep the shape and normalization of the star-forming SMF roughly constant over this redshift range, while there is a significant decrease in the SSFRs, implies a carefully orchestrated balance between galaxy growth and quenching with redshift.  
\newline\indent
Given that with the UltraVISTA data we now have a reasonably well-determined SMF at $z =$ 3.5 for star-forming galaxies, it is interesting to compare this to measurements of the SMFs of UV-selected star-forming galaxies.  In principle, our SMF of rest-frame-optical-selected star-forming galaxies should be more complete, as both UV-bright and UV-faint star-forming galaxies will be selected, where the latter are likely to be highly dust-obscured galaxies.  With few rest-frame-optical-selected SMFs for star-forming galaxies at $z >$ 3 in the literature, it is still unclear how large the population of UV-faint star-forming galaxies is.
\begin{figure}
\plotone{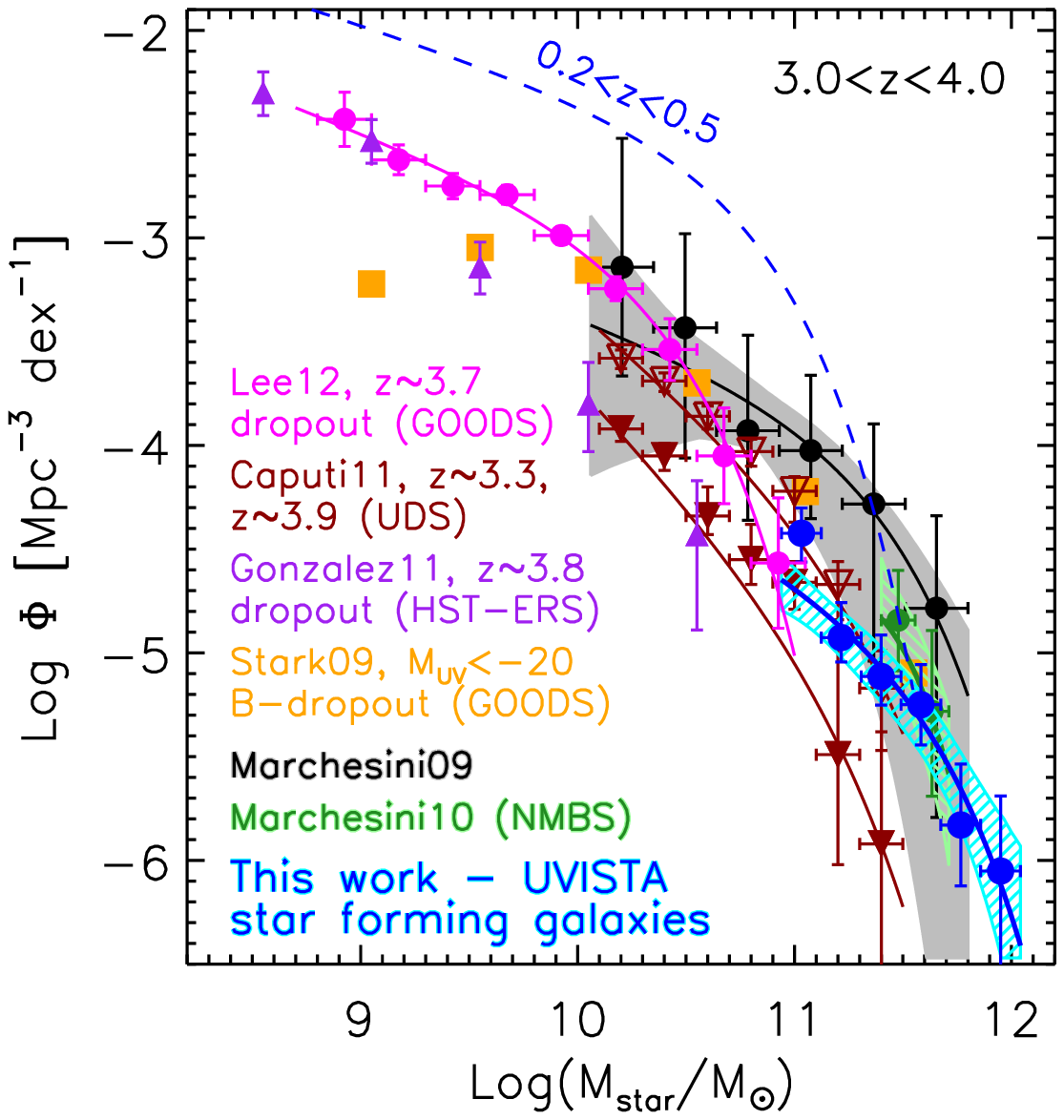}
\caption{\footnotesize Comparison of the K$_{s}$-selected SMF of star-forming galaxies at 3.0 $< z <$ 4.0 from UltraVISTA (blue) and other SMF in the literature.  The \cite{Marchesini2009} and \cite{Marchesini2010} SMFs are also K$_{s}$-selected samples and agree well with the UltraVISTA SMF.  The \cite{Caputi2011} SMFs are IRAC-selected and also agree well with the UltraVISTA SMF.  The \cite{Stark2009,Gonzalez2011}, and \cite{Lee2012} SMFs are UV-selected.  These agree reasonably well with UltrsVISTA at Log(M$_{star}$/M$_{\odot}$) = 11.0, but it appears the UV-selection may miss the most massive galaxies in this redshift range.}
\end{figure}
\newline\indent
In the left panel of Figure 11 we plot the SMF of K$_{s}$-selected star-forming galaxies at 2.0 $< z <$ 2.5, as well as the SMF of BM/BX selected galaxies by \cite{Reddy2009}.  We have converted the \cite{Reddy2009} SMF from a Salpeter IMF to a Kroupa IMF to match UltraVISTA (N. Reddy, private communication).  The BM/BX galaxies typically span the redshift range 1.9 $< z <$ 2.7 \citep{Reddy2009}, which is a reasonable match to the redshift range of the K$_{s}$-selected galaxies.  
\newline\indent
Considering the sizable systematic differences that can arise from different ways of measuring the SMF \cite[see e.g., the inter-comparison of SMFs in the literature in the appendix of][]{Marchesini2009}, the SMFs do agree reasonably well, particularly at the highest masses.  This suggests either that the BM/BX selection effectively selects the majority of massive star-forming galaxies, regardless of their UV-luminosity, or that massive dusty star-forming galaxies are less abundant than UV-bright ones at $z \sim$ 2.3.  Qualitatively speaking, the UVJ diagram (Figure 3) suggests that the colors of star-forming galaxies above the M$_{star}$-completeness limit are primarily red as compared to blue, so the high completeness of the BM/BX selection for this population is somewhat unexpected.  
\newline\indent
The low-mass-end of the SMFs do not agree well, with the BM/BX population being more abundant than the rest-frame-optical-selected population.  This is difficult to reconcile given that we have adopted strict M$_{star}$-completeness limits and that K$_{s}$-selection should be more robust than UV-selection.  Although we can not be conclusive, the discrepancy could arise from systematic effects due to the fact that these SMFs are determined in quite different ways.  \cite{Reddy2009} determine the BM/BX SMF from the UV luminosity function (LF) by using a subsample of galaxies for which they have spectroscopic redshifts and SED-determined M$_{star}$.  They convert the UV LF to a SMF assuming that the distribution of M$_{star}$ in the subsample is representative of the entire population.  This is different than our approach of fitting $z_{phot}$ and M$_{star}$ for each galaxy individually.  Interestingly, \cite{Reddy2009} also determine a much steeper faint-end slope ($\alpha$ = -1.73 $\pm$ 0.07) for the SMF than other UV-selected samples which measure the SMF with similar $z_{phot}$ and SED-fitting techniques as used for the UltraVISTA catalog \citep[e.g.,][]{Stark2009,Gonzalez2011,Lee2012}.  
\newline\indent
In the right panel of Figure 11 we compare the K$_{s}$-selected star-forming SMF at 2.5 $< z <$ 3.0 to the SMF of Lyman Break Galaxies (LBGs) from \cite{Reddy2009}.  Down to the mass limit of UltraVISTA, the shapes of those SMFs agree reasonably well, although the number density of K$_{s}$-selected star-forming galaxies is a factor of $\sim$ 2 higher than the number density of LBGs at Log(M$_{star}$/M$_{\odot}$) = 11.0.  Taken at face value, it suggests the LBG selection may miss approximately half massive galaxies at $z >$ 2.5.   A similar result was also obtained by \cite{Marchesini2010}, who found that only 8/14 galaxies in their mass-complete sample (Log(M$_{star}$/M$_{\odot}$) $>$ 11.4) of K$_{s}$-selected galaxies at 3.0 $< z <$ 4.0 would be selected with U- and B-dropout selection.  Using the VVDS spectroscopic sample \cite{Lefevre2005} and \cite{Cucciati2012} have also found that $\sim$ 50\% of the star-forming population at $z >$ 3 may obey the LBG selection criteria, although we note that that result is based on a comparison of luminosity functions, not SMFs.
\newline\indent
In Figure 12 we expand the comparison of the K$_{s}$-selected SMFs and UV-selected SMF using more recent determinations in the literature from \cite{Stark2009,Gonzalez2011} and \cite{Lee2012}.   We also compare to the K$_{s}$-selected total SMF at 3.0 $< z <$ 4.0 determined from the MUSYC/FIREWORKS/FIRES surveys \citep{Marchesini2009} and the NMBS \citep{Marchesini2010}, as well as the IRAC-selected SMFs in the UDS from \cite{Caputi2011}.  In general, most K$_{s}$-selected galaxies at this redshift are star-forming galaxies (e.g., Figure 6), so comparing the total SMFs from \cite{Marchesini2009,Marchesini2010} and \cite{Caputi2011} to the star-forming SMFs is a reasonable comparison.  We note that all of the SMFs in Figure 12 have been determined with a method similar to the UltraVISTA K$_{s}$-selected catalog, e.g., $z_{phot}$ from broadband photometry, and M$_{star}$ from SED-fitting with similar assumptions about star formation histories.  These SMFs have a slightly higher median redshift than the BM/BX and LBG SMFs, so we compare to the highest-redshift SMF in UltraVISTA, 3.0 $< z <$ 4.0.  
\newline\indent
Beginning with the K$_{s}$-selected samples, we find that within the errors, the K$_{s}$ and IRAC-selected SMFs agree reasonably well, although the region of comparison is limited to fairly high M$_{star}$.  Interestingly, all three show little evolution in the number density of the most massive galaxies (at Log(M$_{star}$/M$_{\odot}$) $>$ 11.5) from $z =$ 3.5 to $z =$ 0.  The largest discrepancy between the K$_{s}$-selected SMFs is for UltraVISTA and MUSYC at Log(M$_{star}$/M$_{\odot}$) = 11.0, where the number density from UltraVISTA is a factor of $\sim$ 3 less abundant in such galaxies than MUSYC.  Comparison of the 1/V$_{max}$ points shows that this difference is not significant.  
\newline\indent
If confirmed, the result that the abundance of Log(M$_{star}$/M$_{\odot}$) $>$ 11.5 star-forming galaxies is unchanged since $z =$ 3.5 is quite interesting.  We note that at present we cannot be 100\% confident of how real this population is.  Examination of their SEDs shows that the fits are reasonable; however, they are almost all extremely red and dusty.  It is also difficult to tell if emission from an AGN causes their M$_{star}$ to be inflated.  
\newline\indent
At present there are no spectroscopic redshifts for such galaxies so we cannot rule out the possibility that they are extremely dusty galaxies from lower-$z$ \citep[see also the discussion in][]{Marchesini2010}.  We explore this possibility further in the appendix and find that the use of an additional very dusty template in $z_{phot}$ fitting can reduce the number density of this population by a factor of $\sim$ 2.  Given this, at present it may be best to consider their abundance an upper limit to the true abundance.
\newline\indent
Returning to the UV-selected samples, comparison of the K$_{s}$-selected SMFs to the UV-selected SMFs also shows reasonable agreement in the limited mass range where the surveys overlap.  At Log(M$_{star}$/M$_{\odot}$) = 11.0 the K$_{s}$-selected SMF lies between the \cite{Stark2009} and \cite{Lee2012} SMFs.  Extrapolation of the best-fit Schechter functions of both of those samples does not agree well with the abundance of massive star-forming galaxies in UltraVISTA.  A simultaneous Schechter function fit to the \cite{Lee2012} data and the UltraVISTA data does not produce a good fit.  There seem to be two possible reasons for this.  Either UV-selection misses most of the most massive star-forming galaxies at $z >$ 3 due to dust obscuration, or that the abundance of these galaxies is overestimated in the K$_{s}$-selected sample due to a very dusty population at lower redshift.
\newline\indent
Overall, the SMFs of rest-frame-optical-selected samples and UV-selected samples compare reasonably well in the mass range where they overlap.  There is some evidence that the UV-selection is not complete for galaxies with Log(M$_{star}$/M$_{\odot}$) $>$ 11.0; however, spectroscopic verification of the massive population K$_{s}$-selected population will be important to verify this result.  
\begin{figure}
\plotone{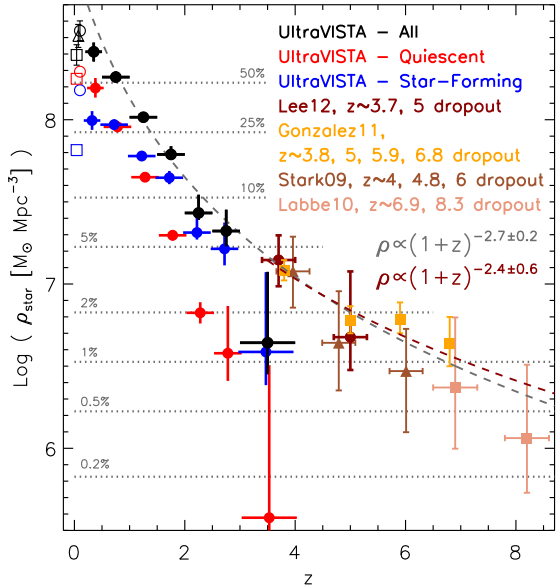}
\caption{\footnotesize Evolution of the stellar mass density in the universe between $z =$ 0 -- 8.5.  The SMDs determined from UV-selected samples are shown at $z >$ 3.5.  Below $z <$ 3.5 the K$_{s}$-selected SMDs from UltraVISTA are shown for the total (black), star forming (blue) and quiescent (red) populations.  The $z \sim$ 0 data from \cite{Cole2001} (triangle), \cite{Bell2003} (circles) and \cite{Baldry2012} (squares) are also shown.  The SMD in star-forming galaxies from the K$_{s}$-selected and UV-selected samples agrees to within 1$\sigma$ suggesting that UV-selected samples account for most of the SMD at $z >$ 3.5.  The dashed gray curve shows a simultaneous fit to the total SMD from UltraVISTA at $z >$ 1.5 and the UV-selected samples, and the dashed maroon shows a fit to just the UV-selected samples, both of which agree well.}
\end{figure}
\begin{figure*}
\plotone{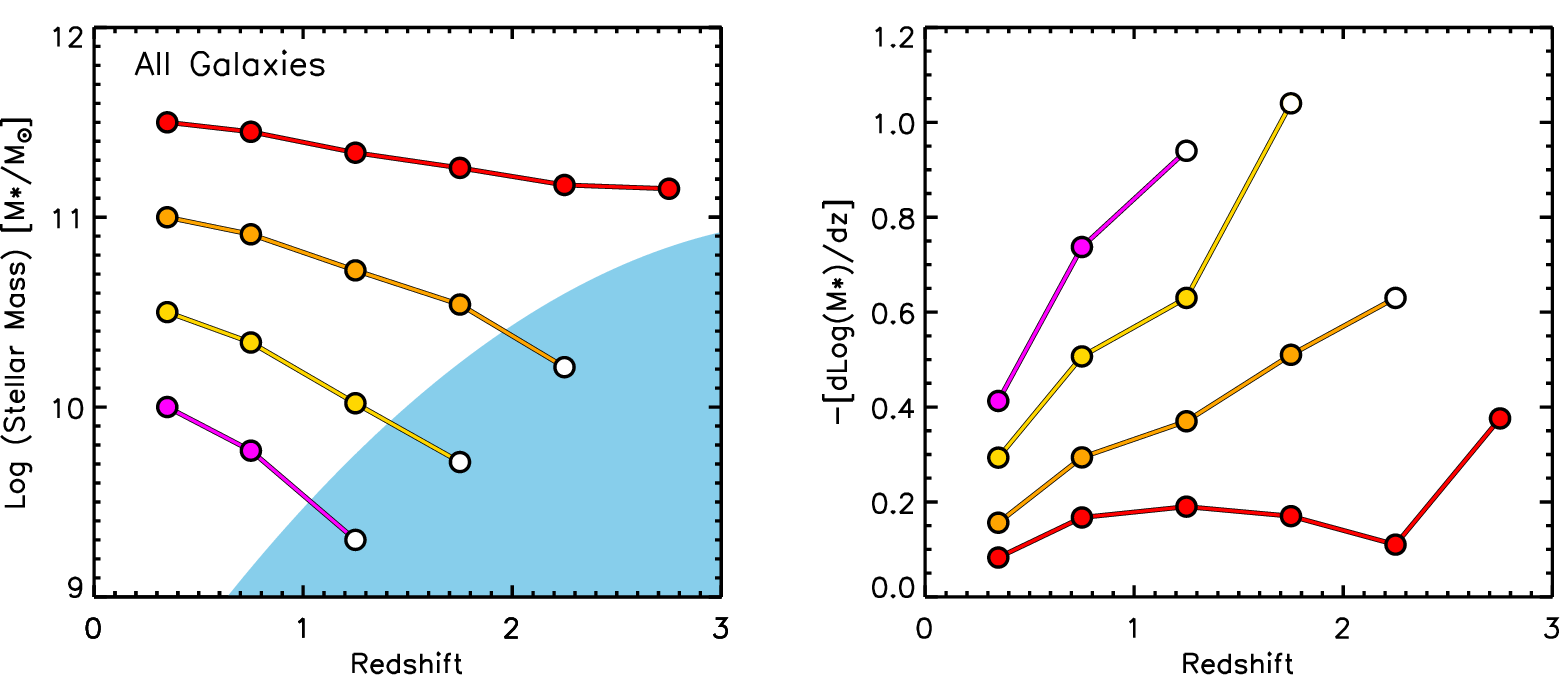}
\caption{\footnotesize Left panel: Average stellar mass of galaxies chosen at a fixed cumulative number density from the Schechter function fits.  The cumulative number densities are chosen so that they correspond to galaxies with Log(M$_{star}$/M$_{\odot}$) = 11.5, 11.0, 10.5, and 10.0 in the lowest-redshift SMF.  The shaded blue region represent regions that require extrapolating the Schechter function beyond the depth of the data.  Right panel: Derivatives of the growth in stellar mass as a function of redshift.  The derivatives separate into a sequence showing that the rate mass growth is always larger for lower-mass galaxies at $z <$ 2.}
\end{figure*}
\subsection{Evolution of the Stellar Mass Density}
\indent
The left panel of Figure 7 shows that the measured evolution of the stellar mass density (hereafter, SMD) for the combined population is well-determined and in excellent agreement with previous determinations.  Our data show that the SMD of the universe was only 50\%, 10\% and 1\% of its current value at $z \sim$ 1.0, 2.0, and 3.5, respectively.
\newline\indent
The left panel of Figure 8 shows the evolution of the SMD of the quiescent and star-forming populations separately.  The SMD in these populations evolves quite differently since $z =$ 3.5, a conclusion that was already apparent from the comparison of the SMFs themselves.  The SMD in star-forming galaxies increases at a rate of $\rho_{star}$ $\propto$ (1 + z)$^{-2.3\pm0.2}$, whereas the SMD in quiescent galaxies evolves much faster, as $\rho_{star}$ $\propto$ (1 + z)$^{-4.7\pm0.4}$.  This strong differential evolution in the SMD can also be seen in the inset panel of Figure 8 where we plot the fraction of the total SMD in star-forming and quiescent galaxies as a function of redshift.
\newline\indent 
As Figure 7 shows, at $z =$ 3.5 the universe contained only $\sim$ 1\% of the total M$_{star}$ formed by $z =$ 0.  The inset of Figure 8 shows that at that time the fraction of quiescent galaxies was small and approximately 90\% of the total SMD was contained within star-forming galaxies.  Since $z =$ 3.5, the fraction of the SMD in quiescent galaxies has grown continuously and at $z \sim$ 0.75 the SMD in quiescent galaxies became approximately equal to the SMD in star-forming galaxies.  Perhaps coincidently, this equality in SMD between the two types also occurs precisely at the redshift when $\sim$ 50\% of the total SMD of the universe has formed.  Thereafter, the SMD in quiescent galaxies exceeds the SMD in star-forming galaxies, although we note that the details of this statement depend on the low-redshift comparison samples, as there are notable differences between the local studies of \cite{Bell2003} and \cite{Baldry2012}.
\newline\indent
Due to the superior area and depth of the UltraVISTA data, we can also compare SMDs determined from K$_{s}$-selected samples at $z =$ 3.5 to those determined from UV-selected samples at the same redshift.  In Figure 13 we plot the SMD from the UltraVISTA data as well as from the UV-selected samples of \cite{Stark2009}, \cite{Labbe2010b}, \cite{Gonzalez2011}, and \cite{Lee2012}.  We have also included the SMDs at $z =$ 0 from \cite{Cole2001,Bell2003} and \cite{Baldry2012}, which means that the measurements of the SMD in Figure 13 span an impressive redshift baseline of $z =$ 0.0 -- 8.5.
\newline\indent
If we compare the K$_{s}$-selected SMD at $z =$ 3.5 to the SMDs of the UV-selected SMDs at $z \sim$ 3.7 determined from \cite{Stark2009}, \cite{Gonzalez2011}, and \cite{Lee2012} there is reasonable agreement.  The SMD from the UV-selected samples agree well with each other and are systematically $\sim$ 0.5 dex higher than the K$_{s}$-selected measurement.   The uncertainties in the K$_{s}$-selected SMD at $z =$ 3.5 are quite large because the data just reaches M$^{*}_{star}$ and requires a substantial extrapolation of the Schechter function.  Although lower, the K$_{s}$-selected SMD does agree with the UV-selected SMD within the 1$\sigma$ uncertainties.  As shown in Figure 12, the UV-selected SMFs under predict the number density of the most massive galaxies, so the agreement of the total SMDs demonstrates that the total M$_{star}$ contained in the most massive galaxies is negligible compared to lower-mass galaxies.
\newline\indent
The reasonable agreement between the total SMD of the K$_{s}$-selected sample and the UV-selected sample illustrates a key point, namely that at $z >$ 3.5, UV-selected samples do select samples that account for most of the SMD in the universe.  Even though direct comparison of the SMFs in $\S$ 5.3 showed that UV-selection misses approximately half of the massive star-forming galaxies and all of the massive quiescent galaxies, Figure 13 shows that the SMD in such galaxies at $z =$ 3.5 is fairly small.  Therefore, UV selection appears to be quite complete for the majority of star-forming galaxies at these redshifts.
\newline\indent
Given that our results suggest that UV-selected samples should be representative of the SMD at $z >$ 3.5, it seems reasonable to fit the SMD over a large redshift baseline.  The star formation rate density shows a clear decline at $z <$ 1.5 \citep[e.g.,][]{Hopkins2006,Bouwens2011,Sobral2013}, so we fit the data down to that redshift.  Including the UV-selected samples with UltraVISTA we find that the total SMD evolves as $\rho_{star}$ $\propto$ (1 + $z$)$^{-2.7\pm0.2}$ from $z =$ 8.5 -- 1.5.  This fit is also in good agreement with a fit to just the UV-selected samples which evolve as $\rho_{star}$ $\propto$ (1 + $z$)$^{-2.4\pm0.6}$.
\newline\indent
The exponent in the overall fit is tantalizingly similar to the volume growth of the universe.  Why the density of stars in galaxies increases at such a rate may be purely coincidental, but obviously needs to be investigated in more detail and in the context of models of galaxy evolution.  
\subsection{The Average Mass Growth of Galaxies}
The evolution of the SMFs and SMDs illustrate the cosmological evolution of the distribution of galaxies as a function of M$_{star}$, and the integrated M$_{star}$ within the overall galaxy population.  While important quantities, from the standpoint of modeling the process of galaxy evolution, measurements of how individual galaxies assemble their mass may be more useful.  Such measurements are difficult, as they requires the non-trivial task of linking galaxies and their descendants through cosmic time.
\newline\indent
Several studies have argued that one approach for linking galaxies to their descendants is to select galaxies at a fixed constant number density \citep[e.g.,][]{vandokkum2010,Papovich2011}.  More recently, \cite{Brammer2011} argued that selecting galaxies at a fixed cumulative number density is a better approach, as it is single-valued at all M$_{star}$.  In a recent paper, \cite{Leja2013} tested this method on semi-analytic models (SAM) of galaxy formation.  They found that selection at fixed cumulative number density recovered the mass evolution of the population to an accuracy $\sim$ 0.15 dex.  They found that the fixed cumulative number density approach may underpredict the mass growth of high-mass galaxies, and overpredict the mass growth of lower-mass galaxies, although they note that this result depends on the evolution of galaxies in the SAM model, which does not properly reproduce the SMF as a function of redshift.
\newline\indent
Here we perform fixed cumulative number density selection using the UltraVISTA SMFs in order to measure the average mass evolution of the population.  We have chosen four fixed cumulative number densities to follow.  These cumulative number densities are chosen to correspond to the cumulative number density for galaxies with Log(M$_{star}$/M$_{\odot}$) = 11.5, 11.0, 10.5, and 10.0, in the lowest-redshift SMF bin (0.2 $< z <$ 0.5).  
\newline\indent
In Figure 14 we plot the M$_{star}$ at these four fixed cumulative number densities out to $z =$ 3.  The solid shaded region in Figure 14 represents the M$_{star}$-completeness limits of the survey.  Cumulative number densities that require extrapolation of the Schechter function to M$_{star}$ below the completeness limits are shown as open circles.
\newline\indent
The UltraVISTA data is sufficiently complete, and the M$_{star}$ evolution sufficiently slow that we measure the mass growth of the Log(M$_{star}$/M$_{\odot}$) = 11.5 population out to $z \sim$ 3.0.  This population demonstrates a remarkably slow growth in M$_{star}$, increasing by only by 0.3 dex since $z \sim$ 3, and 0.2 dex since $z \sim$ 2.  Using the same method \cite{Brammer2011} measured a growth of 0.17 dex for this population since $z \sim$ 2, in excellent agreement with our finding.
\newline\indent
With the caveat that increasingly larger extrapolations are required, Figure 14 suggests that at $z <$ 2 the growth of galaxies with lower M$_{star}$ is always faster than galaxies with higher M$_{star}$.  This can be seen clearer in the right panel of Figure 14 where we plot the derivative of the M$_{star}$-growth curves.  These curves present a remarkable sequence at all $z <$ 3, where the derivatives are always higher for lower-mass galaxies.
\newline\indent
Figure 14 encapsulates what has already been shown by many previous studies, namely that there is a ``downsizing" of the galaxy population such that the highest-mass galaxies assemble most of their M$_{star}$ at high-redshift, whereas lower-mass galaxies grow more slowly over cosmic time.  One of the most interesting implications of the curves in Figure 14 is that although low-mass galaxies grow at higher rates than higher-mass galaxies at $z <$ 2, they do not ``overtake" their more massive counterparts.  This is simply because the most massive galaxies have assembled such significant amounts of M$_{star}$ at very early times.  It also implies that at redshifts higher than probed by the current data, the mass assembly rate of the most massive galaxies must be extremely rapid.  
\newline\indent
For example, galaxies with Log(M$_{star}$/M$_{\odot}$) = 11.5 at $z \sim$ 0.35 have to assemble as much mass at $z <$ 3 as they do at $z >$ 3.  The amount of cosmic time between 0 $< z <$ 3 is $\sim$ 5$\times$ more than at 3 $< z <$ $\infty$, which implies that even if those galaxies begin forming star shortly after the big bang, their mass assembly rates must be substantially higher in the past, and must be substantially higher than those of low-mass galaxies at any epoch.
\section{Summary}
In this paper we have presented the SMFs of star-forming and quiescent galaxies to $z =$ 4 using a K$_{s}$-selected catalog of the COSMOS/UltraVISTA field.  The catalog is unique in terms of the large areal coverage (1.62 deg$^2$) with a significant depth (K$_{s,tot}$ $=$ 23.4, 90\% completeness).  The high quality of the data allows for arguably the best measurement of the SMFs over a large redshift baseline to date.  
\newline\indent
The total SMFs agree well with previous measurements at 0.2 $< z <$ 3.5, particularly at the high-mass end where the wide-field UltraVISTA data provides a substantial improvement in the statistical uncertainties.  We find no significant evolution in M$^{*}_{star}$ out to $z =$ 3.5, although the uncertainties in M$^{*}_{star}$ are large at $z >$ 2.5.  There is also a significant evolution in $\Phi^{*}$ out to $z =$ 3.5.  These results are also consistent with the results of \cite{Ilbert2013}, who also computed the SMFs using the UltraVISTA data.  Most of the evolution in the total SMFs is driven by the quiescent population, which grows approximately twice as fast as the star forming population at all redshifts.  Integrating the SMFs we find that the SMD of the universe was only 50\%, 10\%, and 1\% of its current value at $z \sim$ 1.0, 2.0, and 3.5, respectively.  
\newline\indent
Classification of star-forming and quiescent galaxies was performed using the rest-frame UVJ diagram.  The SMFs of these populations evolve quite differently out to $z =$ 4.  The quiescent population evolves much faster than the star-forming population, and its growth drives most of the growth in the combined SMF.  The SMD contained in star-forming galaxies grows as (1 + $z$)$^{-2.3\pm0.2}$ whereas the SMD in quiescent galaxies grows much faster, as (1 + $z$)$^{-4.7\pm0.4}$.  The fraction of the total SMD contained in quiescent galaxies increases with decreasing redshift, and at $z \sim$ 0.75, the SMD in quiescent galaxies becomes equal to that in star-forming galaxies.  This equivalence in SMD occurs at the redshift where $\sim$ 50\% of the current SMD has formed.  
\newline\indent
Starting from low-redshift we find that quiescent galaxies dominate the SMF at high-masses, but that dominance declines to higher masses with increasing redshift.  At $z >$ 2.5, star-forming galaxies dominate the SMF at all stellar masses.  
\newline\indent
Comparison of the SMFs of the K$_{s}$-selected star-forming galaxies to the SMFs of UV-selected star-forming galaxies at 2.5 $< z <$ 4.0 shows reasonable agreement.  It appears that the UV-selected samples do miss approximately half of the population of Log(M$_{star}$/M$_{\odot}$) $>$ 11.0 star-forming galaxies at $z >$ 2.5 due to the fact that they are very dusty; however, we note that this population of massive dusty star-forming galaxies does need to be spectroscopically confirmed.  
\newline\indent
Comparison of the SMD for the K$_{s}$-selected and UV-selected samples shows that they agree within the uncertainties, which implies that even if the massive dusty population exists, it contributes relatively little to the total SMD at $z >$ 3.5.  This suggests that UV-selected samples at $z >$ 3.5 are likely to be representative of the majority of the SMD in the universe.  Given this consistency we combined the UltraVISTA SMDs with UV-selected SMDs from the literature and fit the evolution from 1.5 $< z <$ 8.5.  We find that the SMD evolves as (1 + $z$)$^{2.7\pm0.2}$, similar to the volume growth of the universe.
\newline\indent
We also perform selection at fixed cumulative number density to measure the average growth in the M$_{star}$ of galaxies.  We find that at $z <$ 2, the derivatives of the M$_{star}$ growth are always larger for lower-mass galaxies, which shows that since that time galaxy growth is mass-dependent and primarily bottom-up.  In the following appendix we take a closer look at the effects of the assumptions made in SED modeling on the SMFs.  We also examine the effect of the different definitions of star-forming and quiescent galaxies on the SMF.  
\acknowledgements
We would like to thank Henk Hoekstra and Remco van der Burg for generously making available the computing time for Monte Carlo simulations used in this paper.  DM acknowledges support from program HST\_AR\_12141.01, provided by NASA through a grant from the Space Telescope Science Institute, which is operated by the Association of Universities for Research in Astronomy Incorporated under NASA contract NAS-26555.  DM also acknowledges support from Tufts University Mellon Research Fellowship in Arts and Sciences.  BMJ and JPUF acknowledges support from the ERC-StG grant EGGS-278202.  The Dark Cosmology Centre is funded by the Danish National Research  Foundation.  JSD acknowledges the support of the European Research Council through the award of an Advanced Grant, and the support of the Royal Society via a Wolfson Research Merit Award.
\begin{appendix}
\begin{center}
THE EFFECT OF SPS MODELS, METALLICITY AND STAR FORMATION HISTORY ON THE SMF
\end{center}
\begin{figure}
\plotone{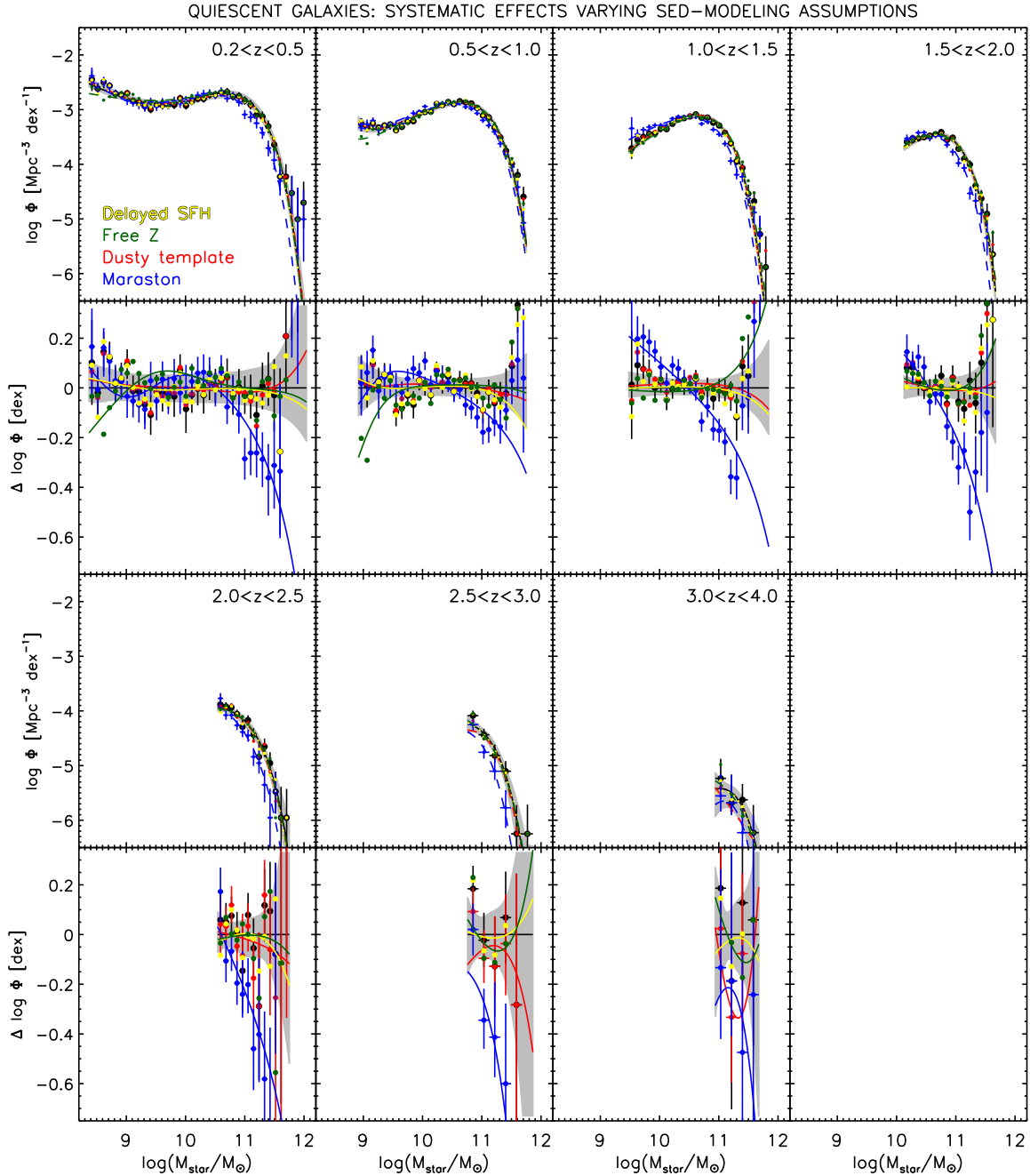}
\caption{\footnotesize First and third rows: Stellar mass functions for quiescent galaxies in different redshift bins.  The stellar mass functions have been determined using the different SED modeling assumptions in the legend (see text for details).  Second and fourth rows: Difference in measured number density at a given stellar mass compared to the default SMF.  The shaded region represents the formal uncertainty (including Poisson noise, cosmic variance, and modeling errors) in the default mass function.  The use of the Maraston models and a free metallicity affect the quiescent SMFs at levels larger than the formal uncertainties, but the dusty template and different star formation history do not.}
\end{figure}
\begin{figure}
\plotone{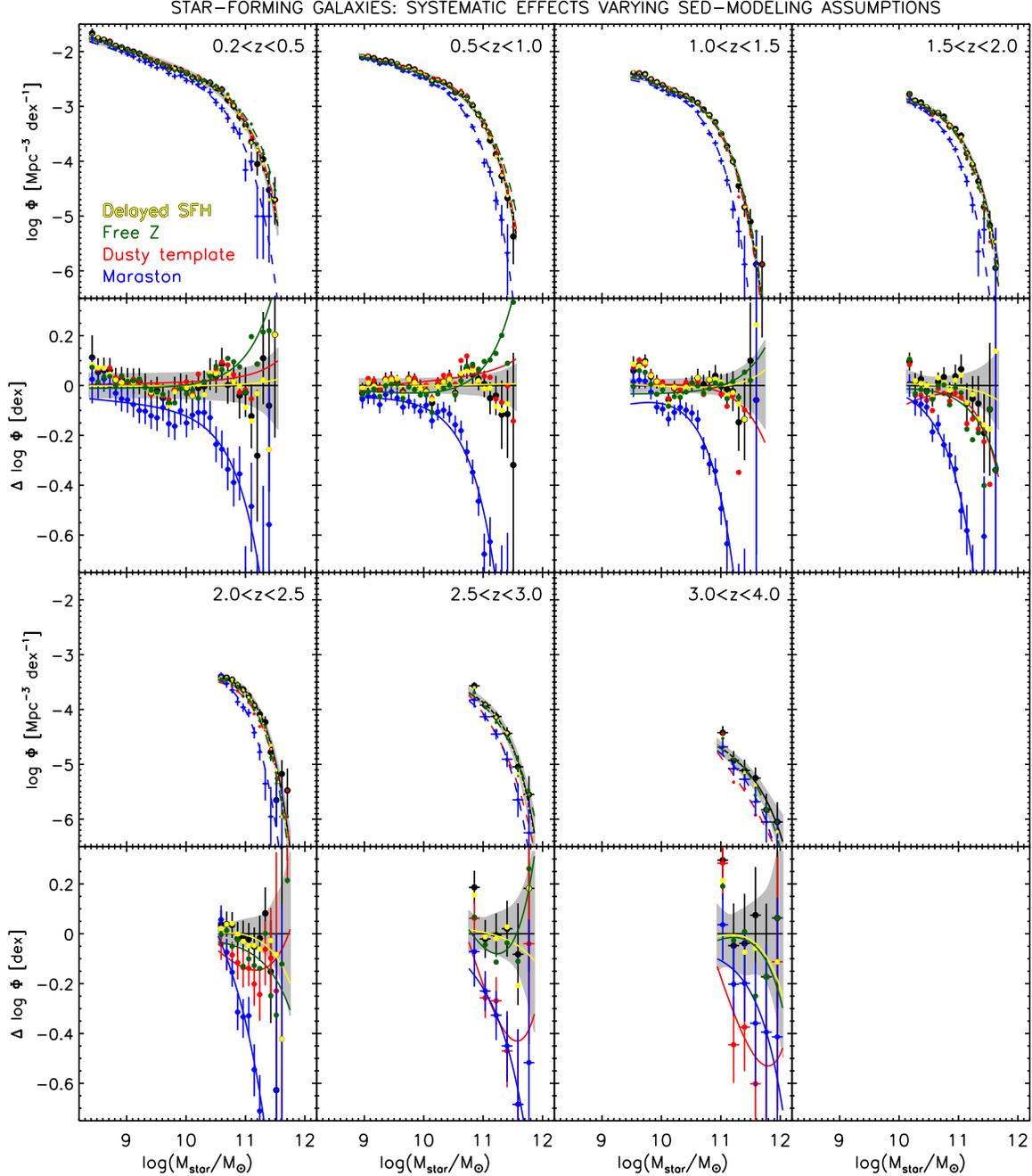}
\caption{\footnotesize As Figure 15, but for star-forming galaxies.  The stellar mass function of star-forming galaxies is not affected by free metallicity or a different star formation history.  The SMF is affected by the use of the Maraston models at all redshifts, and at the highest-redshift the use of the dusty template reduces the number density of massive star-forming galaxies.}
\end{figure}
\begin{figure}
\plotone{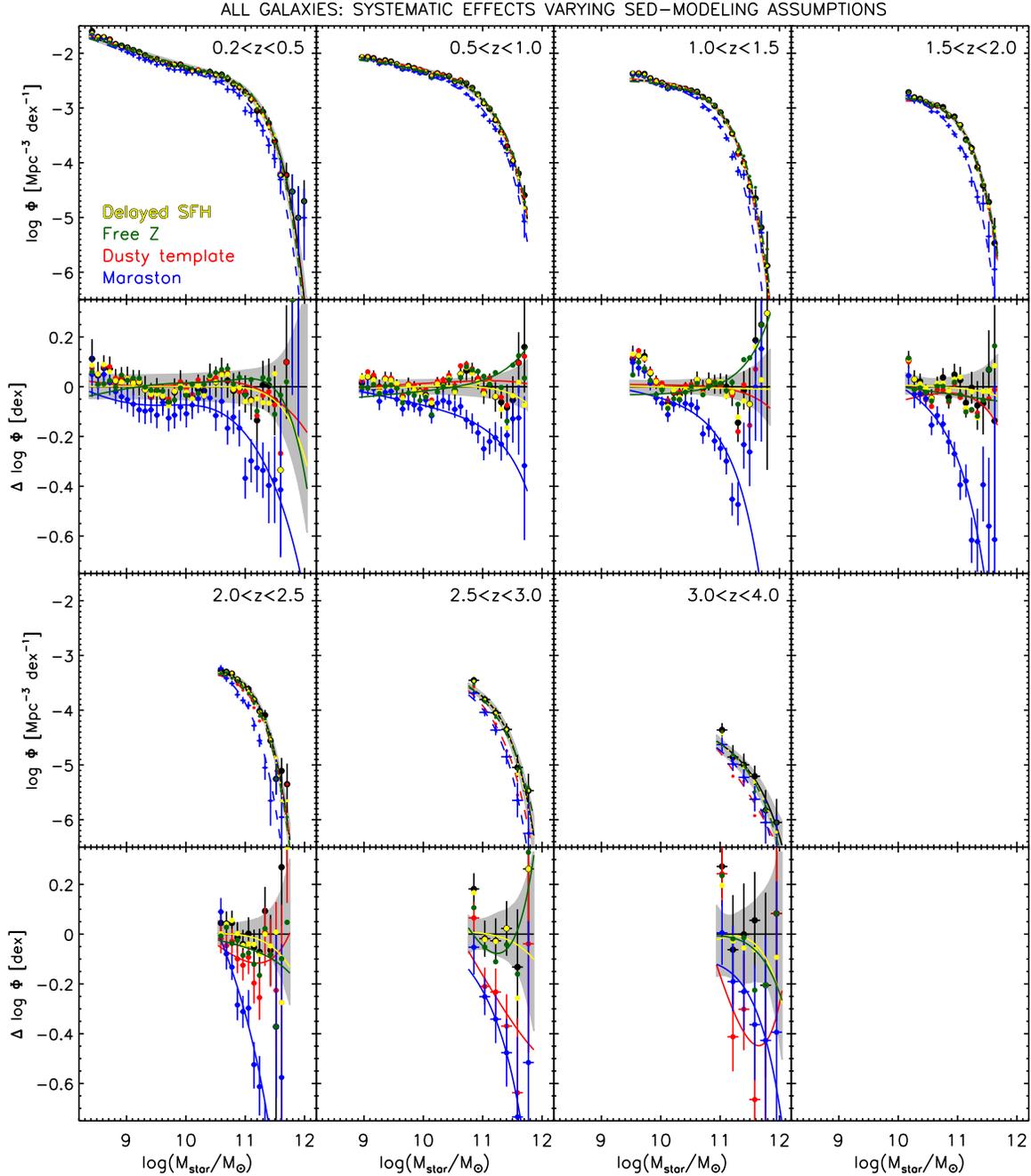}
\caption{\footnotesize As Figure 15, but for the combined population.  Again, the Maraston models and the dusty template have the largest impact on the SMFs.  }
\end{figure}
\begin{deluxetable}{lccccccc}
\tabletypesize{\scriptsize}
\centering
\tablecaption{Best-fit Schechter Function Parameters of the SMFs: Different SED-modeling assumptions\label{smfsys1}}
\tablehead{\colhead{Redshift} & \colhead{Sample} & \colhead{Number} &  
           \colhead{$\log{M^{\star}_{\rm star}}$} & \colhead{$\Phi^{\star}$} & 
           \colhead{$\alpha$} &  \colhead{$\Phi^{\star}_{\rm 2}$} &  \colhead{$\alpha_{\rm 2}$} \\
                     &   &   &  ($M_{\odot}$) & (10$^{-4}$~Mpc$^{-3}$) 
                     &   &  (10$^{-4}$~Mpc$^{-3}$)  &   }
\startdata
$0.2\leq z <0.5$ & A,SFH      & 18511 &  11.22$^{+0.02}_{-0.03}$ & 12.13$^{+0.83}_{-0.45}$ & -1.29$\pm$0.01 & \nodata & \nodata \\
$0.2\leq z <0.5$ & A,MET      & 18644 &  11.20$^{+0.03}_{-0.02}$ & 13.98$^{+0.79}_{-0.55}$ & -1.26$\pm$0.01 & \nodata & \nodata \\
$0.2\leq z <0.5$ & A,DUSTY    & 18885 &  11.20$^{+0.02}_{-0.03}$ & 12.80$^{+0.81}_{-0.47}$ & -1.28$\pm$0.01 & \nodata & \nodata \\
$0.2\leq z <0.5$ & A,MA05     & 15543 &  11.10$^{+0.03}_{-0.03}$ & 10.65$^{+0.46}_{-0.65}$ & -1.30$\pm$0.01 & \nodata & \nodata \\
$0.2\leq z <0.5$ & A,SFH      & 18511 &  11.06$\pm$0.01        & 18.98$\pm$0.14         & -1.2           & \nodata & \nodata \\
$0.2\leq z <0.5$ & A,MET      & 18644 &  11.10$\pm$0.01        & 18.63$\pm$0.14         & -1.2           & \nodata & \nodata \\
$0.2\leq z <0.5$ & A,DUSTY    & 18885 &  11.05$\pm$0.01        & 19.51$^{+0.07}_{-0.14}$  & -1.2           & \nodata & \nodata \\
$0.2\leq z <0.5$ & A,MA05     & 15543 &  11.93$\pm$0.01        & 17.58$\pm$0.14         & -1.2           & \nodata & \nodata \\
$0.2\leq z <0.5$ & A,SFH      & 18511 &  10.94$^{+0.11}_{-0.05}$ & 17.21$^{+3.29}_{-2.55}$  & -0.47$^{+0.16}_{-0.37}$ & 10.02$^{+2.12}_{-5.66}$ & -1.37$^{+0.03}_{-0.10}$ \\
$0.2\leq z <0.5$ & A,MET      & 18644 &  10.91$^{+0.08}_{-0.03}$ & 17.80$^{+2.43}_{-4.24}$  & -0.27$^{+0.11}_{-0.32}$ & 12.94$^{+3.35}_{-4.20}$ & -1.31$^{+0.03}_{-0.03}$ \\
$0.2\leq z <0.5$ & A,DUSTY    & 18885 &  10.96$\pm$0.08        & 19.56$^{+1.87}_{-4.61}$  & -0.68$^{+0.32}_{-0.22}$ &  7.25$^{+5.22}_{-2.69}$ & -1.42$\pm$0.06 \\
$0.2\leq z <0.5$ & A,MA05     & 15543 &  10.94$\pm$0.05        & 15.23$^{+1.64}_{-2.44}$  & -0.93$^{+0.21}_{-0.11}$ &  3.23$^{+4.05}_{-0.52}$ & -1.53$^{+0.10}_{-0.05}$ \\
$0.2\leq z <0.5$ & Q,SFH      &  4360 &  11.20$\pm$0.03     & 10.09$^{+0.56}_{-0.52}$     & -0.92$\pm$0.02 & \nodata & \nodata \\
$0.2\leq z <0.5$ & Q,MET      &  4505 &  11.14$\pm$0.03     & 12.21$^{+0.77}_{-0.55}$     & -0.87$\pm$0.02 & \nodata & \nodata \\
$0.2\leq z <0.5$ & Q,DUSTY    &  4362 &  11.20$\pm$0.03     &  9.97$\pm$0.54            & -0.93$\pm$0.02 & \nodata & \nodata \\
$0.2\leq z <0.5$ & Q,MA05     &  4135 &  11.07$\pm$0.03     & 10.18$^{+0.67}_{-0.51}$     & -0.92$\pm$0.02 & \nodata & \nodata \\
$0.2\leq z <0.5$ & Q,SFH      &  4360 &  10.75$\pm$0.01     & 30.63$^{+0.03}_{-0.02}$      & -0.4              & \nodata & \nodata \\
$0.2\leq z <0.5$ & Q,MET      &  4505 &  10.75$\pm$0.02     & 31.66$\pm$0.02             & -0.4              & \nodata & \nodata \\
$0.2\leq z <0.5$ & Q,DUSTY    &  4362 &  10.74$\pm$0.01     & 30.67$^{+0.03}_{-0.02}$      & -0.4              & \nodata & \nodata \\
$0.2\leq z <0.5$ & Q,MARASTON &  4135 &  10.63$\pm$0.01     & 29.38$^{+0.02}_{-0.03}$      & -0.4              & \nodata & \nodata \\
$0.2\leq z <0.5$ & Q,SFH      &  4360 &  10.92$^{+0.05}_{-0.03}$  & 19.73$^{+1.23}_{-1.56}$  & -0.38$\pm$0.11         & 0.48$^{+0.36}_{-0.21}$ & -1.57$\pm$0.11 \\
$0.2\leq z <0.5$ & Q,MET      &  4505 &  10.92$^{+0.05}_{-0.03}$  & 18.25$^{+1.76}_{-1.90}$  & -0.32$^{+0.11}_{-0.16}$  & 2.26$^{+0.29}_{-1.16}$ & -1.22$\pm$0.11 \\
$0.2\leq z <0.5$ & Q,DUSTY    &  4362 &  10.94$^{+0.03}_{-0.05}$  & 18.62$^{+1.63}_{-1.13}$  & -0.43$^{+0.11}_{-0.05}$  & 0.45$^{+0.34}_{-0.20}$ & -1.57$\pm$0.11 \\
$0.2\leq z <0.5$ & Q,MARASTON &  4135 &  10.86$^{+0.03}_{-0.05}$  & 17.25$^{+1.82}_{-0.60}$  & -0.59$\pm$0.11         & 0.06$^{+0.13}_{-0.01}$ & -1.95$\pm$0.21 \\
$0.2\leq z <0.5$ & SF,SFH      & 14151 &  10.82$\pm$0.03     & 11.26$^{+0.70}_{-0.72}$ & -1.34$\pm$0.01 & \nodata & \nodata \\
$0.2\leq z <0.5$ & SF,MET      & 14139 &  10.91$\pm$0.03     & 10.28$^{+0.65}_{-0.67}$ & -1.34$\pm$0.01 & \nodata & \nodata \\
$0.2\leq z <0.5$ & SF,DUSTY    & 14523 &  10.83$\pm$0.03     & 11.64$^{+0.93}_{-0.68}$ & -1.33$\pm$0.02 & \nodata & \nodata \\
$0.2\leq z <0.5$ & SF,MA05     & 11408 &  10.62$\pm$0.04     & 10.47$^{+0.93}_{-0.66}$ & -1.36$\pm$0.02 & \nodata & \nodata \\
$0.2\leq z <0.5$ & SF,SFH      & 14151 &  10.76$\pm$0.02 & 13.42$^{+0.20}_{-0.13}$ & -1.3              & \nodata & \nodata \\
$0.2\leq z <0.5$ & SF,MET      & 14139 &  10.85$\pm$0.02 & 12.35$^{+0.18}_{-0.12}$ & -1.3              & \nodata & \nodata \\
$0.2\leq z <0.5$ & SF,DUSTY    & 14523 &  10.78$\pm$0.02 & 13.57$^{+0.13}_{-0.20}$ & -1.3              & \nodata & \nodata \\
$0.2\leq z <0.5$ & SF,MARASTON & 11408 &  10.54$\pm$0.02 & 13.49$^{+0.21}_{-0.14}$ & -1.3              & \nodata & \nodata \\
\enddata
\tablecomments{This table is available in its full form in the electronic journal.  A subsample of the data is shown here to represent its form.} 
\end{deluxetable}
\indent
The determination of stellar masses from SED modeling requires making assumptions about various quantities that are not well-constrained by broadband photometry alone, such as the SFH, dust attenuation law, and metallicity.  Depending on the assumptions made, systematically different estimates of M$_{star}$ \citep[e.g.,][]{Maraston2006,Muzzin2009b,Muzzin2009c,Conroy2009} and the SMF \citep[e.g.,][]{Marchesini2009} can result for a given dataset.  Previous work has shown that typically the largest systematic uncertainties in M$_{star}$ and the overall SMF arise from the choice of the SPS model \citep[e.g.,][]{Marchesini2009, Muzzin2009b}, with metallicity and the choice of dust attenuation law being lesser, although not unimportant sources of systematic error.  In one of the first works to do a careful study of these systematic errors, \cite{Marchesini2009} found that systematic uncertainties in the SMF from the MUSYC survey were approximately as large a contribution to the overall error budget as the random errors.  Given that the UltraVISTA dataset covers an area $>$ 10 times larger than the MUSYC survey, and has superior photometric data, in this appendix we re-examine possible sources of systematic uncertainties on the SMF, with the expectation is that they are likely to dominate the overall error budget.
\newline\indent
We explore four different possible sources of systematic error.  For each case considered we re-perform the SED fitting, generating a completely new catalog of M$_{star}$ for each galaxy.  The first source of systematic error we test is to change the SPS model from BC03 to the models of \cite{Maraston2005}, hereafter M05.  The M05 models have a different treatment of the thermally-pulsating asymptotic horizontal branch stars (TP-AGB) which, amongst other differences with BC03 causes them to produce M$_{star}$ that are typically a factor of $\sim$ 0.65 lower than those from the BC03 models \citep[e.g.,][]{Wuyts2007,Muzzin2009b,Marchesini2009}.  Whether these models are a better treatment of this phase remains an open issue \citep[e.g.,][]{Kriek2010,Zibetti2013}.  
\newline\indent
We also explore the effect of leaving metallicity as a free parameter.  The FAST code allows four different metallicities for the BC03 models, $Z$ = 0.004, 0.008, 0.02, and 0.05, which span a range of sub-solar to super-solar.  We also explore the effect of using a different SFH.  \cite{Maraston2010} showed that exponential-declining models were likely an overestimate of the M$_{star}$ for star-forming galaxies.  In order to allow for an increasing SFH, we explore the effect of the ``Delayed $\tau$-model" option in FAST.  The Delayed-$\tau$ is a SFH of the form SFR $\propto$ t$\times$$e^{-t/\tau}$, which begins with a smoother growth in the SFH before the exponential decline.
\newline\indent
Lastly, we explore the effect of using an additional very red, ``old-and-dusty" template in EAZY when fitting the $z_{phot}$.  Using NIR medium band data from the NMBS, \cite{Marchesini2010} found that the inclusion of such a template caused approximately half of the massive galaxies population at $z >$ 3 to be consistent with a somewhat lower $z_{phot}$ in the range 2 $< z <$ 3.  At present there is no strong evidence from spectroscopic samples that such an old-and-dusty population exists; however, there are clearly strong selection biases against obtaining successful spectroscopic redshifts for such a population if it were to exist (i.e., both age and dust make the detection of emission lines unlikely).  Given the prevalence of such red SEDs at the high-mass end of the SMF at high-redshift, such a template could be important and is worth examining as a potential systematic effect on the SMFs.  
\newline\indent
In Figures 15, 16, and 17 we plot the SMFs derived using these different modeling assumptions for the quiescent, star forming, and combined galaxy populations, respectively.  The panels are organized by row in increasing redshift bins, and below the SMFs in each redshift bin we plot the residuals compared to the default SMF.  Within those panels the shaded region represents the formal uncertainties in the default SMF.  The data points are the SMFs determined with the 1/V$_{max}$ method, and the solid curves represent the best-fit maximum-likelihood Schechter functions.  The parameters from all SMFs are listed in Table 3.
\newline\indent
Starting with a comparison of the SMFs for the quiescent galaxies in Figure 15, it is clear that overall, the SMFs with the different models are reasonably comparable.  There are some clear differences for given assumptions and in specific ranges of M$_{star}$ and $z$ which stand out in the plots of the residuals.  Near the high-mass end, the residuals in $\Phi^{*}$ can be large, but it is important to note that because we have compared the residuals as a function of $\Phi^{*}$ and not M$_{star}$, these can appear extremely large at the exponential tail where a small change in M$_{star}$ results in a significant change in $\Phi^{*}$.  
\newline\indent
The two parameters that have little effect on the quiescent SMFs are the use of the dusty template, and the different SFH history. Quiescent galaxies are typically quite old, and mostly dust-free, so it is unsurprising that the precise choice of SFH at early times, and the allowance of extra dust are not important in the SED modeling.  
\newline\indent
The allowance of metallicity as a free parameter also has little effect at the high-mass end of the SMF, but does reduce the number density of very low-mass quiescent galaxies.  This has the effect to reduce the significance of the upturn in the quiescent SMF seen at the lowest masses in the low-redshift bins.  
\newline\indent
As expected, the most significant systematic effect on the SMF comes from using the M05 models instead of the BC03 models.  At high-redshift the effect is extremely pronounced, and values of M$_{star}$ are typically a factor of $\sim$ 2 lower.  The effect at higher redshift is expected given that quiescent galaxies are younger there, and the TP-AGB phase is most apparent when stellar populations are 0.5 -- 2.0 Gyr in age (M05).  The effect on the SMF from the SPS models diminishes with redshift; however, quite surprisingly the M05 models still produce lower number densities at the high-mass end at low-redshift.  Given that at old ages the BC03 and M05 models are fairly similar, it suggests that the best-fit M05 models fit substantially lower ages for massive low-redshift galaxies than the BC03 models.  
\newline\indent
In Figure 16 the SMFs for the star forming population are plotted.  Figure 16 shows that similar to the quiescent galaxies, the choice of SFH has little effect on their SMFs.  Allowing metallicity as a free parameter also does not change the SMFs within the random uncertainties.  As expected, including the dusty template does have a noticeable effect at the high-mass end of the star forming SMF at high redshift.  In both the 2.5 $< z <$ 3.0, and the 3.0 $< z <$ 4.0 bins, including the dusty templates reduces the number density of Log(M$_{star}$/M$_{\odot}$) $>$ 11.0 galaxies by 0.2 -- 0.4 dex.  This is fully consistent with \cite{Marchesini2010}, who found approximate half of their sample of such galaxies was consistent with a lower-redshift solution.  If these galaxies truly are very dusty galaxies at lower-redshift, it could alleviate some of the tension between the measured SMF and the predicted SMFs from models at the high-mass end.  It is clear that obtaining spectroscopic redshifts and confirmation of the M$_{star}$ for this population is an important task for future observations.  Unfortunately, as discussed in the text, these galaxies are faint and have extremely red SEDs.  They are usually only well-detected in bands redward of the H-band.  Unless they have very strong emission lines that are not completely obscured, obtaining redshifts will be challenging.
\newline\indent
In Figure 17, the SMFs for the combined population are plotted.  These reflect most of the same trends as were already pointed out in the star forming and quiescent SMFs.  In summary, we find that for the four sources of systematic error we have considered, it appears that two have an impact on the SMFs which is larger than the random uncertainties, whereas the other two do not.  The choice of SFH appears to have little effect on the SMFs of any type.  Furthermore, allowing metallicity as a free parameter does not affect the SMFs in any notable way other than to reduce the size of the upturn of the quiescent SMF at very low M$_{star}$.  This has very little effect on the combined SMF, because star-forming galaxies dominate at low M$_{star}$.  
\newline\indent
The two areas for concern for the SMFs in terms of systematic uncertainties are the potential dusty massive galaxies at high-redshift, and the SPS models.  Because star-forming galaxies dominate the SMF at all M$_{star}$ at $z >$ 2.5, the possibility that some of these galaxies are lower-redshift dusty sources has serious consequences of the high-mass end of the SMF at high-redshift.  Lastly, it has been well-known for several years now that the choice of SPS model is the most significant systematic uncertainty in the determination of M$_{star}$ and the SMF.  Using the best data available at the time, \cite{Marchesini2009} showed that this uncertainty was at least as large as the random uncertainties in the data, and \cite{Marchesini2010} showed that it was the dominant source of error in the NMBS SMF.  With a substantially larger survey like UltraVISTA, it is clear that we are now completely in the regime where the SPS uncertainties dominate the total error budget in the SMFs.  Resolving this issue will be critical in order to better test models of galaxy formation {\it at all redshifts}.
\begin{center}
THE EFFECT OF DIFFERENT SEPARATION BETWEEN STAR FORMING AND QUIESCENT GALAXIES ON THE SMF
\end{center}
\begin{figure}
\plotone{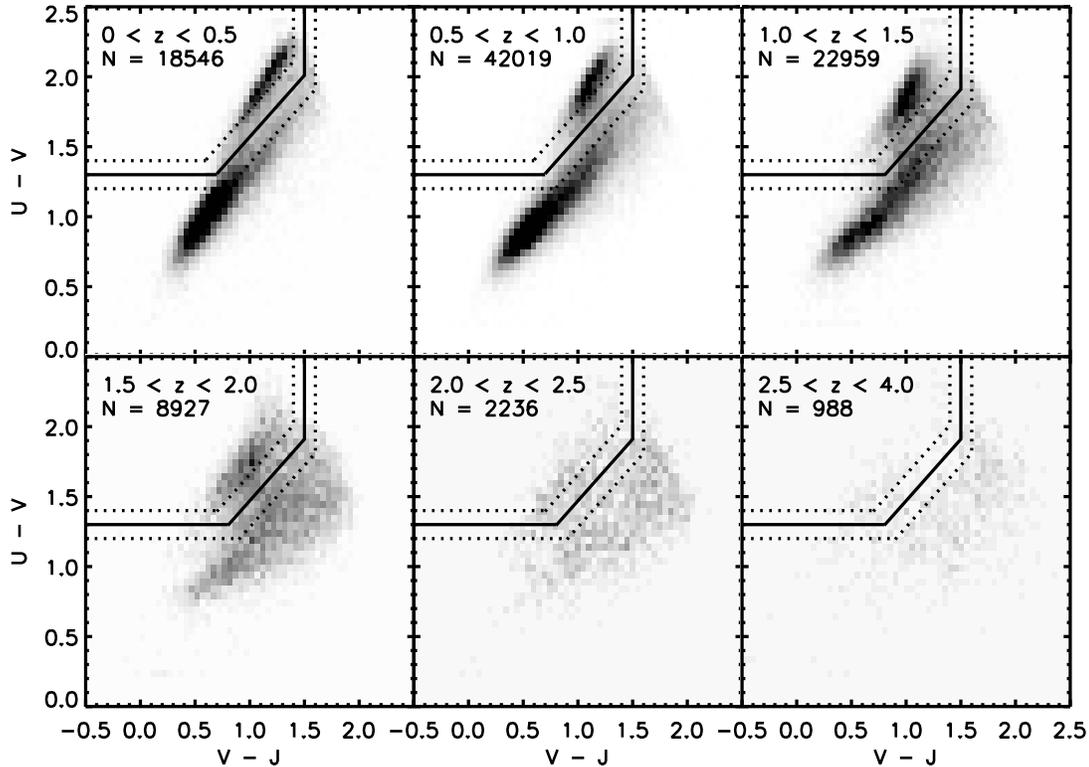}
\caption{\footnotesize UVJ diagram as Figure 3, but with the varied definitions of quiescent galaxies shown as the dotted line.  The dotted lines represent extreme definitions of star-forming and quiescent galaxies, but therefore bracket the full range of possible stellar mass functions for these populations.}
\end{figure}
\begin{figure}
\plotone{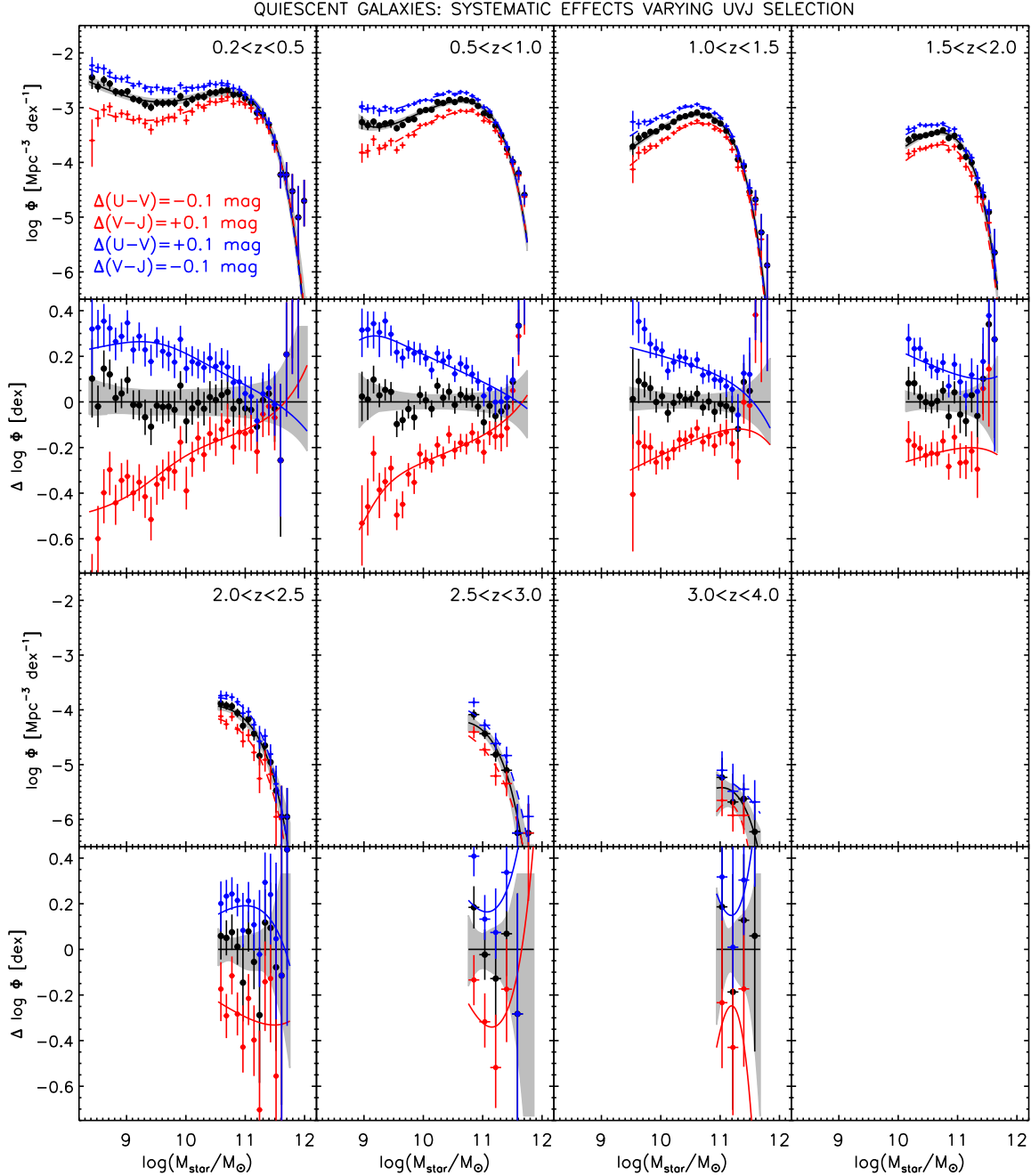}
\caption{\footnotesize As Figure 15 but for quiescent galaxies with two different definitions of star-forming and quiescent galaxies. Varying the selection has a much stronger affect on the derived mass functions for high masses than low masses because high-mass galaxies are the most-quiescent.}
\end{figure}
\begin{figure}
\plotone{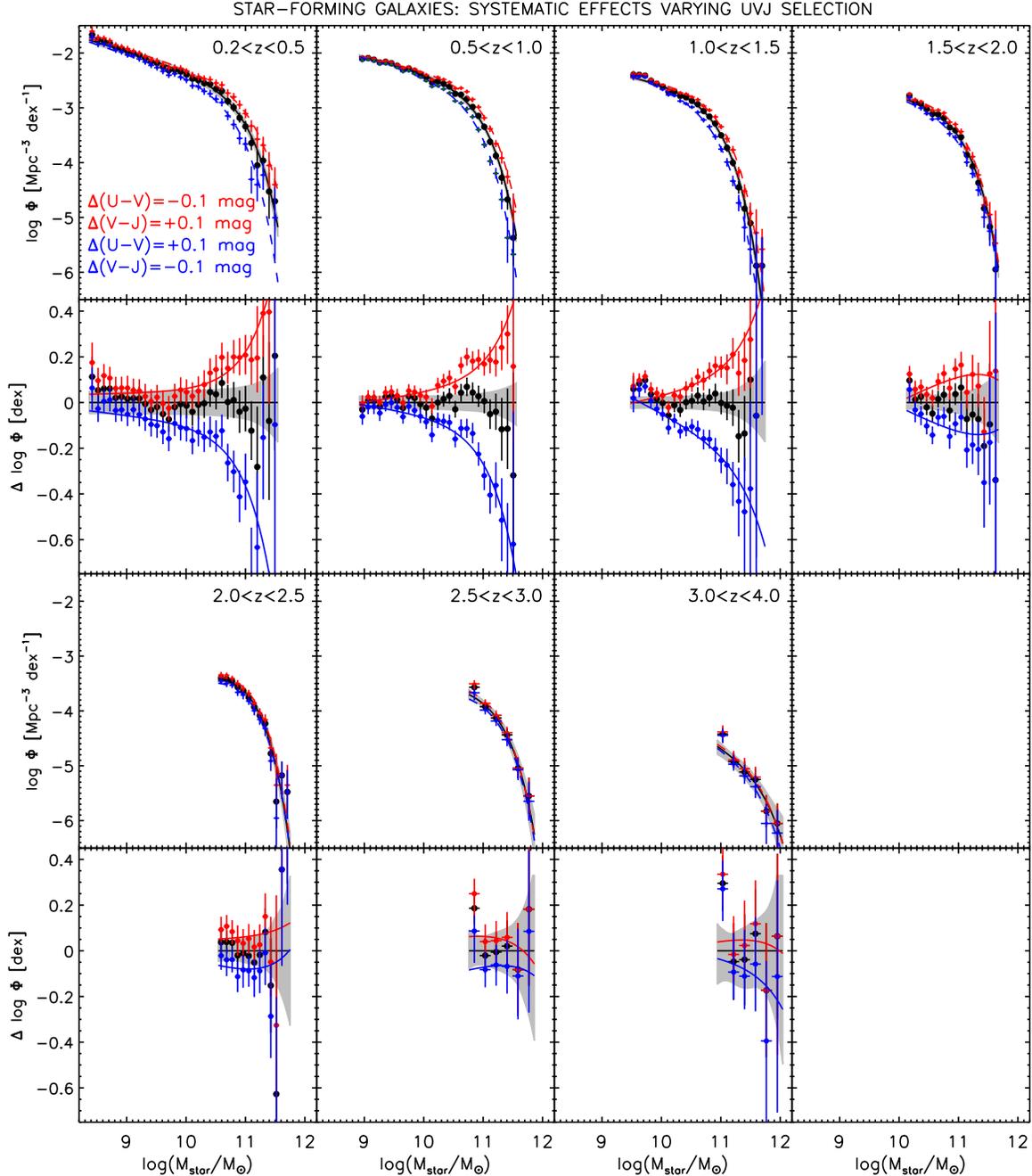}
\caption{\footnotesize As Figure 15 but for star-forming galaxies with two different definitions of star-forming and quiescent galaxies.  The effect of different definitions of quiescent on the star-forming population is opposite to quiescent galaxies.  Low-mass galaxies are not affected because they have the highest specific star formation rates, whereas the most massive star-forming galaxies have much lower specific star formation rates.}
\end{figure}
\begin{figure}
\plotone{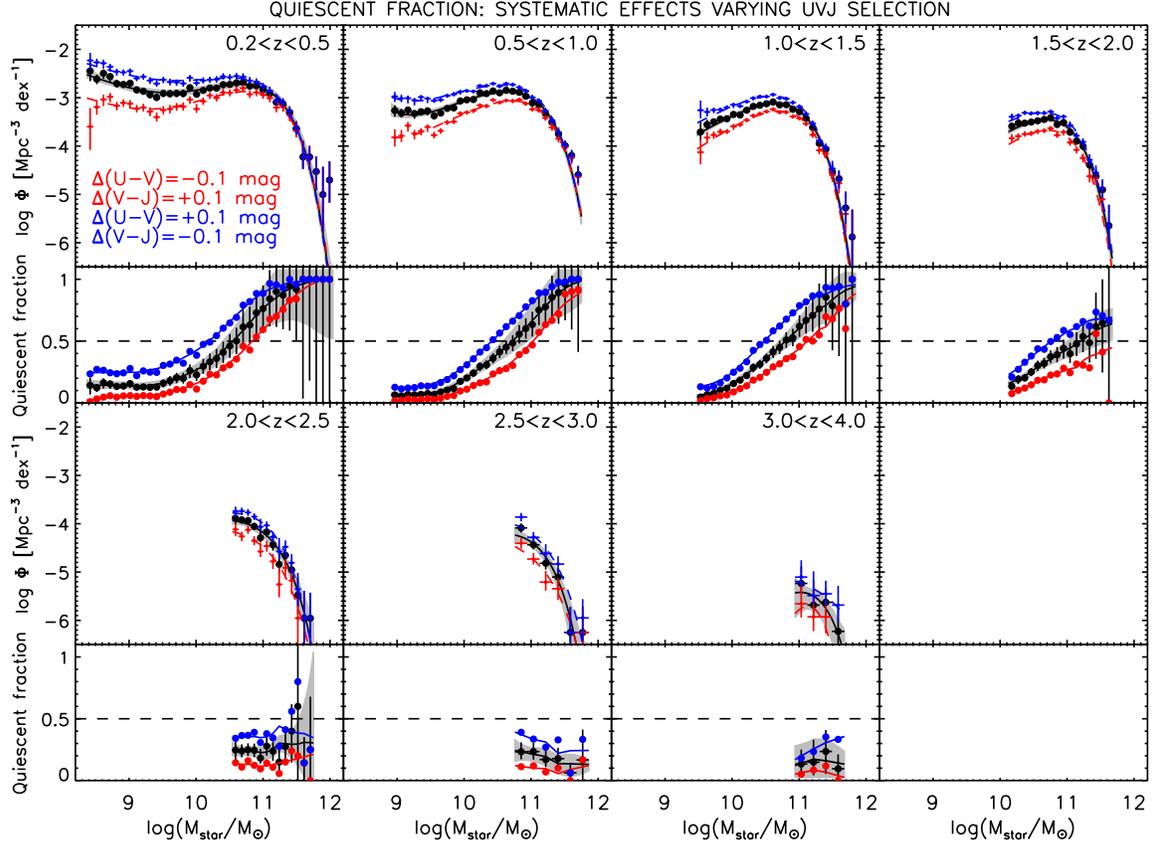}
\caption{\footnotesize As Figures 19 and 20, but in the middle panels the fraction of quiescent galaxies as a function of stellar mass is shown.  Although the absolute fraction of quiescent galaxies changes with the definition of quiescence, the trend that more massive galaxies are more frequently quiescent is robust.  The result that star-forming galaxies dominate the SMF at all masses at $z >$ 2.5 is also robust to the definition of quiescence.}
\end{figure}
\begin{deluxetable}{lccccccc}
\tabletypesize{\scriptsize}
\centering
\tablecaption{Best-fit Schechter Function Parameters of the SMFs: Different UVJ selection\label{smfsys11}}
\tablehead{\colhead{Redshift} & \colhead{Sample} & \colhead{Number} &  
           \colhead{$\log{M^{\star}_{\rm star}}$} & \colhead{$\Phi^{\star}$} & 
           \colhead{$\alpha$} &  \colhead{$\Phi^{\star}_{\rm 2}$} &  \colhead{$\alpha_{\rm 2}$} \\
                     &   &   &  ($M_{\odot}$) & (10$^{-4}$~Mpc$^{-3}$) 
                     &   &  (10$^{-4}$~Mpc$^{-3}$)  &   }
\startdata
$0.2\leq z <0.5$ & Q,UVpVJm  &  2489 &  11.13$\pm$0.03     &  9.86$^{+0.49}_{-0.58}$     & -0.72$\pm$0.03 & \nodata & \nodata \\
$0.2\leq z <0.5$ & Q,UVmVJp  &  6769 &  11.18$\pm$0.03     & 12.38$^{+0.70}_{-0.52}$     & -1.00$\pm$0.02 & \nodata & \nodata \\
$0.2\leq z <0.5$ & Q,UVpVJm  &  2489 &  10.89$\pm$0.01     & 17.30$\pm$0.02      & -0.4              & \nodata & \nodata \\
$0.2\leq z <0.5$ & Q,UVmVJp  &  6769 &  10.64$\pm$0.01     & 48.05$\pm$0.03      & -0.4              & \nodata & \nodata \\
$0.2\leq z <0.5$ & Q,UVpVJm  &  2489 &  10.94$^{+0.05}_{-0.03}$  & 14.56$^{+0.77}_{-1.19}$  & -0.33$\pm$0.11         & 0.24$^{+0.49}_{-0.19}$ & -1.48$^{+0.21}_{-0.26}$ \\
$0.2\leq z <0.5$ & Q,UVmVJp  &  6769 &  10.92$^{+0.05}_{-0.03}$  & 23.55$^{+1.91}_{-1.83}$  & -0.48$\pm$0.11         & 1.29$^{+1.03}_{-0.61}$ & -1.48$\pm$0.11 \\
$0.2\leq z <0.5$ & SF,UVpVJm & 16057 &  10.94$\pm$0.03     & 11.48$^{+0.72}_{-0.69}$ & -1.33$\pm$0.01 & \nodata & \nodata \\
$0.2\leq z <0.5$ & SF,UVmVJp & 11777 &  10.68$\pm$0.04     &  9.86$^{+0.83}_{-0.69}$ & -1.37$\pm$0.02 & \nodata & \nodata \\
$0.2\leq z <0.5$ & SF,UVpVJm & 16057 &  10.89$\pm$0.01 & 13.51$^{+0.15}_{-0.13}$ & -1.3              & \nodata & \nodata \\
$0.2\leq z <0.5$ & SF,UVmVJp & 11777 &  10.58$\pm$0.02 & 13.43$^{+0.14}_{-0.21}$ & -1.3              & \nodata & \nodata \\
\enddata
\tablecomments{This table is available in its full form in the electronic journal.  A subsample of the data is shown here to represent its form.  ``UVpVJm'' corresponds to shifting the UVJ selection box by 
$+$0.1~mag in $U-V$ and $-$0.1~mag in $V-J$; ``UVmVJp'' corresponds to 
shifting the UVJ selection box by $-$0.1~mag in $U-V$ and $+$0.1~mag in $V-J$. 
The listed errors are the 1$\sigma$ Poisson errors.} 
\end{deluxetable}
Although there are now several determinations of the SMFs of star forming and quiescent galaxies in the literature \cite[e.g.,][]{Ilbert2010,Brammer2011,Dominguezsanchez2011,Moustakas2013}, the systematic uncertainties in these SMFs due to the definition of ``quiescence" has not been examined in detail.  Indeed, given that galaxies exhibit a range of SSFRs \citep[e.g.,]{Muzzin2013a}{Noeske2007,Whitaker2012}, and that there is a ``green valley' in the color-magnitude relation, placing galaxies within a binary definition is implicitly an oversimplification of the problem.  Here we test how varying the definition of star forming and quiescent galaxies affects the SMFs of these types.
\newline\indent
In order to provide this test we vary the bounds of the quiescent population within the UVJ diagram by $\pm$ 0.1 mag in both U - V and V - J.  Figure 18 shows an illustration of how the variation of these bounds appear in the UVJ diagram.  As Figure 18 shows, an alteration of $\pm$ 0.1 mag in color is a rather extreme test; however, we have chosen to do so in order to see what the maximum systematic uncertainties will be, and how robust the results from the default model are.
\newline\indent
In Figures 19 and 20 we plot the SMFs generated using the new UVJ definition for the quiescent and star-forming galaxies respectively.  Again, similar to the figures about the assumptions in SED modeling we have included middle panels that show the change in $\Phi^{*}$ relative to the default model.  All SMFs derived with the different UVJ selection are listed in Table 4.
\newline\indent
Examining the SMFs for the quiescent population it is clear that the definition of quiescence is quite important in the determination of the SMF.  Interestingly, it is much more important for galaxies with Log(M$_{star}$/M$_{\odot}$) $<$ 11.0 than galaxies with Log(M$_{star}$/M$_{\odot}$) $>$ 11.0.  This is because the most massive galaxies are the most unambiguously quiescent.  Typically they are the ones with the reddest U - V colors within the quiescent box in the UVJ diagram, and therefore variations in the box do not change their classification or number density.  Figure 19 shows that the change in the number density of galaxies with Log(M$_{star}$/M$_{\odot}$) $<$ 11.0 is a clear function of M$_{star}$, and can reach as high as 0.2 -- 0.4 dex at the lowest masses probed.  This change is a reflection of the colors of this population, which are bluer than the more massive population, and hence their definition as quiescent is more ambiguous.  
\newline\indent
Examination of the SMFs for the star forming population with the different UVJ selection (Figure 20) shows opposite trends as the quiescent population.  The number densities of the lowest-mass galaxies are mostly unchanged, whereas the number densities at the highest masses can change by 0.2 -- 0.6 dex.  This is again a reflection of the fact that the colors of the lowest-mass galaxies are typically very blue, making their classification as star-forming galaxies unambiguous; whereas the most massive star-forming galaxies have more intermediate colors making their classification less clear.  This comparison of SMFs makes clear that there are issues of interpretation when dividing the spectrum of galaxy SSFRs into a binary classification scheme of star forming and quiescent.  It shows that comparison of the observed SMFs of star forming and quiescent galaxies to those from models of galaxy formation would require a careful matching of definitions.
\newline\indent
One result that appears to be robust to the definition of quiescent is the trend of an increasing fraction of quiescent galaxies as a function of M$_{star}$, and a decrease with $z$.  In Figure 21 we again plot the SMFs of quiescent galaxies with the different UVJ selection, but show the quiescence fractions in the middle panels.  No matter how quiescence is defined, the trend that more massive galaxies are more frequently quiescent than lower mass galaxies at $z <$ 2.0 holds.  Furthermore, the fact that this trend seems to disappear, and our conclusion that star-forming galaxies dominate the SMF at all M$_{star}$ at $z >$ 2.5 appears to both hold, no matter the definition of quiescence.  It appears that at the $z >$ 2.5, there are very few galaxies consistent with being quiescent, no matter what the definition.
\begin{center}
COMPARISON TO ILBERT ET AL. STELLAR MASS FUNCTIONS
\end{center}
\begin{figure}
\plotone{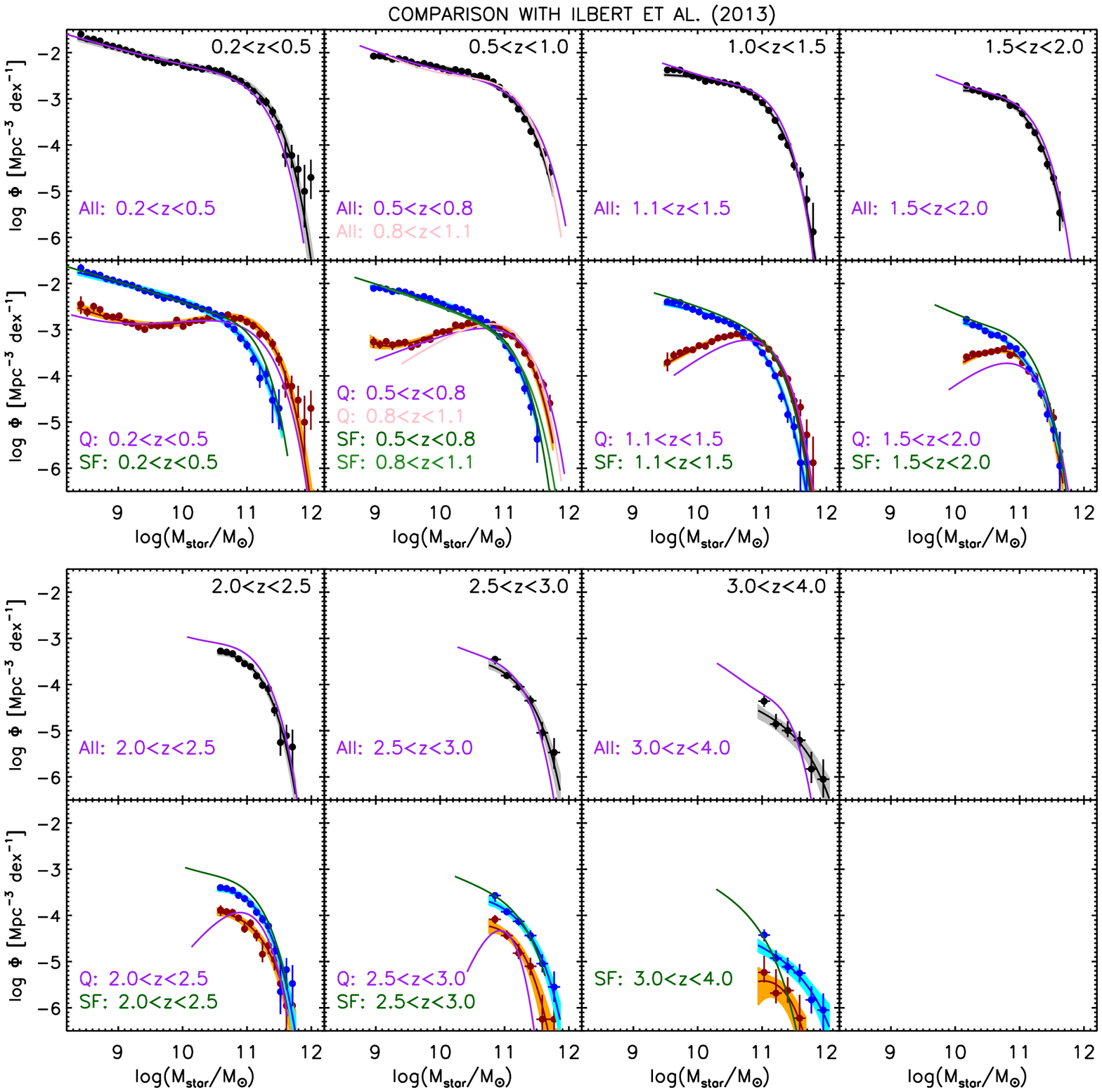}
\caption{\footnotesize Comparison of SMFs of the star-forming, quiescent, and combined populations from this paper to those from \cite{Ilbert2013}.  The shaded regions represent the best-fit maximum-likelihood Schechter functions and the associated uncertainties and the solid points represent the 1/V$_{max}$ SMFs.  The colored curves in each panel are the best-fit maximum-likelihood SMFs from \cite{Ilbert2013}. The combined SMFs show good agreement in most places.  There is some disagreement between the SMFs of the star-forming and quiescent populations, which is likely from the different definitions for these used (see text). }
\end{figure}
\begin{figure}
\plotone{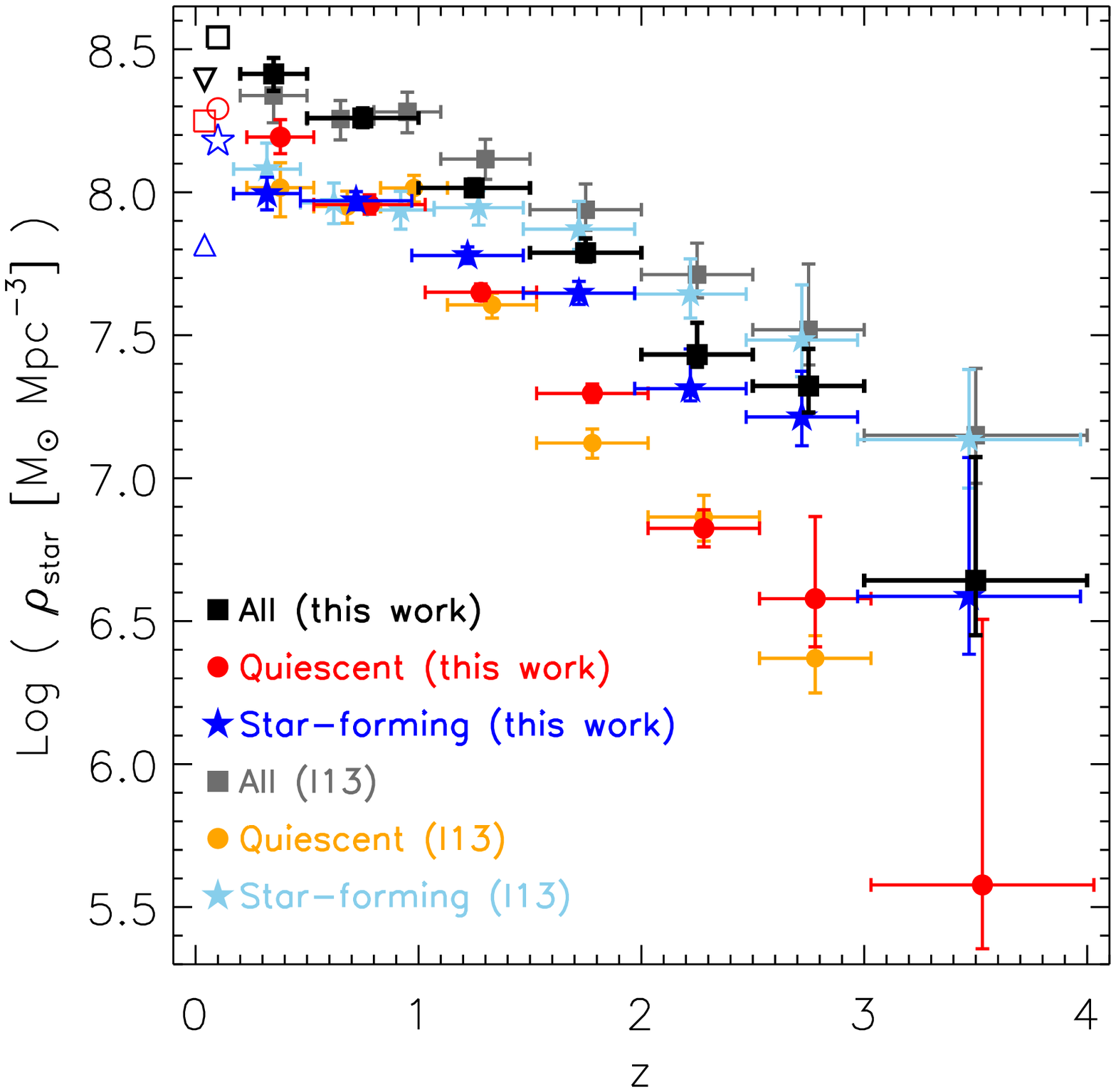}
\caption{\footnotesize Comparison of SMDs of the star-forming, quiescent, and combined populations from this paper to those from \cite{Ilbert2013}.  In general there is reasonable agreement at most redshifts.  Notable exceptions are the SMDs in star-forming at high-redshift which is primarily due to a steeper $\alpha$ in \cite{Ilbert2013}.  There is also some discrepancy in the SMD in quiescent galaxies at low-redshift, and the reason for this is not immediately clear given that the data quality is best for bright galaxies at low redshift and both catalogs agree very well with the zCOSMOS spectroscopic redshifts.}
\end{figure}
Recently Ilbert et al. (2013, hereafter I13) have also computed the SMFs of star-forming and quiescent galaxies up to $z =$ 4 based on an independently-generated 30 band PSF-matched catalog of the COSMOS/UltraVISTA field.  In this appendix we make a comparison between those SMFs and the ones derived in this paper.  In Figure 22 we plot both our SMFs and the I13 SMFs for the total, star-forming, and quiescent populations in the different redshift bins.  In Figure 23 a comparison of the SMDs derived from the SMFs of the various populations as a function of redshift is shown.
\newline\indent
In general there is reasonable agreement between our SMFs and the I13 SMFs for the combined population at all redshifts, particularly at the high-mass end.  There is some tension on the low-mass end slopes, and this can also be seen in the SMDs where the I13 total SMDs are systematically 0.1 -- 0.2 dex higher than our derivation, mostly a result of a slightly steeper low-mass end in I13, but also partially due to slightly higher overall number densities at all masses in several redshift bins.  
\newline\indent
Comparing the SMFs of the star-forming and quiescent galaxies between the surveys shows more mixed agreement than the total SMFs.  In almost all redshift bins the high-mass end of the SMFs of both types do agree reasonably well.  The exceptions to this are the lowest-redshift bin, 0.2 $< z <$ 0.5, and an intermediate-redshift bin, 1.5 $< z <$ 2.0.  The disagreement at 0.2 $< z <$ 0.5 is surprising and it is not obvious what its origin is.  Both catalogs show excellent agreement between the $z_{phot}$ and the zCOSMOS spectroscopic redshifts, which are most complete for high-mass galaxies at low redshift.  The difference could result from the definitions of a quiescent galaxy; however, it is surprising that it should matter as it is at the lowest-redshifts where the definition of a quiescent galaxy is least ambiguous.  A better understanding of this discrepancy will require an object-by-object comparison between catalogs.
\newline\indent
The agreement at the low-mass end of the star-forming and quiescent SMFs is not as good as the high-mass end, with generally the I13 SMFs having shallower $\alpha$ for the quiescent galaxies and steeper $\alpha$ for the star-forming galaxies.  This is most likely the result of the different definitions of a quiescent galaxy between the studies.  I13 define quiescent galaxies using a  NUV - M(r) vs. M(r) - M(J) color-color space, which is similar, although not identical to the UVJ selection that we use.  As we showed in the previous appendix, if we adjust the location of the UVJ box to a more conservative cut we would produce quiescent SMFs that are in better agreement with those from I13.  Likewise, I13 could most likely accommodate our SMFs if they were to move the location of their UV - optical color box to a less conservative cut.  
\newline\indent
Comparison of the SMDs in Figure 23 shows that there is reasonable agreement between the surveys for the quiescent population at all redshifts.  This is because the $\alpha$ is fairly shallow so the SMD is dominated by massive galaxies.  For star-forming galaxies I13 derive SMDs that are a factor of $\sim$ 2 higher than ours at high-redshift.  Again, this is partially due to the steeper $\alpha$ they derive, but is also partially because their overall number densities are slightly higher at all M$_{star}$.
It may not be a surprise that the largest difference is the low-mass-end slopes, as $\alpha$ is always the most difficult part of the SMF to constrain. The low-mass-end slope has frequently been a point of controversy in previous measurements of the SMFs as it is the location where the photometry is of the poorest quality, hence the M$_{star}$ and $z_{phot}$ are the poorest-constrained.  
\newline\indent
Overall, although there are some specific differences, the comparison between the two sets of SMFs and SMDs derived from identical data using different methods shows more consistency than discrepancy up to $z =$ 4.  This shows that there is a reasonable consensus in the SMFs determined with NIR-selected samples, particularly on the high-mass end where the S/N of the photometry is highest.  The comparison does illustrate two outstanding issues in the accuracy of the SMFs that warrant further investigation.  Firstly, it is clear that the definition of star-forming and quiescent galaxies needs to be defined in a careful and consistent way.  A detailed study of both the UVJ diagram and the NUV - optical diagram and the locations of galaxies of a given SSFR in those diagrams would be useful for choosing boundaries in both that not only correspond better to each other, but also correspond to the best-possible separation of star-forming and quiescent galaxies.  Secondly, better measurements of $\alpha$ will be important to resolve differences in the SMDs.  Given that measuring $\alpha$ requires pushing the data to the lowest S/Ns, it is not surprising that discrepancies between measurements at the low-mass end are common in this type of work \citep[e.g., the discussion in][]{Reddy2009}.  One obvious step forward in measuring $\alpha$ will be the DR2 UltraVISTA data which should offer a substantial improvement in S/N for the faintest sources. 
\end{appendix}

\bibliographystyle{apj}
\bibliography{apj-jour,myrefs}




\end{document}